\documentclass[fleqn,usenatbib]{mnras}

\usepackage{newtxtext,newtxmath}

\usepackage[T1]{fontenc}

\DeclareRobustCommand{\VAN}[3]{#2}
\let\VANthebibliography\thebibliography
\def\thebibliography{\DeclareRobustCommand{\VAN}[3]{##3}\VANthebibliography}
\usepackage{enumitem}
\usepackage{graphicx}
\graphicspath{{figures/}}
\usepackage{amsmath}
\usepackage{booktabs}
\usepackage{tabularx}
\usepackage{ragged2e}
\usepackage{array}
\usepackage{float}
\usepackage[table]{xcolor}
\usepackage{bbm} 

\newcommand{\Msun}{\ensuremath{{\rm M}_{\odot}}}
\newcommand{\Zsun}{\ensuremath{{\rm Z}_{\odot}}}
\newcommand{\muv}{\ensuremath{M_{\mathrm{UV}}}}

\newcommand{\sfrten}{\ensuremath{{\rm SFR}_{10}}}
\newcommand{\sfrhun}{\ensuremath{{\rm SFR}_{100}}}

\newcommand{\Nn}{\ensuremath{\mathcal{N}}}
\newcommand{\Un}{\ensuremath{\mathcal{U}}}
\newcommand{\jwst}{\textit{JWST}}
\newcommand{\bhun}{\ensuremath{b_{10-50}}}
\newcommand{\sfrtenfifty}{\ensuremath{{\rm SFR}_{10-50}}}
\newcommand{\mass}{\ensuremath{\log_{10}(M_\ast/\Msun)}}
\newcommand{\ceri}{\textsc{Ceridwen}}

\newcommand{\burst}{\ensuremath{\mathrm{SFR}_{10}/\mathrm{SFR}_{10-50}}}

\usepackage[normalem]{ulem}

\definecolor{andypink}{HTML}{E0218A}%

\renewcommand{\dblfloatpagefraction}{0.8}
\definecolor{revgreen}{HTML}{0A8A3A}%

\renewcommand{\floatpagefraction}{0.8}

\title[What shapes the bright end of the UV luminosity function?]{What shapes the bright end of the UV luminosity function? Stellar mass, star-formation variability, and dust in JADES}
\author[A. Stoffers et al]{Amanda Stoffers,$^{1, 2}$\thanks{E-mail: aas208@cam.ac.uk}
Sandro Tacchella,$^{1,2}$ Benjamin D. Johnson,$^{3}$ Brant Robertson,$^{4}$
\newauthor
Roberto Maiolino,$^{1,2,5}$ Andrew J. Bunker,$^{6}$ Emma Curtis-Lake,$^{7}$
\newauthor A. Lola Danhaive,$^{6}$ Qiao Duan,$^{1,2}$ Francesco D'Eugenio$^{1,2}$
\\\\
$^{1}$Kavli Institute for Cosmology, University of Cambridge, Madingley Road, Cambridge CB3 0HA, UK\\
$^{2}$Cavendish Laboratory, University of Cambridge, 19 JJ Thomson Avenue, Cambridge CB3 0HE, UK\\
$^{3}$Center for Astrophysics | Harvard \& Smithsonian, 60 Garden Street, Cambridge, MA 02138, USA\\
$^{4}$Department of Astronomy and Astrophysics, University of California, Santa Cruz, 1156 High Street, Santa Cruz, CA 95064, USA\\
$^{5}$Department of Physics and Astronomy, University College London, Gower Street, London WC1E 6BT, UK\\
$^{6}$Department of Physics, University of Oxford, Denys Wilkinson Building, Keble Road, Oxford OX1 3RH, UK\\
$^{7}$Centre for Astrophysics Research, Department of Physics, Astronomy and Mathematics, University of Hertfordshire, Hatfield AL10 9AB, UK}

\date{Accepted XXX. Received YYY; in original form ZZZ}

\pubyear{\the\year{}}

\begin{document}
\label{firstpage}
\pagerange{\pageref{firstpage}--\pageref{lastpage}}
\maketitle

\begin{abstract}
JWST surveys find more UV-bright galaxies at $z>10$ than galaxy-formation models predict, but the roles of recent star formation, stellar mass and dust remain uncertain. We analyse 1365 spectroscopically confirmed JADES galaxies at $3\leq z\leq9$, fitting NIRCam photometry and NIRSpec emission lines with \ceri. Our model includes flexible star-formation histories (SFHs) under three priors, two-component dust attenuation and nebular emission. We account for spectroscopic selection through population reweighting and assess the reliability of inferred trends.
At $z=3$--5, rising SFHs are more common in low-mass galaxies, whereas most galaxies above $10^{9.5}\,\Msun$ have declining SFHs. The apparent increase in rising fractions with redshift cannot be distinguished from the adopted SFH prior alone. At fixed stellar mass, recent star formation is the strongest predictor of UV brightness. UV-brighter galaxies have higher recent SFRs, younger stellar populations and less attenuation at fixed SFR. Below $10^9\,\Msun$, they are also more likely to have rising SFHs, with odds increasing by up to a factor of 2.3 per magnitude brighter, although the strongest trend requires caution. In contrast, galaxies with $\muv<-20$ at $z=3$--5 are predominantly massive, and approximately half have declining SFHs. Dust partly offsets burst-driven UV brightening: attenuation increases with recent SFR and declines gradually after the inferred SFR peak. In model sensitivity tests, dust has the largest effect on bright-end counts, whereas removing galaxy-to-galaxy SFH scatter changes them by at most 0.17~dex. Thus, bursts can make low-mass galaxies unusually UV-bright for their mass, while the absolute bright end of the UVLF is primarily shaped by stellar mass and dust attenuation.
\end{abstract}
\begin{keywords}
galaxies: evolution -- galaxies: high-redshift -- galaxies: star formation -- galaxies: statistics
\end{keywords}

\section{Introduction}
\label{sec:introduction}
\jwst\ has revealed an unexpectedly high abundance of UV-bright ($\muv<-20$) galaxies at $z>10$ \citep{Naidu:2022aa, Castellano:2022aa, Finkelstein:2023aa, Harikane:2023aa, Donnan:2023aa, Chemerynska:2024aa, McLeod:2024aa, Robertson:2024aa, Finkelstein:2024aa, Harikane:2025aa, Whitler:2025aa}. Spectroscopically confirmed examples include GN-z11 at $z=10.60$ \citep{Bunker:2023aa}, JADES-GS-z14-0 at $z=14.32$ \citep{Carniani:2024aa, Carniani:2025aa, Schouws:2025aa}, and MoM-z14 at $z=14.44$ \citep{Naidu:2026aa}. The observed UV luminosity function (UVLF), which describes the number density of galaxies as a function of UV luminosity, has a brighter and more slowly evolving bright end than predicted by pre-\jwst\ galaxy-formation models \citep{Naidu:2022aa, Harikane:2023aa, Bouwens:2023aa, Mauerhofer:2023aa, Finkelstein:2024aa, Donnan:2024aa, Whitler:2025aa}. However, this relatively weak evolution does not extend indefinitely: the absence of robust candidates at $z\sim17$ implies a decline in the cosmic UV luminosity density by at least a factor of $\sim50$ between $z\sim10$ and $z\sim17$ \citep{Robertson:2024aa, Weibel:2026aa}, indicating rapid growth of the galaxy population during this interval \citep{McLeod:2026aa}.

Several mechanisms have been proposed to explain the abundance of UV-bright galaxies. These include increased star-formation efficiency at high redshift through feedback-free starbursts \citep{Dekel:2023aa, Li:2024aa}, bursty star formation \citep{Shen:2023aa, Sun:2023ab, Kravtsov:2024aa, Gelli:2024aa}, stochastic halo assembly \citep{Mason:2023aa}, a top-heavy initial mass function (IMF) \citep{Inayoshi:2022aa, Harikane:2023aa, Yung:2024aa}, rapidly rotating massive stars undergoing chemically homogeneous evolution, which emit more UV light per unit stellar mass \citep{Liu:2025aa}, and attenuation-free models in which UV-obscuring dust is ejected after an episode of star formation \citep{Ferrara:2023aa, Ferrara:2024aa, Ferrara:2025aa}. These mechanisms have different implications for stellar mass, recent star-formation history (SFH), and dust attenuation. We therefore investigate how these properties jointly shape UV luminosity at $3\leq z<9$, where the galaxy population is better sampled, and discuss the implications for $z>10$.

An important scaling relation describing the star-forming behaviour of galaxies is the star-forming main sequence (SFMS), which relates stellar mass to star-formation rate (SFR). The SFMS has been studied extensively from the local Universe \citep{Brinchmann:2004aa, Noeske:2007aa, Elbaz:2011aa, Leja:2022aa} to $z\sim4$--6 \citep{Schreiber:2015aa, Speagle:2014aa, Popesso:2023aa}. Predictions of the SFMS at $3<z<9$, based on halo-accretion models and simulations \citep{dekel_toy_2013, Tacchella:2016aa, Tacchella:2018ab}, have now been complemented by large photometric \jwst\ surveys \citep{Cole:2025aa, Rinaldi:2025aa, Simmonds:2025aa}. The scatter around the SFMS, however, provides insight into the mechanisms that drive galaxies away from the mean relation. Low-mass galaxies with $M_\ast<10^9\,\Msun$ are expected to show greater variability because episodic stellar feedback has a stronger influence in their shallow potential wells than in more massive systems \citep{Stinson:2007aa, Hopkins:2014aa, Hopkins:2023aa, Dome:2024aa}. Variability is also expected to increase towards higher redshift, driven by fluctuating gas accretion and shorter equilibrium timescales \citep{Tacchella:2020aa, McClymont:2025aa}, as well as short dynamical times in gas-rich galaxies \citep{Faucher-Giguere:2018aa}. Such variability can temporarily increase the UV luminosity of galaxies and potentially contribute to the observed bright-end excess.

Star-formation variability can only be observed directly over the timescales traced by the available indicators. H$\alpha$ emission traces ionising radiation from short-lived, massive stars, probing the SFR over the past $\sim3$--10~Myr, while the UV continuum is sensitive to longer timescales of $\sim10$--100~Myr \citep{Sullivan:2000aa, Caplar:2019aa, McClymont:2025aa}. Together, these indicators provide a measure of recent changes in SFR. However, converting these observables into SFRs depends on dust attenuation, metallicity, and other assumptions underlying their calibration. Spectral energy distribution (SED) fitting combines information across wavelengths to infer the SFH alongside other physical parameters \citep{conroy_modeling_2013, Leja:2019aa, Johnson:2021aa, Stoffers:2026ab}. While this allows us to investigate recent star formation together with stellar mass, age, and dust attenuation, the inferred SFHs can depend strongly on the adopted forward models and priors, both for recent star formation \citep{Leja:2019aa, tacchella_stellar_2022, Wang:2025aa} and for the older stellar populations it outshines \citep{Whitler:2023aa, Narayanan:2024aa}. Assessing this dependence is therefore essential when interpreting the recovered star-formation variability. The choice between a bursty and a smoother continuity prior can change inferred stellar masses and ages substantially, even when both priors fit the data equally well \citep{Whitler:2023aa, tacchella_stellar_2022}. Where the data constrain the recent SFH only weakly, the inferred fraction of bursting galaxies may therefore reflect the rising prior rather than the data.
Previous \jwst\ studies have investigated star-formation variability using photometric \citep{Asada:2024aa, Endsley:2025aa} and spectroscopic H$\alpha$/UV ratios \citep{Clarke:2024aa}, mini-quenched galaxies \citep{Looser:2024aa, Looser:2025aa}, star-formation duty cycles \citep{Cole:2025aa}, and bursty fractions inferred with flexible SFH models \citep{Ciesla:2024aa, Carvajal-Bohorquez:2025aa}. The fraction of galaxies in an extreme emission-line phase, indicating strong ongoing star formation relative to their assembled stellar mass, also increases with redshift \citep{Boyett:2024aa}. At $z\sim6$, \citet{Endsley:2024aa} find that UV-brighter galaxies more frequently show recent SFR upturns, although an \muv-dependent escape fraction could explain their emission-line trend.
Together, these studies reveal diverse recent SFHs, but the contribution of star-formation variability to the abundance of UV-bright galaxies remains uncertain. In particular, it is unclear whether UV-bright galaxies are predominantly experiencing recent bursts or sustaining elevated SFRs, and how dust attenuation modifies their luminosities. 

In this work, we investigate UV-bright galaxies at $z\geq3$ using NIRCam and NIRSpec observations from the \textit{JWST Advanced Deep Extragalactic Survey} (JADES) \citep{ Eisenstein:2026aa}. We infer their recent SFHs under three priors with different degrees of burstiness and quantify how strongly the data constrain them through the information gain relative to each prior, extending previous studies that adopted a fixed prior \citep{Ciesla:2024aa, Cole:2025aa} or selected among models with different burstiness \citep{Carvajal-Bohorquez:2025aa}. Accounting for selection effects and measurement reliability, we then assess the contributions of stellar mass, recent star formation, and dust attenuation to UV luminosity and the bright end of the UVLF.

The paper is organised as follows. \S~\ref{sec:data} describes the data and sample, and \S~\ref{sec:method} presents the modelling and analysis methods. Sections~\ref{sec:results} and \ref{sec:dust} present our results, followed by the discussion (\S~\ref{sec:discussion}) and conclusions (\S~\ref{sec:summary}). The appendices define the derived quantities (Appendix~\ref{app:derived}) and give details of the model implementation (Appendix~\ref{app:implementation}), the fit validation and robustness tests (Appendix~\ref{app:validation}), the sample (Appendix~\ref{app:sample}), the stellar mass--UV magnitude relation (Appendix~\ref{app:mass_muv}), and a test for a top-heavy IMF (Appendix~\ref{sec:imf}). We assume a flat $\Lambda$CDM cosmology with $H_0 = 67.4~{\rm km\,s^{-1}\,Mpc^{-1}}$ and $\Omega_{\rm m} = 0.315$ \citep{Planck-Collaboration:2020aa}.

\section{Data and sample}\label{sec:data}
We analyse galaxies from the JADES \citep{Rieke:2020aa, Bunker:2020aa, Eisenstein:2026aa} Data Release 5 (DR5) photometric catalogue \citep{Robertson:2026aa} and Data Release 4 (DR4) spectroscopic catalogue \citep{Curtis-Lake:2026aa, Scholtz:2026aa}. The reduction, calibration, astrometric alignment, and mosaicking of the DR5 NIRCam imaging are described in \citet{Johnson:2026aa} including the subtraction of scattered-light wisps \citep{Wu:2026aa}, and the processing of the spectroscopic data in \citet{Scholtz:2026aa}. Many of the spectra and redshifts were first presented in the earlier NIRSpec releases DR1 \citep{Bunker:2024aa} and DR3 \citep{DEugenio:2025aa}. We select 4948 galaxies in the two fields of the Great Observatories Origins Deep Survey \citep[GOODS-South and GOODS-North;][]{Giavalisco:2004aa} entries in both catalogues.
The NIRCam imaging covers filters from F070W to F480M. Fluxes are measured on images PSF-matched to the F444W band to mitigate colour biases arising from wavelength-dependent resolution. Given the complex morphologies of high-redshift galaxies, we use Kron convolved-aperture fluxes and uncertainties (\texttt{\{Filter\}\_KRON} and \texttt{\{Filter\}\_KRON\_e}) rather than circular apertures; \citet{Robertson:2026aa} describe the PSF matching and aperture photometry. For emission lines, we use the R1000 grating fluxes from the DR4 catalogue (5-pixel extraction; \citealt{Scholtz:2026aa}). We fit all 18 NIRCam filters with valid Kron fluxes: F070W, F090W, F115W, F150W, F162M, F182M, F200W, F210M, F250M, F277W, F300M, F335M, F356W, F410M, F430M, F444W, F460M and F480M. The emission lines are C\,\textsc{iv}\,$\lambda\lambda1548,1551$, He\,\textsc{ii}\,$\lambda1640$, O\,\textsc{iii}]\,$\lambda\lambda1661,1666$, C\,\textsc{iii}]\,$\lambda\lambda1907,1909$, [Ne\,\textsc{iv}]\,$\lambda2424$, [O\,\textsc{ii}]\,$\lambda\lambda3726,3729$, [Ne\,\textsc{iii}]\,$\lambda3869$, H$\delta$, H$\gamma$, [O\,\textsc{iii}]\,$\lambda4363$, He\,\textsc{ii}\,$\lambda4686$, H$\beta$, [O\,\textsc{iii}]\,$\lambda5007$, He\,\textsc{i}\,$\lambda5876$, [O\,\textsc{i}]\,$\lambda6300$, H$\alpha$, [N\,\textsc{ii}]\,$\lambda6584$, [S\,\textsc{ii}]\,$\lambda\lambda6716,6731$, [S\,\textsc{iii}]\,$\lambda\lambda9069,9532$, Pa$\delta$, He\,\textsc{i}\,$\lambda10830$, Pa$\gamma$ and Pa$\beta$. The C\,\textsc{iv}, O\,\textsc{iii}], C\,\textsc{iii}] and [O\,\textsc{ii}] doublets are reported as single fluxes and fitted as such. We fit all reported R1000 lines and all photometric bands with finite fluxes and positive uncertainties, including bands with zero or negative measured flux.
We limit our sample to a more reliable subset, based on the following criteria. We only include galaxies with a spectroscopic redshift quality flag of \texttt{A}, \texttt{B}, or \texttt{C} \citep{Curtis-Lake:2026aa}, and restrict our sample to $z \geq 3$, below which aperture effects become too large and complex to mitigate with simple slit-loss scaling, leaving 1495 galaxies. Before fitting, we exclude all known AGN \citep{Juodzbalis:2026aa, Scholtz:2025ab} and additional AGN candidates identified using the diagnostics described in \citet{Mazzolari:2024aa} (Appendix~\ref{app:sample}, Fig.~\ref{fig:agn_diagnostics}). We further identify and remove little red dots (LRDs) from our sample based on the compact-red selection criteria of \citet{Labbe:2025aa, Greene:2024aa}, together with the photometrically selected LRDs of \citet{Rinaldi:2026aa} flagged in the NIRCam parent catalogue of \citet{Duan:2026aa}. This removes 107 AGN and LRD candidates, leaving 1388 galaxies.
After fitting (\S~\ref{sec:method}), we remove 5 galaxies for which the observed flux blueward of Ly$\alpha$ exceeds the model prediction by $\geq5\sigma$. At these redshifts ($z\geq6$), the intergalactic medium (IGM) should absorb nearly all of this flux, so an excess may indicate an incorrect redshift, contamination from a neighbouring source, or an unusually transparent sightline through the IGM, as reported for one JADES galaxy at $z=7.57$ \citep{Helton:2026aa}. That galaxy (NIRCam ID 217704) is not among the five: it has no Kron flux in our catalogue and is not fitted. We also remove 18 galaxies with poor continuum fits ($\chi^2_{\rm cont}/N_{\rm cont} \geq 4$), evaluated only in bands free of strong emission lines. Applying the quality cut to $\chi^2_{\rm cont}$ rather than $\chi^2_{\rm phot}$ reduces the risk of preferentially excluding young, bursty, UV-bright galaxies. 

This leaves us with a sample of 1365 galaxies at $3\leq z\leq14.2$, 21 of them at $z\geq9$. Compared with the full NIRCam parent catalogue of \citet{Duan:2026aa}, the sample occupies the more massive, line-bright tail of the population (Appendix~\ref{app:sample}, see Figs.~\ref{fig:sample_histograms} and \ref{fig:mass_z}). We aim to minimise these selection biases in \S~\ref{sec:muv_completeness}, and acknowledge that we cannot fully correct for them.

\begin{table*}
\centering
\caption{Parameters and priors of the fiducial \texttt{bursty} MIST/MILES model. $\Nn(\mu,\sigma)$ is a normal distribution truncated to the quoted interval, $\Un$ a uniform, $\mathcal{LN}$ a log-normal and $\mathcal{T}_\nu$ a Student's-$t$ distribution with $\nu$ degrees of freedom. The prior variants are listed in Table~\ref{tab:variants}.}

\label{tab:sfms12}

\footnotesize
\setlength{\tabcolsep}{3pt}
\renewcommand{\arraystretch}{0.95}

\begin{tabularx}{\textwidth}{
    @{}
    >{\RaggedRight\arraybackslash}p{2.5cm}
    >{\RaggedRight\arraybackslash}p{3.0cm}
    >{\centering\arraybackslash}p{0.75cm}
    >{\RaggedRight\arraybackslash}X
    >{\RaggedRight\arraybackslash}p{3.0cm}
    @{}
}
\toprule
\textbf{Parameter}
    & \textbf{Description}
    & \textbf{Free?}
    & \textbf{Prior / value}
    & \textbf{Reference} \\
\midrule

\multicolumn{5}{@{}l}{
    \cellcolor{gray!15}
    \textit{Star-formation history}
    (non-parametric; 12 age bins, 11 free ratios)
} \\
\addlinespace[2pt]

$\delta_i=\log_{10}(\mathrm{SFR}_i/\mathrm{SFR}_{i+1})-\mu_i$,
$i=1,\ldots,11$
    & Adjacent-bin SFR ratios, sampled as residuals about the rising-SFMS expectation (\S~\ref{sec:sfhprior})
    & free
    & $\delta_i\sim\mathcal{T}_{\nu=2}(\mathrm{loc}=0,\ \mathrm{scale}=\sigma)$;
      the physical ratio is $\delta_i+\mu_i(\log M_\ast\,[,z])$ inside the SFH
      transform, which is identical to $\mathcal{T}_{\nu=2}(\mathrm{loc}=\mu_i)$.
      The location $\mu_i$ is obtained by integrating the \citet{Duan:2026aa} SFMS
      (their eq.~2) backwards from $(z_{\rm obs},M_\ast)$ to $z=30$, binning on
      the grid below and taking the mean SFR per bin, then adjacent log ratios
      (clipped to $[-5,5]$). The fiducial fit is \texttt{bursty} ($\sigma=0.5$);
      the other widths are the \texttt{continuity} ($0.3$) and
      \texttt{hyperburst} ($1.0$) variants (Table~\ref{tab:variants}).
      The residual prior is unbounded, both in the nested-sampling log-prior and for the initial live points; only $\mu_i$ is clipped.
    & \cite{Leja:2019aa} ($\sigma=0.3$);
      \cite{Duan:2026aa} ($\mu$, $\sigma=0.5$); \cite{Danhaive:2025aa} ($\sigma=0.5$);
      \cite{tacchella_stellar_2022}, \cite{Tacchella:2023aa} ($\sigma=1.0$) \\

\addlinespace
Age bins (lookback)
    & Bin edges
    & fixed
    & Six fixed young bins $0$--$3$, $3$--$5$, $5$--$10$, $10$--$15$,
      $15$--$30$, $30$--$50$\,Myr, followed by six bins
      logarithmically spaced between $50$\,Myr and
      $t_{\rm univ}(z_{\rm obs})-t_{\rm univ}(z=30)$.
      
    & \\

\addlinespace
$\log_{10}(M_\ast/\Msun)$
    & Total stellar mass formed
    & free
    & $\Un[6.0,\,12.5]$
    & flat in $\log M_\ast$ \citep{leja_deriving_2017}; bounds this work \\

\midrule

\multicolumn{5}{@{}l}{\cellcolor{gray!15}\textit{Stellar population}} \\
\addlinespace[2pt]

$\log_{10}(Z_\ast/\Zsun)$
    & Stellar metallicity (MIST grid label; sampled as $\log_{10}(Z_\ast/\Zsun)$)
    & free
    & $[{\rm Fe/H}]=\log_{10}(Z_\ast/\Zsun)\sim\Un[-2.5,\,0.5]$, the MIST/MILES grid range, with $\Zsun=0.0185$ (the grid's solar normalisation).
      The \texttt{bursty\_mzr} variant instead couples $Z_\ast$ to $M_\ast$
      (Table~\ref{tab:variants})
    & \\

\addlinespace
SSP / IMF
    & MIST isochrones + MILES stellar library via FSPS;
      \citet{Chabrier:2003aa} IMF, $0.08$--$120\,\Msun$
    & fixed
    & 
    & \cite{choi_mesa_2016, Dotter:2016aa, Sanchez-Blazquez:2006aa,
      conroy_propagation_2009, Chabrier:2003aa}
    \\

\addlinespace
$z$ 
    & Redshift
    & per flag
    & Flags A/B/C: fixed at $z_{\rm spec}$.
      Flags D/E: free, $\Un[0,\,20]$ (a tentative D redshift only seeds
      the sampler); the SFMS $\mu_i$ is then a 2-D $(z,\log M_\ast)$
      table interpolated per sample and the age bins follow the
      sampled $z$.
    & JADES NIRSpec flags:
        \cite{DEugenio:2025aa, Scholtz:2026aa}\\

\midrule

\multicolumn{5}{@{}l}{
    \cellcolor{gray!15}
    \textit{Nebular emission}
    (CLOUDY tables for MIST SSPs \citealt{Byler:2017aa}), \quad Grid:
    $\log(Z_{\rm gas}/\Zsun)\in[-1.98,0.2]$,
    $\log U\in[-4,-1]$,
    SSP age $0.5$--$20$\,Myr
} \\

\addlinespace[2pt]

$\log_{10}(Z_{\rm gas}/\Zsun)$ 
    & Gas-phase metallicity
    & free
    & $\Un[-1.3,\,0.2]$, truncated at the upper edge of the CLOUDY grid
    & \cite{Byler:2017aa} (upper bound = grid edge) \\

\addlinespace
$\log_{10} U$
    & Ionisation parameter
    & free
    & $\Un[-4.0,\,-1.0]$
    & \cite{tacchella_stellar_2022} \\

\addlinespace
$s_{\rm line}$
    & Emission-line (slit/aperture) scaling
    & free\textsuperscript{a}
    & $\mathcal{LN}$ with median $1.0$ and
      $\sigma_{\ln}=0.2$
    & This work (slit-loss calibration) \\

\addlinespace
$f_{\rm esc}$ (\texttt{frac\_obrun})
    & Escaping ionising fraction
    & fixed
    & $0$ in all fits
    & \\

\midrule

\multicolumn{5}{@{}l}{
    \cellcolor{gray!15}
    \textit{Dust attenuation}
    (Charlot \& Fall 2000 two-component model)
} \\
\addlinespace[2pt]

$\tau_V$
    & Diffuse ISM optical depth
      (Prospector \texttt{dust2} equivalent)
    & free
    & $\Nn(0.3,\,1.0)$, truncated to $[0,\,4]$
    & \cite{charlot_simple_2000};
      \cite{tacchella_stellar_2022} (Table~2)\\

\addlinespace
$n$
    & Kriek \& Conroy (2013) slope offset of the diffuse curve
    & free
    & $\Un[-1.0,\,0.4]$
    & \cite{kriek_dust_2013};
       \cite{tacchella_stellar_2022} (Table~2)\\

\addlinespace
$\tau_{\rm BC}/\tau_V$
    & Birth-cloud to diffuse optical-depth ratio
      (SSP ages $<10^{-1.97}$\,Gyr $=10.7$\,Myr)
    & free
    & $\Nn(1.0,\,0.3)$, truncated to $[0,\,2]$
    &\cite{tacchella_stellar_2022} (Table~2, Sec.~3.1)\\

\addlinespace
$\alpha_{\rm BC}$
    & Birth-cloud power-law slope,
      $\tau_{\rm BC}\propto\lambda^{\alpha_{\rm BC}}$
    & fixed
    & $-1$ in all fiducial fits; sampled with $\Un[-2,0]$ in the
      \texttt{bursty\_freealpha} variant (Table~\ref{tab:variants}).
    & \cite{tacchella_stellar_2022}, Sec.~3.1;
      FSPS \texttt{dust1\_index} \\

\midrule

\multicolumn{5}{@{}l}{\cellcolor{gray!15}\textit{Other}} \\
\addlinespace[2pt]

IGM
    & Madau (1995) attenuation
      (FSPS \texttt{igm\_absorb} port)
    & fixed
    & \texttt{igm\_factor} $=1.0$
    & \cite{madau_radiative_1995}; FSPS \\
\addlinespace
$\sigma_{\rm gal}$
    & Velocity dispersion of the model spectrum
    & fixed
    & $300$\,km\,s$^{-1}$
    & \\
\addlinespace
Cosmology
    & Flat $\Lambda$CDM
    & fixed
    & $H_0 = 67.4$\,km\,s$^{-1}$\,Mpc$^{-1}$, $\Omega_{\rm m} = 0.315$, $T_{\rm CMB} = 2.726$\,K
    & \cite{Planck-Collaboration:2020aa} for $H_0$ and $\Omega_{\rm m}$ \\

\bottomrule
\end{tabularx}
\vspace{2pt}\raggedright\scriptsize \textsuperscript{a}Not sampled for the 197 galaxies without line measurements (18 free parameters).
\end{table*}

\begin{figure*}
    \centering
    \includegraphics[width=0.97\linewidth]{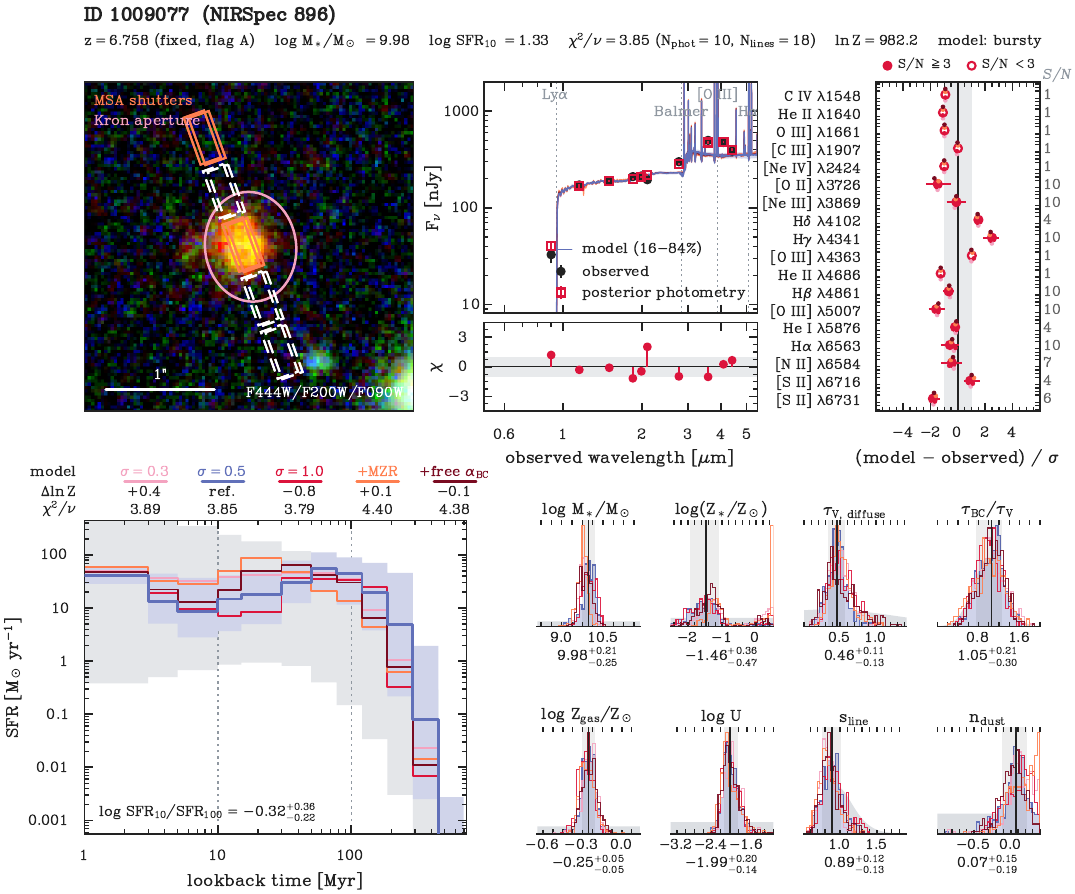}%
    \caption{Example fit of NIRCam ID 1009077 (NIRSpec 896, $z = 6.758$) under the five prior variants of Table~\mbox{\ref{tab:variants}}: \texttt{continuity} ($\sigma = 0.3$, pink), the fiducial \texttt{bursty} ($\sigma = 0.5$, blue), \texttt{hyperburst} ($\sigma = 1.0$, red), \texttt{bursty\_mzr} (coral) and \texttt{bursty\_freealpha} (dark red). \textbf{Top left:} F444W/F200W/F090W colour-composite image ($3 \times 3$~arcsec) with the MSA shutters of the target (coral), neighbouring shutters (white dashed) and the Kron aperture (pink). \textbf{Top centre:} observed photometry (black; errors $\sigma_{\rm eff}$, including the error floor of Appendix~\mbox{\ref{app:implementation}}), fiducial posterior model spectrum (blue: median and 16--84~per~cent), posterior photometry (red squares) and median spectra of the other variants; the NIRSpec spectrum itself is not shown, because only its emission-line fluxes enter the fit. Dotted lines mark Ly$\alpha$, [O\,\textsc{iii}]$\lambda5007$, H$\alpha$ and the Balmer limit. The lower subplot displays the pulls $\chi = (F_{\rm model} - F_{\rm obs})/\sigma_{\rm eff}$ of the photometric data. \textbf{Top right:} emission-line pulls, with the red point showing the fiducial model and the smaller points the other model variants. Filled points have an observed ${\rm S/N} \geq 3$. \textbf{Bottom left:} SFH against lookback time: fiducial posterior (blue, median and 16--84~per~cent), rising main-sequence prior at the posterior-median mass (grey, 16--84~per~cent) and the other variants (coloured lines). The dotted lines mark 10 and 100~Myr in lookback time. The table above shows the priors of each model and how their evidence for the galaxy changed with respect to the fiducial \texttt{bursty} model ($\Delta\ln Z$), as well as the $\chi^2/\nu$ per variant. \textbf{Bottom right:} marginal posteriors (blue: fiducial, with median and 16--84~per~cent in black; outlines: other variants) over their priors (grey). As for about half of the galaxies, the five variants fit this galaxy about equally well ($|\Delta\ln Z| < 1$).}
    \label{fig:example_SED}
\end{figure*}

\section{Modelling, selection and reliability}\label{sec:method}
We use the SED-fitting code \ceri\ \citep{Stoffers:2026ab} to fit the NIRCam photometry and NIRSpec emission-line fluxes simultaneously, following \citet{Tacchella:2023aa}. Five model variants are tested, which are summarised in Table~\ref{tab:variants} in the Appendix. All fits use the \textsc{MIST} isochrones \citep{choi_mesa_2016, Dotter:2016aa} and the \textsc{MILES} stellar library \citep{Sanchez-Blazquez:2006aa}, with a fixed \citet{Chabrier:2003aa} initial mass function (IMF) within a simple stellar population (SSP) grid based on \textsc{FSPS} 
\citep{conroy_propagation_2009}. The whole forward model, including its free parameters and their priors, is summarised in Table~\ref{tab:sfms12}.

We fit for stellar mass, stellar metallicity, and dust attenuation, which is parametrised using a two-component dust model following \citet{charlot_simple_2000}: a diffuse component with a variable slope linked to the strength of the 2175-Å bump strength \citep{kriek_dust_2013} affecting all stars, and a birth-cloud component attenuating only the youngest ($\lesssim10$~Myr) stars. Nebular continuum and emission lines are modelled using the dust-free grids computed by \citet{Byler:2017aa} for the \textsc{MIST} SSPs, as distributed with \textsc{FSPS} (commit \texttt{cbcd0ee}, which includes the updated metallicity and age gridpoints of the MIST nebular line tables; python-\textsc{fsps} v0.5.0), 
We fit for the ionisation parameter, gas-phase metallicity, and a global line scaling factor, $s_{\rm line}$, to correct for slit losses. 
We assume that stars are the only ionising source and that no ionising photons escape, so the predicted emission-line fluxes reflect the full ionising output of the modelled stellar population. Allowing for a non-zero escape fraction would increase the inferred recent SFR, but both the increase and the inferred escape fraction depend strongly on the assumed escape geometry \citep{Stoffers:2026aa}. We fix the redshift to its spectroscopic value and apply redshift-dependent IGM attenuation following \citet{madau_radiative_1995}. Further details of the metallicity priors and birth-cloud attenuation slope are provided in Appendix~\ref{app:implementation}.

\subsection{Fitting the star-formation history under the SFMS prior}
\label{sec:sfhprior}

We adopt a non-parametric, piecewise constant SFH with $N=12$ lookback age bins, with a constant star-formation rate $\mathrm{SFR}_i$ in bin $i$. We do not sample $\mathrm{SFR}_i$ directly, instead we sample the eleven adjacent-bin log ratios, following \citet{Leja:2019aa} \begin{equation}
    r_i \;=\; \log_{10}\!\left(\frac{\mathrm{SFR}_i}{\mathrm{SFR}_{i+1}}\right),
    \qquad i = 1,\ldots,N-1,
    \label{eq:ratios}
\end{equation}
\noindent
where $r_1$ describes the most recent ratio. The ratios only determine the shape of the SFH, the overall normalisation is set by the total stellar mass.
We set the prior to follow the expected shape of the SFH if the galaxy had grown along the SFMS, rising towards the time of observation \citep{Turner:2025aa, Duan:2026aa}. We show an example of the prior in the SFH subplot in Fig.~\ref{fig:example_SED}. The prior centre $\mu_i$ is built by integrating the SFMS backwards in time, following the parametrisation of \citet{Duan:2026aa} (their eq.~2, a \citealt{Schreiber:2015aa} form):
\begin{align}
\log_{10}\!\mathrm{SFR}_{\rm MS}(m,z)
    &= m - m_0 + a_0 d \nonumber\\
    &\quad
    - a_1 \left[\max\!\left(0,\,m-m_1-a_2 d\right)\right]^2,
    \label{eq:sfms}\\[3pt]
m &\equiv \log_{10}\!\left(\frac{M_\ast}{\Msun}\right)-9,
\qquad
d \equiv \log_{10}(1+z),
\end{align}

\noindent with $(m_0,a_0,a_1,a_2,m_1) = (0.5,\,2.0,\,0.3,\,2.5,\,0.4).$ Then, starting from the observed $z_\mathrm{obs} and sampled  M_\ast$ pair, we step backwards in redshift, and subtract at each step the mass formed based on the SFMS, of the running mass, until the galaxy has formed all its mass and the SFR falls to zero. We then average this track over our age bins, and find the location vector $\mu_i(\log M_\ast, z)$ by taking the adjacent log ratios (Eq.~\ref{eq:ratios}). Below the bend of Eq.~(\ref{eq:sfms}) $\mu_i$ is independent of mass, because the normalisation cancels in the ratios.

To simplify the prior geometry for the nested slice sampler \citep{Yallup:2026aa}, we re-parametrise the prior. The rising SFMS depends on the sampled stellar mass and redshift; we instead parametrise the SFR ratios in terms of residuals, $\delta_i \equiv r_i - \mu_i(\log M_\ast,z)$, with an independent, zero-centred prior
\begin{equation}
    \delta_i \sim \mathrm{StudentT}\left(0,\sigma,\nu=2\right).
    \label{eq:resid}
\end{equation}
  
Sampling these residuals is equivalent to the Student's-$t$ prior of \citet{Duan:2026aa}, centred on $\mu_i(\log M_\ast,z)$, for $r_i = \mu_i + \delta_i$.
The main-sequence expectation is given by $\delta_i \simeq 0$, positive and negative values correspond to enhanced and suppressed star formation. The heavy tails of the Student-$t$ prior with $\nu=2$ allow strong deviations where the data require them. To avoid integrating Eq.~(\ref{eq:sfms}) at every likelihood evaluation, we precompute $\mu_i$ on a grid \citep{Duan:2026aa}.

\subsection{Three burstiness models}
We explore the effect of three SFH priors. Each prior is a Student's-$t$ distribution with $\nu=2$, location $\mu_i$ and scale $\sigma$ (Eq.~\ref{eq:resid}); our three priors share $\mu_i$ and differ only in $\sigma$, with larger values allowing for greater deviations from the SFMS and hence more variable star formation. We adopt the widths of the continuity prior of \citet{Leja:2019aa} ($\sigma=0.3$, \texttt{continuity}), the SFMS prior of \citet{Duan:2026aa} ($\sigma=0.5$, \texttt{bursty}), and the ``bursty continuity'' prior of \citet{tacchella_stellar_2022} ($\sigma=1.0$, \texttt{hyperburst}), but keep the SFMS-based location $\mu_i$ for all three; the original continuity and bursty-continuity priors are centred on $\mu_i=0$, i.e.\ a constant SFH. Because our age bins differ from theirs, the same $\sigma$ does not imply the same prior on SFR variability.

The Bayesian evidence disfavors the widest prior: \texttt{hyperburst} has lower evidence than \texttt{bursty} for 78~per~cent of galaxies (median $\Delta\ln\mathcal{Z}=-0.54$). We therefore restrict the choice of fiducial prior to \texttt{continuity} and \texttt{bursty}, retaining \texttt{hyperburst} to test sensitivity to the prior width. The median per-galaxy difference in log Bayesian evidence between \texttt{continuity} and \texttt{bursty} is only $\Delta\ln\mathcal{Z}=+0.24$, providing no strong preference for either model. The two priors also recover the H$\alpha$ flux about equally well: \texttt{bursty} performs better for 54~per~cent of galaxies with constant or declining SFHs and 48~per~cent of those with rising SFHs. We adopt \texttt{bursty} as our fiducial prior for all galaxies. Across the three priors, the inferred $\log{\rm SFR}_{10}$ shifts by $-0.10$ to $+0.09$~dex and \bhun\ by $-0.24$ to $+0.42$ (16th--84th percentiles). We report the main results for all three priors in Table~\ref{tab:robustness}.
\begin{figure*}
    \centering
    \includegraphics[width=\textwidth]{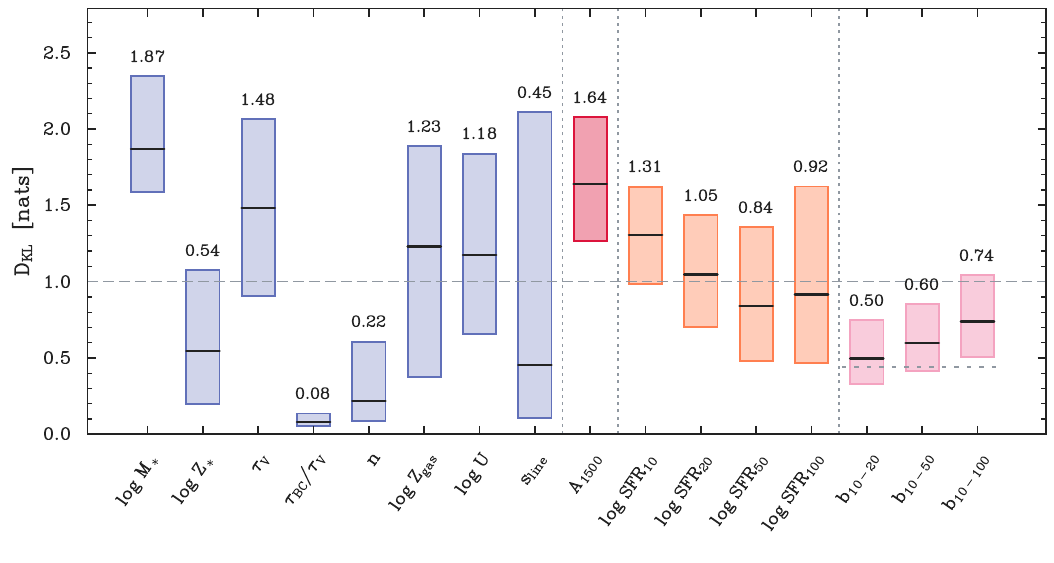}
    \caption{Information gain per parameter, measured by the KL divergence $D_{\rm KL}$ between the posterior and prior distributions. Each box spans the 16th--84th percentiles across 1365 galaxies (1168 for $s_{\rm line}$, which requires emission-line data). The black line and the number above each box indicate the median, the dashed line marks $D_{\rm KL}=1$, and the dotted line over the burst ratios marks the threshold below which we flag results as potentially prior-dominated ($D_{\rm KL}=0.44$~nat; flag K, \S~\ref{sec:reliability}). Blue boxes indicate directly sampled parameters (Table~\ref{tab:sfms12}), while the remaining quantities are derived through the forward model. Red denotes UV attenuation ($A_{1500}$), coral the ${\rm SFR}_t$, and pink the burst ratios $b_{10-20}$, \bhun\ and $b_{10-100}$. Stellar mass, UV attenuation, diffuse dust, ${\rm SFR}_{10}$, gas-phase metallicity, and ionisation parameter show the largest information gains. The ${\rm SFR}_t$ reach $D_{\rm KL}\approx1$, while the burst ratios, each defined as the ratio of two SFRs, lie below this value but above the reliability threshold for most galaxies.}
    \label{fig:kl_divergence}
\end{figure*}

\subsection{Quantities derived from the fits}

We compute derived quantities for every posterior draw of each galaxy. The quoted values are the posterior medians, with the 16--84~per~cent ranges as uncertainties (see Appendix~\ref{app:derived} for definitions).

Some quantities frequently used in the following analysis are the SFR averaged over the most recent $t$ of the SFH, $\mathrm{SFR}_t$, and the burst ratios $b_{10-t} = \log_{10}({\rm SFR}_{10}/{\rm SFR}_{10-t})$ for $t=20$ and 100~Myr, where ${\rm SFR}_{10-t}$ is the mean SFR between 10 and $t$~Myr lookback time, and $\bhun\ = \log_{10}({\rm SFR}_{10}/\sfrtenfifty)$, where \sfrtenfifty\ is the mean SFR between 10 and 50~Myr lookback time. \bhun\ compares the last 10~Myr with the preceding 40~Myr and classifies a posterior draw as rising ($\bhun\ > 0.2$), declining ($\bhun\ < -0.2$), or constant ($|\bhun| \leq 0.2$). The quantity $t_{50}$ is the lookback time within which the most recent half of the stellar mass formed, while $M_{\rm UV}$ is the absolute magnitude of the model spectrum around rest-frame 1500~\AA.

For dust, we fit the optical depth of the diffuse component at 5500~\AA, $\tau_V$ \citep[using the attenuation curve of][with a sampled slope offset $n$]{kriek_dust_2013}. The youngest stars experience additional attenuation from birth clouds with optical depth $\tau_{\rm BC}$. We characterise the combined attenuation from both components by $A_{1500}$, the attenuation at 1500~\AA. We calculate the dust-free UV magnitude, $M_{\rm UV}^{0}$, using the same stellar and gas properties but without dust attenuation. The attenuation is then
\begin{equation}
    A_{1500} \equiv M_{\rm UV} - M_{\rm UV}^{0},
    \label{eq:a1500}
\end{equation}

\subsection{Fit validation and recovery}\label{sec:fit_quality}\label{sec:quality}

We validate our fits by testing whether the models reproduce the observations, whether the inferred quantities are unbiased with realistic uncertainties, and whether quantities with direct observational proxies agree with those measurements.

The models reproduce the observed photometry well, with a median $\chi^2/N_{\rm phot}=0.88$. The emission-line fits show somewhat larger residuals, with a median $\chi^2/N_{\rm EL}=1.72$ (excluding He\,\textsc{i}~10830 and [N\,\textsc{ii}]$\lambda6584$). Detected high-ionisation lines (He\,\textsc{ii}, C\,\textsc{iv}, [Ne\,\textsc{iv}]) and [O\,\textsc{i}] are underpredicted by $\approx3\sigma$ in the 135 affected galaxies, which may indicate harder ionising sources than our stellar models include, such as X-ray binaries or shocks, or a mix of high- and low-ionization regions that cannot be captured by the single-zone cloudy models we employ \citep{Moreschini:2026aa}; they contribute 6~per~cent of the total line $\chi^2$.
The pulls, $\chi=(F_{\rm model}-F_{\rm obs})/\sigma$, give the difference between model and data in units of the measurement uncertainty. Stacked over all galaxies (Figs.~\ref{fig:stacked_pulls} and \ref{fig:fit_quality}), they show that the photometric pulls are centred on zero, whereas the model underpredicts the line fluxes (median pull $-0.8$). This effect is mainly driven by extreme emission lines that lie outside the grids of \citet{Byler:2017aa}. The Balmer lines, which are most relevant for our analysis, are recovered with median pulls within $0.3\sigma$ of zero, whereas we find a slight underprediction of strong ($> 10^{-18.5}\,{\rm erg\,s^{-1}\,cm^{-2}}$) [O\,\textsc{iii}]$\lambda5007$ at $z<4.5$ (Fig.~\ref{fig:lines}).

We compare the SED-based \sfrten\ with SFRs inferred from H$\alpha$ using the calibration of \citet{Kennicutt:2012aa}, restricting the comparison to galaxies with significant (${\rm S/N}\geq3$) H$\alpha$ detections. After scaling the H$\alpha$ fluxes to the Kron photometry, the median ${\rm SFR}_{10}^\mathrm{SED}$ is lower than the H$\alpha$-based estimate by $0.15$~dex, increasing to $0.28$~dex when H$\alpha$ is corrected for dust using the SED-fit attenuation (Fig.~\ref{fig:sfr10_halpha}). These H$\alpha$-based SFRs use the solar-metallicity calibration of \citet{Kennicutt:2012aa}. For metal-poor stellar populations with binaries, the conversion is 0.3--0.4~dex lower \citep{Reddy:2022aa, Clarke:2025aa}, which would bring the dust-corrected estimate into agreement with ${\rm SFR}_{10}^\mathrm{SED}$. The SED fits instead use the conversion implied by each galaxy's fitted stellar population and nebular parameters.  Because H$\alpha$ is included in the SED fits, this comparison tests internal consistency rather than providing an independent validation.

To assess parameter recovery, we perform injection--recovery tests using JADES-like data for a subset of galaxies (Appendix~\ref{app:validation}). As shown in Fig.~\ref{fig:injection_recovery}, $M_\ast$, ${\rm SFR}_t$, and $\bhun$ are recovered with median offsets of at most 0.06~dex, although the recovery of $\bhun$ is less accurate for galaxies with photometry alone. For most galaxies, the true attenuation lies within the 95~per~cent posterior credible interval.

\subsection{Information gain}
\label{subsec:information_gain}
The extent to which a posterior is constrained by the data rather than dictated by the prior can be quantified using the Kullback--Leibler (KL) divergence of the marginal posterior $p$ from the prior $\pi$,
\begin{equation}
    D_{\rm KL}(p\,\|\,\pi) = \int p(\theta)\,\ln\frac{p(\theta)}{\pi(\theta)}\,{\rm d}\theta \;\simeq\; \sum_{k} p_k \ln\frac{p_k}{\pi_k} \quad [{\rm nats}],
    \label{eq:kl}
\end{equation}
which we evaluate for each galaxy using histograms of posterior and prior draws. We do not specify priors directly for derived quantities such as ${\rm SFR}_t$, $b_{10-t}$, and $A_{1500}$, but obtain their prior distributions by passing draws from the adopted parameter priors through the forward model at each galaxy's redshift and posterior-median stellar mass (for $A_{1500}$, stellar mass is also drawn from its prior). A value of $D_{\rm KL}=0$ indicates that the posterior and prior are identical, i.e.\ that the data provide no information beyond the prior. We consider $D_{\rm KL}>1$~nat to indicate a large marginal information gain \citep[][Sect.~4.4]{Duan:2026aa}. Because \bhun\ is the ratio of two SFRs, it gains less information than either of them, so we do not apply this criterion to it. Instead, we flag for caution the mass--redshift bins in which \bhun\ is least constrained, i.e.\ in which its median information gain lies in the lowest fifth of the sample ($D_{\rm KL}<0.44$~nat; flag K, \S~\ref{sec:reliability}). For a Gaussian, this threshold corresponds to a posterior about 2.3 times narrower than its prior.

Fig.~\ref{fig:kl_divergence} shows the $D_{\rm KL}$ of the main sampled and derived quantities. Stellar mass, the UV attenuation $A_{1500}$, diffuse dust, \sfrten, gas-phase metallicity and ionisation parameter show the largest marginal information gain. The birth-cloud fraction $\tau_{\rm BC}/\tau_V$, the diffuse slope $n$ and, for the median galaxy, the slit-loss factor $s_{\rm line}$ are set mostly by the prior. The ${\rm SFR}_t$ lie close to $D_{\rm KL} = 1$, with the shortest timescales better constrained ($D_{\rm KL,~\sfrten} = 1.31$) than the longer ones ($D_{\rm KL,~\sfrhun} = 0.92$). The burst ratios $b_{10-20}$, \bhun\ and $b_{10-100}$ fall below $D_{\rm KL}=1$ (medians 0.50, 0.60 and 0.74). The posterior widths are about twice those of the SFRs. This motivates us to specifically compare the burst ratio result of each galaxy to its own SFH priors (flag Q in the following sections). We flag the mass--redshift bins in which the population statistics of \bhun\ follow the prior in \S~\ref{sec:reliability}.

\subsection{Selection function and weights}\label{sec:muv_completeness}

The sample is a spectroscopic subsample of the JADES NIRCam population, shaped by target prioritisation, spectroscopic success and the quality cuts of \S~\ref{sec:data}. We model the probability $C$ that a galaxy is observed and enters our sample as a function of \muv, redshift and stellar mass, and weight each galaxy by the inverse of this probability, $1/C$. $C$ describes the inclusion probability that a galaxy of the reference population (the \citealt{Bouwens:2021aa} UVLF and the NIRCam parent catalogue of \citealt{Duan:2026aa}) enters the sample. However, this is not the same as correcting to completeness. The weights can only correct the differences in number density depending on \muv, $z$ and mass, which is based on the approximation that the inclusion does not depend on any other properties. They cannot recover populations absent from the sample, such as quenched galaxies, which are not represented in the spectroscopic catalogue. The weighted sample is not complete: our population statistics only rescale the importance of the galaxies it contains. 

\paragraph*{UV selection function.}
We count the galaxies, $n_j$, in bins $j$ of \muv\ in six redshift slices, the last open-ended at $z \geq 9$, and compare them with the number expected over the survey area $\Omega = 465.4$~arcmin$^2$ (the JADES NIRCam area in GOODS-S and GOODS-N, measured from the DR5 catalogue footprint; cf. 469~arcmin$^2$ in \citealt{Johnson:2026aa}):

\begin{equation}
    N^{\rm exp}_j = \Omega\,\Delta M \int_{z_{\rm lo}}^{z_{\rm hi}} \frac{{\rm d}V_{\rm c}}{{\rm d}z\,{\rm d}\Omega}\,\Phi(M_j, z)\,{\rm d}z,
    \label{eq:nexp}
\end{equation}

where $M_j$ is the bin centre and $\Phi(M_j, z)$ the UVLF of \citet{Bouwens:2021aa} at $z<9$ and of \citet[][their table~3]{Donnan:2024aa} at $z \geq 9$. We fit the selection function by maximising the Poisson likelihood of $n_j$ given $C(M_j)N_j^{\rm exp}$, with
\begin{equation}
    C(M_{\rm UV}, z) = \frac{C_0}{1 + 10^{\kappa\,(M_{\rm UV} - M_{\rm UV,50})}}
    \label{eq:completeness}
\end{equation}

\begin{figure*}
    \centering
    \includegraphics[width=\textwidth]{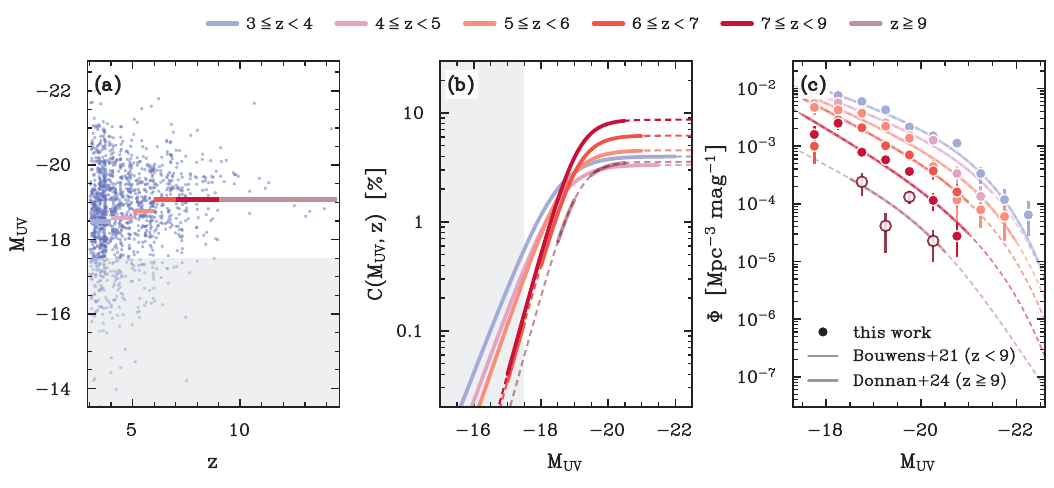}
    \caption{Reweighting of the sample to be more representative of the UVLF.
    Only a small fraction of the UV-selected galaxy population enters the sample, and the inclusion probability falls steeply faintward of a turnover magnitude that brightens with redshift. Panel (a) shows \muv\ against redshift for all galaxies and marks $M_{\rm UV,50}$, the magnitude at which the inclusion probability has halved, as a horizontal segment in each redshift slice. Colours mark the six redshift slices, and grey shading in (a) and (b) marks $M_{\rm UV}>-17.5$. Panel (b) shows $C(M_{\rm UV},z)$ (see Eq.~\ref{eq:completeness}), dashed where the data do not constrain it. Panel (c) compares the $w'$-weighted UV luminosity functions (circles, open for the low-number slice at $z\geq9$) with the reference functions of \protect\citet{Bouwens:2021aa} at $z<9$ and \protect\citet{Donnan:2024aa} at $z\geq9$, drawn as faint lines in the colour of each slice and dashed where they extend beyond the magnitude range covered by the data.}
    \label{fig:sample}
\end{figure*}

$C_0$ is the fraction of galaxies brighter than the turnover that enter the sample, $M_{\rm UV,50}$ the magnitude at which this fraction has halved, and $\kappa$ the steepness of the decline. In the low-number slice at $z \geq 9$ only $C_0$ is free, with $M_{\rm UV,50}$ and $\kappa$ fixed to their $7\leq z<9$ values, and between slice centres the parameters are interpolated linearly in $z$ (Fig.~\ref{fig:sample}). Even well above the turnover, only a few per cent of the parent galaxies enter the sample, so at fixed $M_{\rm UV}$ and $z$ we model it as a sparse subsample of the bright population that is uniform apart from a mass tilt (Eq.~\ref{eq:sel_mass}). Faintward of the turnover, which brightens with redshift, the sample thins rapidly: the galaxies we miss are mainly UV-faint galaxies at $z\gtrsim6$, where weighted statistics just above the magnitude cut rest on few galaxies.

\paragraph*{Mass dependence of the selection at fixed $M_{\rm UV}$.} $C(\muv, z)$ assumes that only \muv\ and redshift decide whether a galaxy ends up in the sample. We test whether that is the case by comparing the sample to the masses and Kron-based \muv\ of the NIRCam parent catalogue \citep{Duan:2026aa}, for members and non-members alike. Using logistic regression, 
\begin{align}
    {\rm logit}\,P({\rm sample}) &= f(M_{\rm UV}, z) + \beta_M\,x, \notag\\
    x &= \log_{10}(M_\ast/\Msun) - 9,
    \label{eq:sel_mass}
\end{align}

with $f$ free in each $(\muv, z)$ bin gives $\beta_M = 0.62 \pm 0.07$. This means that at fixed \muv, more massive galaxies are more likely to be selected, and the sample galaxies are slightly more massive than the parent galaxies of the same \muv\ and $z$.

\paragraph*{Weights.}
Each galaxy with $M_{\rm UV} < -17.5$ receives the weight
\begin{equation}
    w' = \frac{1}{C(M_{\rm UV}, z, M_\ast)}
      = \frac{\langle e^{\beta_M x}\rangle_{M_{\rm UV}, z}}{C(M_{\rm UV}, z)\,e^{\beta_M x}}.
    \label{eq:wmass}
\end{equation}
For galaxies absent from the parent catalogue, $x$ uses our stellar mass shifted by its offset from the parent-catalogue masses, fitted on the galaxies in both. The weights are capped at their 99th percentile. Galaxies fainter than the threshold are assigned zero weight. We choose this cut because the faint end of the UVLF is dominated by a few galaxies with very large weights. Galaxies between the turnover and the cut contribute most of the total weight $\sum w'$ and largely determine the bright-tail slope $k$ (Eq.~\ref{eq:logit_rise}; \S~\ref{sec:bright_tail}). Applying brighter cuts weakens the slope at $z=3$--5 to values consistent with zero (Table~\ref{tab:robustness}). The 1163 weighted galaxies have an effective sample size of $n_{\rm eff} = (\sum_i w'_i)^2/\sum_i (w'_i)^2 = 653$. Using weights $w = 1/C(M_{\rm UV}, z)$, which exclude the mass correction, changes the weighted medians of \bhun, the rising fractions of the UV-brightest galaxies, and the weighted rank correlation $\rho_{\rm S}(-M_{\rm UV}, b)$ (Eq.~\ref{eq:weighted_rank}) by at most 1.27 times their bootstrap uncertainties.

\paragraph*{Treatments.} Depending on the quantity we want to infer, a different implementation of the weights, or none, is appropriate. The default is W (\S~\ref{sec:conventions}):

\begin{itemize}
    \item \textbf{N (unweighted):} all 1365 galaxies, for quantities conditioned on $M_{\rm UV}$ and relations not involving $M_{\rm UV}$.
    \item \textbf{W (weighted):} population statistics at fixed $(M_\ast, z)$ with weights $w'$.
    \item \textbf{R (rank-weighted):} samples defined by rank in $M_{\rm UV}$ (e.g.\ the brightest 10~per~cent at fixed stellar mass), with threshold and fractions from the $w'$-weighted distribution.
    \item \textbf{S (selection-aware):} When $M_{\rm UV}$ is the dependent variable, we explicitly account for the fact that our sample is selected in $M_{\rm UV}$. At fixed regressors $\mathbfit{x}$, we assume that the \emph{intrinsic} distribution of $M_{\rm UV}$ is Gaussian,
    \begin{equation}
        M_{\rm UV} \sim
        \Nn\!\left(\mu(\mathbfit{x}),\sigma\right),
        \qquad
        \mu(\mathbfit{x})=\mathbfit{x}\cdot\bmath{\beta},
    \end{equation}
    where $\bmath{\beta}$ describes the intrinsic relation and $\sigma$ its intrinsic scatter. Not every galaxy drawn from this distribution enters the sample. A galaxy with magnitude $M$ and redshift $z$ is included with probability
    \begin{equation}
        S(M,z)
        =
        C(M,z)\,\Theta(-17.5-M),
    \end{equation}
    where $C(M,z)$ is the inclusion probability from Eq.~(\ref{eq:completeness}) and the Heaviside function imposes the \muv\ cut. The distribution that we expect to observe is therefore the intrinsic Gaussian multiplied by this selection probability and renormalised:
    \begin{equation}
        p(M \mid \mathbfit{x}, {\rm selected})
        =
        \frac{
            \Nn\!\left(M;\mu(\mathbfit{x}),\sigma\right)\,S(M,z)
        }{
            \displaystyle
            \int
            \Nn\!\left(M';\mu(\mathbfit{x}),\sigma\right)\,
            S(M',z)\,{\rm d}M'
        }.
        \label{eq:sel_like}
    \end{equation}
    Rather than fitting a Gaussian directly to the observed $M_{\rm UV}$ distribution, we account for the selection of the observed sample in the model. This allows us to recover the intrinsic relation without assuming that the observed sample is fully representative of the underlying population. We fit $\bmath{\beta}$ and $\sigma$ by maximising the joint likelihood of the galaxies in each mass bin. For this, we use the population-averaged $C(M,z)$, without the mass term of Eq.~(\ref{eq:wmass}). For the intrinsic scatter in $M_{\rm UV}$ at fixed stellar mass (\S~\ref{sec:fixedmass}), the mean varies linearly with stellar mass within each bin, with a slope shared across the bins of a redshift slice.
\end{itemize}

\begin{figure}
    \centering
    \includegraphics[width=\columnwidth]{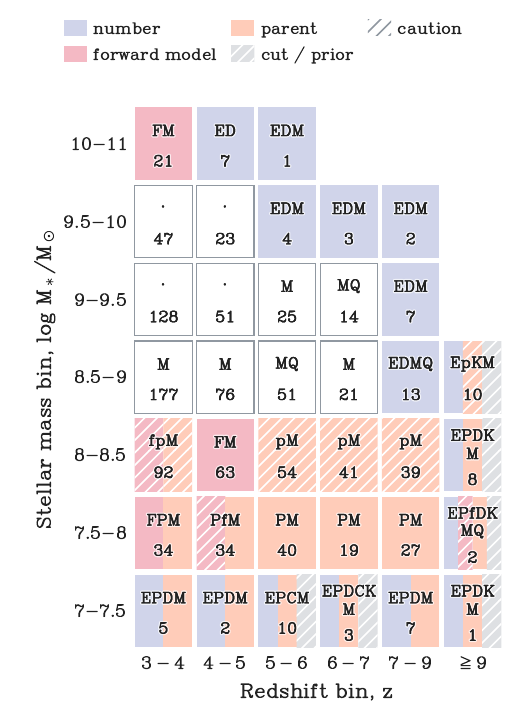}
    \caption{Reliability of population measurements and sensitivity of inferred burst ratios to the SFH prior. Each square represents a stellar mass--redshift bin, labelled with the number of galaxies at $M_{\rm UV}<-17.5$ (the selection-weight limit; \S~\ref{sec:muv_completeness}). White bins are reliable; solid stripes indicate unreliable bins, while hatched stripes indicate caution. Each failed test (\S~\ref{sec:reliability}) adds a colour: blue for sample size and weighting (E, D), pink for the selection forward model (F, f), coral for parent-catalogue completeness (P, p), and grey for proximity to the selection cut or prior sensitivity (C, K). Lower-case letters indicate cautionary rather than failing levels of the corresponding tests. E denotes $N<10$, $n_{\rm eff}<10$, or $z\geq9$; D indicates domination by a few large weights. F and f indicate that the selection forward model excludes $>30$~per~cent or 15--30~per~cent of mock galaxies, respectively; F also applies if it shifts the \bhun\ distribution by $>0.05$~dex. P and p indicate that the UV cut removes $>30$~per~cent or 15--30~per~cent of the parent catalogue, respectively; P also applies below the mass limit of the SFMS sample of Duan et al. (in prep.), and p to bins straddling that limit. C marks bins with most galaxies near the UV cut, and K those with burst ratios closely following the prior. No bin fails test B (selection probability depending on \bhun\ at fixed \muv). Tests M ($>14$~per~cent of stellar mass formed in the last 50~Myr, making \sfrtenfifty\ potentially prior-sensitive) and Q (\bhun\ distribution consistent with the prior) are diagnostic only and do not affect the reliability classification.}
    \label{fig:reliability_map}
\end{figure}

\subsection{Where the measurements are reliable}\label{sec:reliability}
The biggest limitation of our correction for selection effects is that the weights are not able to correct for any type of galaxy that was not at all included in our observed sample. We assess the robustness of our results in bins of stellar mass and redshift. Reliable means that the bin can be trusted for the galaxies that the sample contains, and is not a claim of completeness for that bin. The results are shown in Fig.~\ref{fig:reliability_map}. 
We classify each bin as reliable, requiring caution, or unreliable based on its sample statistics, a forward model of the selection, coverage of the more complete NIRCam parent population, and the information gain of \bhun. The unreliability criteria are as follows:

\begin{itemize}
   \item \textbf{E:} The effective sample size is below ten, or the bin lies at $z \geq 9$.
   \item \textbf{F:} More than 30~per~cent of the mock galaxies fall below the selection cut, or selection shifts the mean or width of the inferred \bhun\ distribution by more than 0.05~dex.
   \item \textbf{P:} Fewer than 70~per~cent of the NIRCam parent galaxies are brighter than the cut, or the bin lies below the mass limit, defined as the first percentile of stellar mass of the SFMS sample of Duan et al. (in prep.) in each redshift slice.
\end{itemize}

\noindent The following criteria call for caution when interpreting the results; f and p are weaker levels of the tests F and P:

\begin{itemize}
  \item \textbf{D:} Three or fewer galaxies contribute more than half of the total weight.
  \item \textbf{f:} Between 15~per~cent and 30~per~cent of mock galaxies fall below the cut, or more than 30~per~cent do when the mock $M_{\rm UV}$ are made 0.3~mag fainter.
  \item \textbf{p:} Between 70~per~cent and 85~per~cent of parent galaxies are brighter than the cut, or the bin lies at that mass limit.
  \item \textbf{C:} Most galaxies lie within 0.5~mag of the cut.
  \item \textbf{K:} The median information gain of \bhun\ is below 0.44~nat.
  \item \textbf{B:} In the NIRCam parent catalogue, the probability of entering the sample depends on \bhun, computed from the catalogue SFRs of \citet{Duan:2026aa}, at fixed \muv\ (logistic slope differing from zero by more than $3\sigma$).
\end{itemize}

\begin{figure*}
    \centering
    \includegraphics[width=\textwidth]{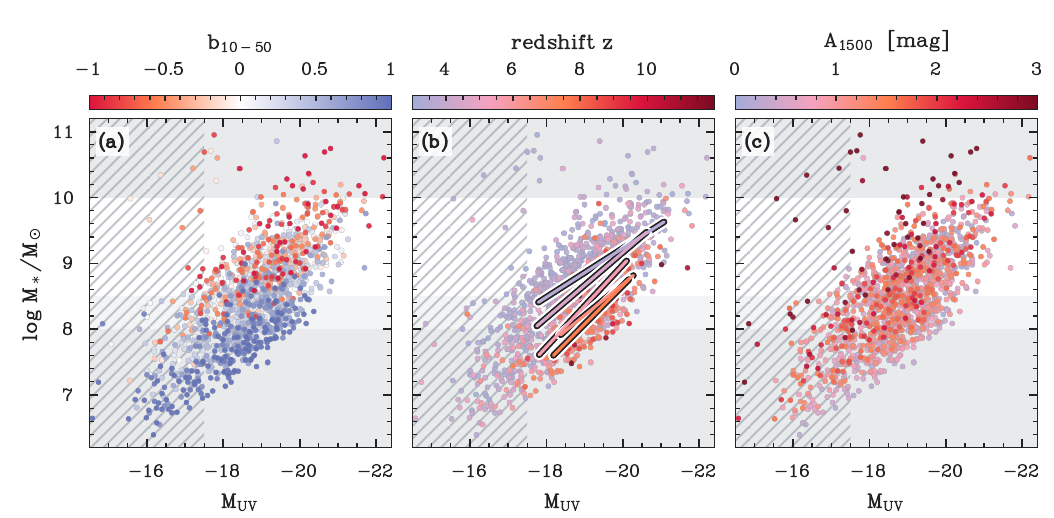}
    \caption{Stellar mass as a function of UV brightness for all galaxies, coloured by (a) burst ratio \bhun, (b) redshift, and (c) dust attenuation $A_{1500}$. The lines in panel (b) show fits of $\log_{10}(M_\ast/\Msun)=a(M_{\rm UV}+19.5)+\log_{10}M_{\ast,0}(z)$ (Eq.~\ref{eq:mass_muv}) in redshift slices at $z<9$ (Appendix~\ref{app:mass_muv}), coloured by redshift. The hatched region marks $M_{\rm UV}>-17.5$, and grey bands mark stellar-mass ranges in which most galaxies lie in unreliable (dark) or caution (light) bins. The strong correlation between stellar mass and UV brightness means that apparent trends with $M_{\rm UV}$ are entangled with trends in stellar mass and should therefore be examined at fixed stellar mass.} 
    \label{fig:naive}
\end{figure*}

Our results are reliable for $8.5 \leq \log_{10}(M_\ast/\Msun) < 10$ at $3 \leq z < 5$. At $5 \leq z < 7$, the reliable range is $8.5 \leq \log_{10}(M_\ast/\Msun) < 9.5$. The bin $8 \leq \log_{10}(M_\ast/\Msun) < 8.5$ requires caution at $3 \leq z < 4$ and $5 \leq z < 9$ and is unreliable at $4 \leq z < 5$, while all bins below $10^8\,\Msun$ or at $z \geq 9$ are unreliable. We introduce two additional flags to identify bins where the SFH prior may influence the results. These flags do not change the reliability classification:

\begin{itemize}
    \item \textbf{M:} The median galaxy formed more than 14~per~cent of its stellar mass within the last 50~Myr. In such bins \sfrtenfifty\ can depend on the prior, which flag Q tests.
    \item \textbf{Q:} The population distribution of \bhun\ is consistent with the SFH prior alone. In this prior-only null test, the $w'$-weighted fraction of draws with $\bhun\ > 0.2$ lies within $2\sigma$ of the prior-only fraction, and the fraction of mass formed in the last 50~Myr differs from the prior prediction by less than 0.05.
\end{itemize}

Flag Q applies to two of the four reliable bins at $z \geq 5$ ($5\leq z<6$, $8.5\leq\log_{10}(M_\ast/\Msun)<9$ and $6\leq z<7$, $9\leq\log_{10}(M_\ast/\Msun)<9.5$), indicating that population statistics of \bhun\ at these redshifts should be interpreted with caution. 

\subsection{Conventions}\label{sec:conventions}
Unless stated otherwise, the following conventions apply throughout the rest of the paper. Captions and text only specify deviations from them.
\emph{Sample and weights:}
Population statistics are based on the galaxies with $M_{\rm UV} < -17.5$, weighted by $w'$ (Eq.~\ref{eq:wmass}; treatment W). These weights correct for differences in number density caused by the selection, but cannot account for types of galaxies missing from the sample (\S~\ref{sec:muv_completeness}).
\emph{Per-draw statistics:}
Individual galaxy quantities are given as medians over their posterior draws (Appendix~\ref{app:derived}). Population fractions account for the full posterior of each galaxy. For example, the rising fraction is
\begin{equation}
    f = \frac{\sum_i w'_i P_i}{\sum_i w'_i},
\end{equation}
where $P_i$ is the fraction of posterior draws of galaxy $i$ with $\bhun\ > 0.2$. Population medians are the $w'$-weighted medians of the per-galaxy values.
\emph{Uncertainties:}
We estimate uncertainties as half the 16--84~per~cent range from 1000 bootstrap resamplings of the galaxies, using their weights in the resampling. Where each resampling requires refitting the selection function or a selection-aware likelihood, we use fewer resamplings (40--300).

\emph{Reliability:}
In figures showing stellar mass or redshift, unreliable ranges are shaded dark, while ranges requiring caution are shaded light (\S~\ref{sec:reliability}, Fig.~\ref{fig:reliability_map}).
\emph{Low-number regimes:}
Bins with $N < 10$ or $n_{\rm eff} < 10$, as well as all bins at $z \geq 9$, are shown with open or semi-transparent symbols (hatched in maps). Magnitudes fainter than the cut are shaded grey. For results based on these regimes, we explicitly quote $N$ (and $n_{\rm eff}$). We do not draw conclusions from these regimes alone.

\section{Results}\label{sec:results}

We first examine the relation between stellar mass and UV luminosity (\S~\ref{sec:naive}), then investigate which galaxy properties correlate with UV brightness at fixed stellar mass (\S~\ref{sec:fixedmass}). We next explore how the prevalence of rising and declining star-formation histories depends on stellar mass and redshift (\S~\ref{sec:variability}), before asking whether UV-bright galaxies preferentially exhibit rising star formation (\S~\ref{sec:bright_tail}). Throughout, we distinguish galaxies that are unusually UV-bright for their mass from those that populate the absolute bright end of the UV luminosity function.

\subsection{The stellar mass--UV relation}
\label{sec:naive}

UV-brighter galaxies tend to be more massive, with an unweighted Spearman correlation coefficient of $\rho(-\muv, \log M_\ast) = 0.58$ for galaxies with $\muv<-17.5$ across all redshifts (Fig.~\ref{fig:naive}). We fit the stellar mass--UV relation following \citet{Tacchella:2018ab} and recover their slope (Appendix~\ref{app:mass_muv}). The slope steepens with redshift in the full sample, but this trend disappears when the fit is restricted to reliable stellar-mass and redshift bins.

Figure~\ref{fig:naive} shows the unweighted sample, coloured by burst ratio \bhun, redshift, and dust attenuation. From the faintest to the brightest \muv\ bin, the median \bhun\ changes from positive to negative, the rising-SFH fraction halves, and declining SFHs become dominant. Dust attenuation, by contrast, shows no apparent trend with UV brightness: the median remains nearly constant across \muv\ bins (Fig.~\ref{fig:naive}c).

The apparent trend in burst ratio is largely driven by stellar mass. The burst ratio correlates more strongly with stellar mass, $\rho(M_\ast, \bhun) = -0.63$, than with UV brightness, $\rho(-\muv, \bhun) = -0.22$. Selection weights $w'$ amplify the apparent UV trend because the faintest bin is dominated by up-weighted low-mass galaxies, about half of which lie in unreliable bins. Restricting the analysis to reliable bins removes the correlation between \bhun\ and \muv, and faint galaxies are no longer predominantly rising.

At $z=3$--5, the median \muv\ brightens with stellar mass up to $10^{10}\,\Msun$, but becomes fainter at higher masses as the median dust attenuation rises to $A_{1500}\approx3$~mag. However, this highest-mass regime contains only 33 galaxies and lies in unreliable bins. Together, these results show that the mass--UV relation can produce or obscure trends with UV brightness. We therefore examine these relations at fixed stellar mass in the following sections.

\begin{figure*}
    \centering
    \includegraphics[width=\textwidth]{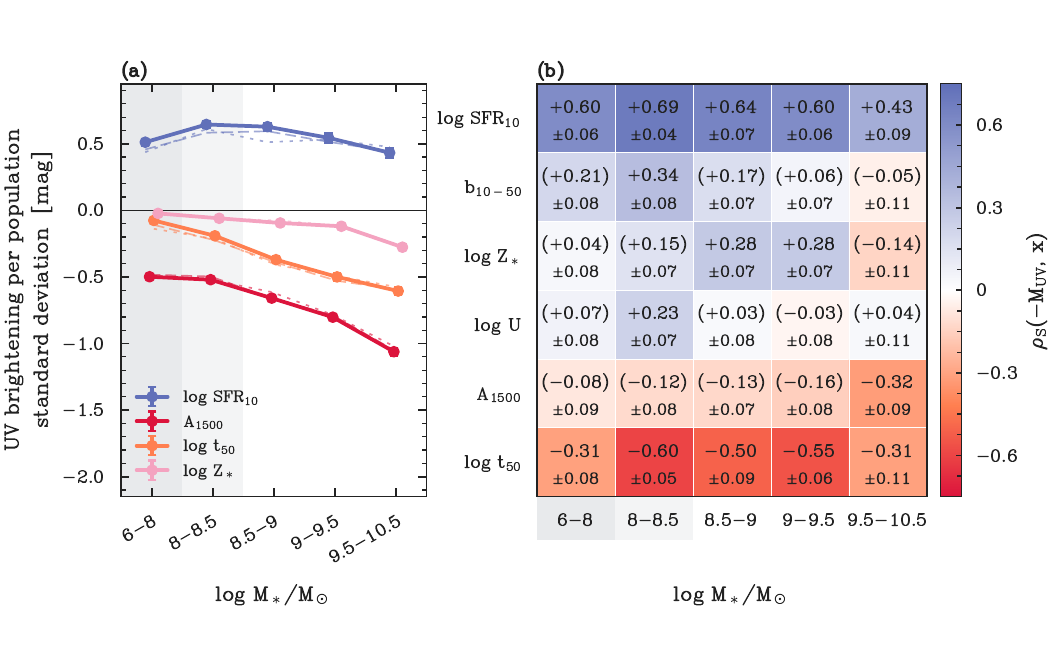}
    \caption{The dependence of UV brightness on galaxy properties at fixed stellar mass. Panel (a) shows the effect of each quantity in Eq.~(\ref{eq:fixedmass}), $-\beta_x$, as the change in UV magnitude per standard deviation of that quantity (the regressors are standardised), holding the others fixed, using the selection-aware likelihood (symbols), weights $w'$ (dashed lines), and the unweighted sample including $M_{\rm UV}>-17.5$ (dotted lines). Panel (b) shows the $w'$-weighted Spearman correlations between $-M_{\rm UV}$ and each quantity, with bootstrap uncertainties below each value, which do not control for the other quantities. Blue cells mean that UV-brighter galaxies have higher values of the quantity, red cells lower values. Values in parentheses have a significance below $2.6\sigma$. Grey shading in (a) and grey strips behind the mass labels in (b) mark unreliable (dark) and caution (light) mass bins. At fixed stellar mass, UV-brighter galaxies have higher recent SFRs, younger stellar populations, and less dust attenuation. Across the reliable mass bins, recent SFR has the strongest positive association with UV brightness, although this association weakens with increasing mass. Stellar age becomes more important at higher masses, while dust attenuation is the strongest negative term in the joint regression and accounts for more of the variation in UV brightness there.}
    \label{fig:fixed_mass}
\end{figure*}

\subsection{What correlates with $M_{\rm UV}$ at fixed stellar mass}
\label{sec:fixedmass}

To identify which galaxy properties are associated with UV brightness at fixed stellar mass, we divide the galaxies into five stellar-mass bins (Fig.~\ref{fig:fixed_mass}). In each bin, we quantify these associations using both a multivariate regression and a weighted rank correlation. For the multivariate analysis, we model the intrinsic mean UV magnitude as
\begin{align}
    \mu &= \beta_0 + \sum_x \beta_x \frac{x-\langle x\rangle_w}{\sigma_w(x)} \notag\\
    &\quad\quad\, + \beta_M \left(\log_{10}M_\ast-\log_{10}M_{\rm c}\right) + \beta_z(z-5), \label{eq:fixedmass}\\[5pt]
    & x\in\left\{\log{\rm SFR}_{10},A_{1500},\log t_{50},\log Z_\ast\right\}. \notag
\end{align}
Here, $\langle x\rangle_w$ and $\sigma_w(x)$ are the $w'$-weighted mean and standard deviation of each quantity among galaxies in the mass bin with $M_{\rm UV}<-17.5$, and $M_{\rm c}$ is the bin centre. We fit the model using the selection-aware likelihood of Eq.~(\ref{eq:sel_like}). Since the regressors $x$ are standardised, $-\beta_x$ gives the change in UV brightness, in magnitudes, associated with a one-standard-deviation increase in $x$, holding the other regressors, stellar mass, and redshift fixed. Positive $-\beta_x$ therefore corresponds to brighter UV magnitudes.

To measure the marginal associations with UV brightness, we also compute the weighted Spearman rank correlation, $\rho_{\rm S}$, between each quantity $x$ and $-M_{\rm UV}$ which, like UV luminosity, increases towards brighter galaxies. We use weighted mid-ranks with weights $w'$,
\begin{equation}
    R_i(x) = \sum_{j:x_j\leq x_i} w'_j-\frac{w'_i}{2},
    \label{eq:weighted_rank}
\end{equation}
and define $\rho_{\rm S}$ as the weighted Pearson correlation coefficient between the ranks of $x$ and $-M_{\rm UV}$. Thus, $\rho_{\rm S}>0$ indicates that larger values of $x$ are associated with brighter UV magnitudes. Unlike the multivariate regression, these correlations do not control for other galaxy properties.

Fig.~\ref{fig:fixed_mass} shows the multivariate regression coefficients and weighted rank correlations. The regression results agree closely across the three selection treatments, W, S, and N, indicating that they are not strongly affected by selection in \muv. The lowest mass bin is unreliable, and the next bin, $\log M_\ast/\Msun=8$--8.5, requires caution. We therefore focus on $\log_{10}(M_\ast/\Msun)=8.5$--10.5, with reliable constraints up to $10^{10}\,\Msun$ at $z<5$.

Both estimators show that \sfrten\ has the strongest positive association with UV brightness in every mass bin, with the largest positive $-\beta_x$ and $\rho_{\rm S}$ of all quantities (Fig.~\ref{fig:fixed_mass}). However, this association weakens with increasing stellar mass. The brightening per dex of \sfrten\ decreases towards higher masses and remains below the 2.5~mag per dex expected for $L_{\rm UV}\propto\sfrten$ in every bin (Table~\ref{tab:robustness}).

In contrast, stellar age becomes more important at higher masses. Older galaxies tend to be UV-fainter at fixed stellar mass, with age showing the strongest negative rank correlation in every bin except the most massive, where dust is equally important (Fig.~\ref{fig:fixed_mass}b). Its contribution to the joint regression also increases with mass (Fig.~\ref{fig:fixed_mass}a; Table~\ref{tab:robustness}), although the amplitude depends on the SFH prior. These opposing trends may reflect the increasing contribution of stars formed more than 10~Myr ago to the UV emission of massive galaxies. At fixed stellar mass, younger galaxies have formed their stars over a shorter period and therefore contain a larger fraction of young, UV-luminous stars. Stellar age thus captures differences in the older stellar populations that are not described by \sfrten.

Dust attenuation is the strongest negative contributor to UV brightness in the joint regression, and its contribution increases with stellar mass (Fig.~\ref{fig:fixed_mass}a). At fixed SFR, age, and metallicity, one magnitude of attenuation makes \muv\ one magnitude fainter in every mass bin (Table~\ref{tab:robustness}), as expected from its additive effect on the UV magnitude. The increasing contribution at higher masses therefore reflects the larger spread in dust attenuation among massive galaxies, rather than a stronger effect per magnitude of attenuation. As a result, dust accounts for more of the variation in UV brightness at higher stellar masses.

However, this strong dust dependence is not apparent in the marginal correlations. The overall relation between dust attenuation and UV brightness is nearly flat (\S~\ref{sec:naive}), and the rank correlations at fixed stellar mass are weak, reaching significance only in the most massive bin (Fig.~\ref{fig:fixed_mass}b). This is because galaxies with higher \sfrten\ also tend to be dustier. The covariance between \sfrten\ and attenuation is positive in every mass bin and exceeds the mean posterior covariance, which means that the relation is not solely a fitting degeneracy. The brightening from an elevated \sfrten\ and the dimming from dust attenuation partially cancel in the marginal relation, but can be separated in the joint regression. We quantify how much of the UV brightening from recent star formation is offset by dust in \S~\ref{sec:selfobscured}.

Stellar metallicity has little effect on UV brightness in the joint regression: metal-rich galaxies are slightly fainter at fixed SFR, age, and dust (Fig.~\ref{fig:fixed_mass}a). The marginal rank correlation is instead positive in the two lower bins of the reliable mass range and insignificant elsewhere (Fig.~\ref{fig:fixed_mass}b), reflecting covariance with the other quantities. These results should be interpreted cautiously, however, because the available data constrain stellar metallicity poorly. Unlike gas-phase metallicity, which is relatively well constrained, stellar metallicity shows little information gain relative to its prior (Fig.~\ref{fig:kl_divergence}).

We also find that galaxies with rising SFHs tend to be UV-brighter at fixed stellar mass. The burst ratio \bhun\ correlates positively with UV luminosity in the four lower mass bins, but significantly only at $8\leq\log_{10}(M_\ast/\Msun)<8.5$ (Fig.~\ref{fig:fixed_mass}b). The correlation is weaker than for \sfrten\ because \bhun\ also depends on \sfrtenfifty, which is closely linked to stellar mass. We exclude \bhun\ from the regression because it is strongly correlated with \sfrten\ at fixed mass. The ionisation parameter correlates significantly with \muv\ only in the cautionary mass bin (Fig.~\ref{fig:fixed_mass}b).

The selection-corrected intrinsic scatter in \muv\ at fixed stellar mass and redshift is 0.46--0.83~mag in the reliable and cautionary bins, with no significant trend with mass or redshift. This estimate uses treatment S and accounts for posterior uncertainties in \muv\ and $\log M_\ast$.

Dust also affects the scatter in UV luminosity. At fixed \sfrten, attenuation has a larger scatter than the unattenuated \muv\ in every mass bin, particularly at low masses. The two residuals are weakly anti-correlated, partly because uncertainty in attenuation shifts the inferred unattenuated magnitude in the opposite direction while the observed \muv\ remains well constrained. This reduces the scatter in observed \muv\ compared with uncorrelated residuals (treatment S, Eq.~\ref{eq:bivariate}).

Overall, UV-brighter galaxies at fixed stellar mass have higher \sfrten, younger stellar populations, and less dust attenuation. The relative importance of these properties changes with mass: the contribution of \sfrten\ decreases, while those of age and dust increase. Above $\log_{10}(M_\ast/\Msun)=9$, age and dust each contribute about as much as \sfrten\ or more to the variation in UV brightness.

\begin{figure*}
    \centering
    \includegraphics[width=\textwidth]{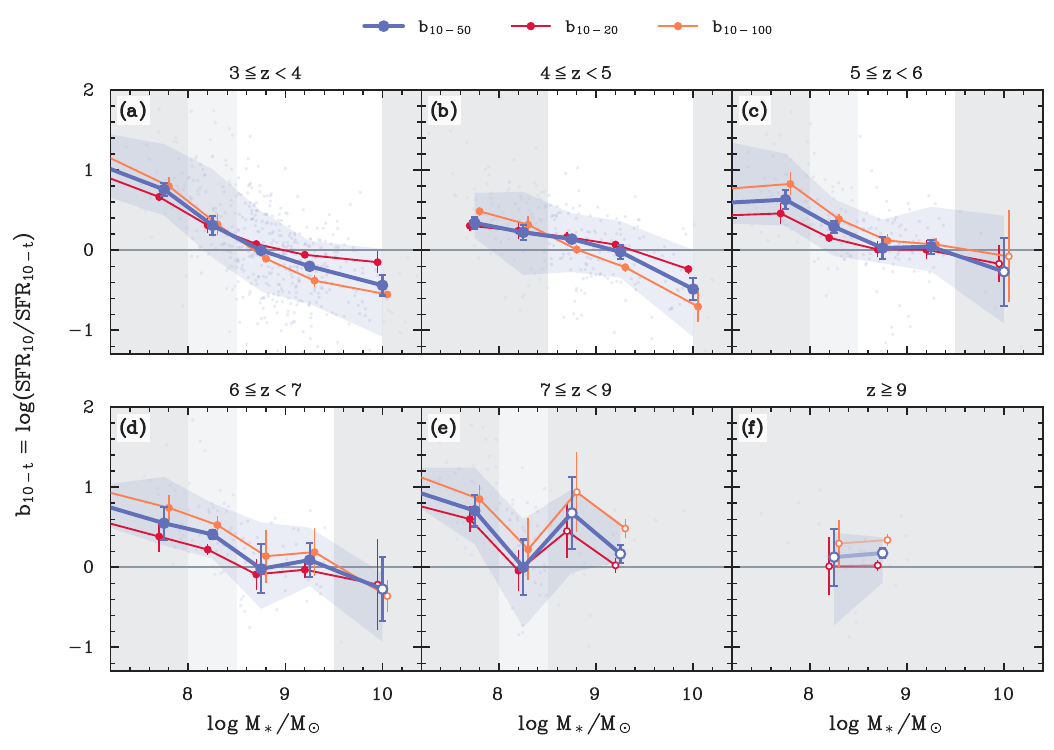}
    \caption{Burst ratios as a function of stellar mass in six redshift slices (a--f), for galaxies with $M_{\rm UV}<-17.5$. Thick blue lines and large symbols show the $w'$-weighted median \bhun, and thin red and coral lines with small symbols the median $b_{10-20}$ and $b_{10-100}$, with bootstrap uncertainties. Open symbols mark bins with $N<10$ or $n_{\rm eff}<10$ and all bins at $z\geq9$. The shaded band shows the 16th--84th percentiles of \bhun, and the small points show individual \bhun\ values. Unreliable mass bins are shaded dark grey, bins that should be interpreted with caution are shaded in light grey. In the reliable mass bins, burst ratios decrease with increasing stellar mass, with lower-mass galaxies more often rising and galaxies above $10^{9.5}\,\Msun$ predominantly declining. This trend is clearest at $3\leq z<5$; at $z\geq5$, evidence for a decline relies on the cautionary bin $8\leq\log_{10}(M_\ast/\Msun)<8.5$.}
    \label{fig:variability}
\end{figure*}

\subsection{The mass dependence of recent star formation}
\label{sec:variability}

Galaxies with higher \sfrten\ and \bhun\ are UV-brighter at fixed stellar mass (\S~\ref{sec:fixedmass}). We therefore investigate how recent star formation depends on stellar mass and redshift, and whether lower-mass galaxies show greater variability. Fig.~\ref{fig:variability} shows the burst ratios ($b_{10-20}$, \bhun\ and $b_{10-100}$) as a function of stellar mass in redshift slices, while Fig.~\ref{fig:variability_z} (Appendix~\ref{app:validation}) shows their evolution with redshift at fixed stellar mass. We restrict the analysis to bins classified as reliable or requiring caution in the reliability map (Fig.~\ref{fig:reliability_map}), which also gives the number of galaxies in each bin. The reliable mass range shrinks with increasing redshift, and no bins remain at $z\geq9$. The bin $8\leq\log_{10}(M_\ast/\Msun)<8.5$ requires caution at $3\leq z<4$ and $5\leq z<9$ and is unreliable at $4\leq z<5$; all bins with a lower stellar mass are unreliable.

The clearest result is that galaxies with lower stellar masses are more likely to have rising SFHs, while more massive galaxies tend to have declining SFHs. At $3\leq z<5$, all three burst ratios decrease with stellar mass across the reliable bins (Fig.~\ref{fig:variability}), and galaxies above $\log_{10}(M_\ast/\Msun)=9.5$ are predominantly declining. In particular, \bhun\ decreases by 0.5--0.7~dex per dex of stellar mass (Table~\ref{tab:robustness}), more steeply than can be explained by correlated posterior errors in \bhun\ and stellar mass alone. The decrease is present in all three baselines, $b_{10-20}$, \bhun\ and $b_{10-100}$ (Fig.~\ref{fig:variability}).

The median burst ratios at fixed stellar mass show a less clear evolution with redshift. For galaxies with $\log_{10}(M_\ast/\Msun)>8.5$, where the measurements are more reliable, $b_{10-100}$ appears to increase with redshift (Fig.~\ref{fig:variability_z}). However, \bhun\ increases less, which does not depend on the SFR more than 50~Myr before observation. The apparent redshift evolution of $b_{10-100}$ is therefore likely driven by the prior on ${\rm SFR}_{10-100}$, rather than by a physical change in the SFH.

\begin{figure*}
    \centering
    \includegraphics[width=\linewidth]{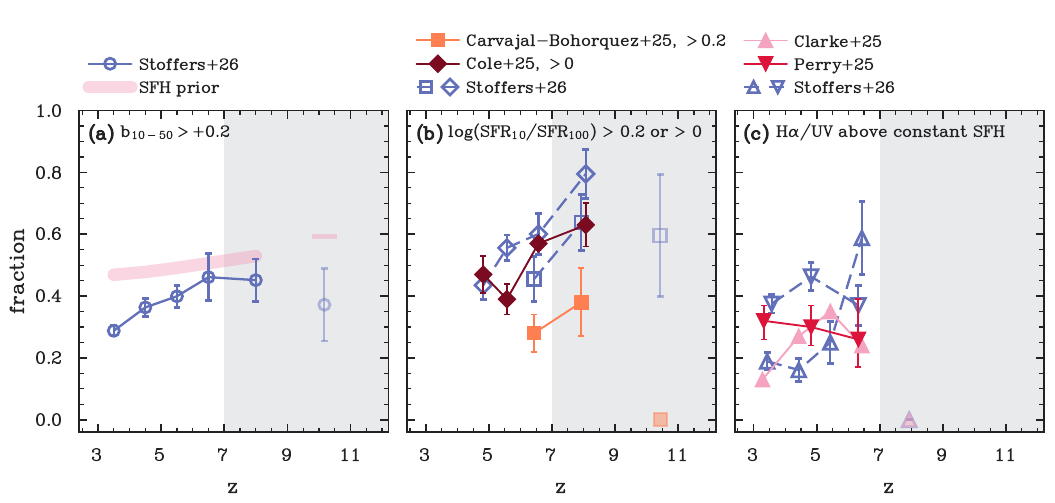}
    \caption{Fraction of galaxies with rising or bursty SFHs as a function of redshift. Panel (a) shows our $w'$-weighted fraction of galaxies with $\bhun>0.2$ at $8.8\leq\log_{10}(M_\ast/\Msun)<9.5$ (blue open circles and solid line), alongside the fraction expected from their SFH priors (pink band). Panels (b) and (c) compare our results with four literature studies using their respective definitions: \protect\citet{Carvajal-Bohorquez:2025aa} (squares) and \protect\citet{Cole:2025aa} (diamonds) use ${\rm SFR}_{10}/{\rm SFR}_{100}$ (b), while \protect\citet{Clarke:2025aa} (triangles) and \protect\citet{Perry:2025aa} (inverted triangles) use dust-corrected H$\alpha$/UV ratios above the range expected for a constant SFH (c). Filled symbols and solid lines show the published measurements, with a different colour for each study; open blue symbols and dashed lines show our results using the corresponding definitions, indicated next to the panel labels (with the thresholds in panel (b) given in the legend above it). \protect\citet{Clarke:2025aa} report no uncertainties. Grey shading indicates unreliable redshift ranges, while semi-transparent symbols mark bins with few galaxies and all bins at $z\geq9$. Overall, we find that the fraction of galaxies with rising SFHs increases with redshift up to our highest reliable redshift bin, approaching the prior expectation at $z\gtrsim6$.}
    \label{fig:bursty_fraction}
\end{figure*}

We also examine the fraction of galaxies with rising SFHs. Because we adopt a rising SFMS prior, a preference for rising SFHs is expected even without constraints from the data. We therefore compare the inferred rising fractions with those expected from each galaxy's own prior. At $z<5$, the measured fractions lie well below the prior expectations (Fig.~\ref{fig:bursty_fraction}a; Table~\ref{tab:robustness}), indicating that they are constrained by the data. Above $\log_{10}(M_\ast/\Msun)=9$, this holds for every analysis cut. At lower masses, the difference becomes smaller for brighter cuts, since UV-brighter galaxies at fixed stellar mass are more likely to have rising SFHs (\S~\ref{sec:bright_tail}; Table~\ref{tab:robustness}). The rising fraction nevertheless appears to increase with redshift. In the mass bins shown in Fig.~\ref{fig:bursty_fraction}, chosen to match the literature, it increases from $0.29\pm0.02$ at $z=3$--4 to $0.46\pm0.08$ at $z=6$--7. However, at $z=6$--7, the measured fraction is consistent with the prior expectation, and two of the four reliable bins at $z\geq5$ are flagged Q, where the rising fraction and the fraction of mass formed in the last 50~Myr match the SFH prior (\S~\ref{sec:reliability}). In the reliable mass range at these redshifts, the rising fraction is also close to the prior expectation, but responds only weakly to shifts in the prior centre, as in the data-driven bins at lower redshift. Their increase with redshift, however, is small (from 0.37 at $z=3$--5 to 0.41 at $z=5$--7 at $8.5\leq\log_{10}(M_\ast/\Msun)<9.5$) and absent for the \texttt{hyperburst} prior. Nevertheless, the small number of galaxies in these bins and the prior sensitivity of the broader mass range prevent us from establishing a robust physical redshift trend.

At $z=6$--7, our inferred rising fraction ($0.46\pm0.08$; $0.46\pm0.07$ in their definition) is higher than that reported by \citet{Carvajal-Bohorquez:2025aa}, who fitted a photometric-redshift sample with \textsc{cigale}, for the same $b_{100}$ threshold, mass range, and redshift (Fig.~\ref{fig:bursty_fraction}). In their definition, our fraction at $z=7$--9 is $0.64\pm0.09$, also above theirs, but this rests on few galaxies in unreliable bins. However, their Bayes-factor classification also shows an increase with redshift, as does the stochastic fraction measured by \citet{Ciesla:2024aa}, who find that more than 80~per~cent of galaxies with $\log_{10}(M_\ast/\Msun)>9$ have stochastic SFHs at $z>9$, while secular star formation dominates at $z\lesssim7$ (their fig.~9). Our sample cannot test this transition, because no bins at $z\geq9$ are reliable. Similar trends are reported by \citet{Langeroodi:2024aa}, in whose redshift bins and without mass selection our median $b_{100}$ rises from $-0.01$ at $z=3$--3.5 to $+0.56$ at $z=7$--10, and by \citet{Looser:2025aa}. \citet{Clarke:2025aa} find tentatively more galaxies above the constant-SFH H$\alpha$/UV ratio at $4<z<7$ than at lower redshift (Fig.~\ref{fig:bursty_fraction}). With their constant-SFH range, dust corrections, sample cuts and mass cut, and with our H$\alpha$ fluxes scaled to the photometry, a similar fraction of our galaxies lies above this range at $z=3$--6 (16--25~per~cent, compared with their 13--35~per~cent), with a higher fraction at $z=6$--7 that rests on 17 galaxies, and our median ratios agree with theirs within 0.03~dex at $z=4$--6. Our median $\log_{10}({\rm SFR}_{10}/{\rm SFR}_{100})$ above $10^{8.5}\,\Msun$ increases with redshift, as theirs does up to $z=7$, but lies 0.1--0.4~dex below their \textsc{Prospector} values. In our mass range, \citet{Cole:2025aa} find a duty cycle with little change in redshift (Fig.~\ref{fig:bursty_fraction}), decreasing with redshift only above $\log_{10}(M_\ast/\Msun)=9.3$; in their definition, ours increases with redshift and exceeds theirs at $z=5$--6. Similarly, \citet{Perry:2025aa} find no evolution of the duty cycle at $1\lesssim z<7$, and in their definition our fraction shows no monotonic trend either, although it is somewhat higher, while \citet{Mitsuhashi:2026aa} find no evolution in the scatter of line-to-UV ratios at $3<z<7$, although it may increase at $z\gtrsim8$. We discuss how SFH priors may affect these inferred trends in \S~\ref{sec:limitations}.

The burst ratios indicate whether a galaxy's recent SFR is rising or declining relative to its past average, but do not directly measure the amplitude of its star-formation variability. Greater variability in low-mass galaxies would produce a larger spread in \bhun\ at fixed stellar mass. After accounting for posterior uncertainties, we find an intrinsic scatter of $\sigma_b=0.14$--$0.51$~dex in the well-constrained bins at $3\leq z<5$, with no clear dependence on stellar mass (Table~\ref{tab:sigma_b}). However, these bins span only a limited mass range. At lower masses, the inferred scatter is comparable to or larger than that at higher masses, but the measurements become increasingly uncertain. Below $10^8\,\Msun$, posterior uncertainties generally exceed the observed spread, preventing a reliable estimate of the intrinsic scatter. We therefore cannot rule out greater variability in lower-mass galaxies.

For comparison, \citet{Simmonds:2025aa} infer a short-term burstiness of 0.2--0.4~dex from the difference in quadrature between the intrinsic main-sequence scatter on 10 and 50~Myr timescales. Applying a similar estimator to our fits gives 0.25--0.41~dex at $3\leq z<5$ and $10^8\leq M_\ast/\Msun<10^9$, consistent with their results. At higher masses and redshifts, posterior uncertainties prevent a reliable measurement.

Finally, we use strong emission-line ratios to investigate whether our sample contains a rejuvenating galaxy population, following \citet{Witten:2025aa}. We find that O32 closely tracks \bhun\ and separates galaxies with rising and declining SFHs, as expected because these lines enter the fits (Fig.~\ref{fig:observable}). However, we find no corresponding trend with R23, the ratio of the combined [O\,II] and [O\,III] emission-line fluxes to H$\beta$.

In summary, the most robust trend is a decrease in burst ratios with increasing stellar mass: lower-mass galaxies are more likely to have rising SFHs, while massive galaxies tend to have declining SFHs. This does not imply that low-mass galaxies have more variable star formation, since the intrinsic scatter in \bhun\ shows no clear stellar mass dependence between $10^{8.5}$ and $10^{10}\,\Msun$. Although both $b_{10-100}$ and the rising fraction appear to increase with redshift, limitations from the SFH prior and sample reliability prevent us from establishing a physical redshift trend.

\begin{figure*}
    \centering
    \includegraphics[width=0.86\textwidth]{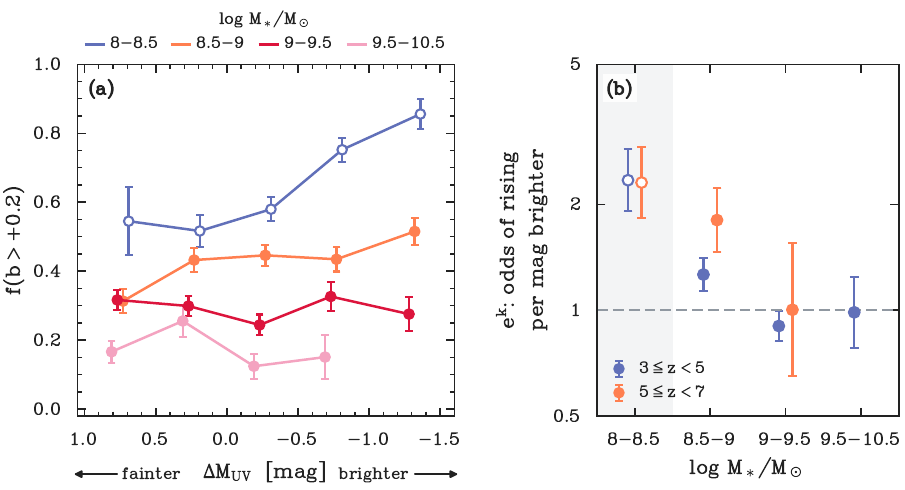}
    \caption{Dependence of the rising SFH fraction on UV brightness at fixed stellar mass and redshift. Panel (a) shows the fraction of galaxies with rising SFHs ($b>0.2$) as a function of their offset from the median $M_{\rm UV}$ of each mass and redshift bin at $3\leq z<7$, including only galaxies in reliable (filled symbols) and caution bins (open symbols), with colours marking the mass bins. Panel (b) shows the factor $e^k$, the odds ratio for a 1-mag brightening; $e^k=1$ indicates no dependence on UV brightness (Eq.~\ref{eq:logit_rise}), including all galaxies. Blue symbols show $3\leq z<5$ and coral symbols $5\leq z<7$. Open symbols and light-grey shading mark mass bins that require caution, and the dashed line marks $e^k=1$. The highest-mass point is based on about 15 rising galaxies. Below $10^9\,\Msun$, UV-brighter galaxies are more likely to have rising SFHs, most clearly at $8\leq\log_{10}(M_\ast/\Msun)<8.5$. Thus, rising SFHs help populate the UV-bright tail at fixed mass in low-mass galaxies, but do not account for the absolute bright end, which at $z=3$--5 is dominated by massive galaxies, about half of which have declining SFHs (Fig.~\ref{fig:fraction_muv}).}
    \label{fig:bright_tail}
\end{figure*}

\begin{figure*}
    \centering
    \includegraphics[width=\textwidth]{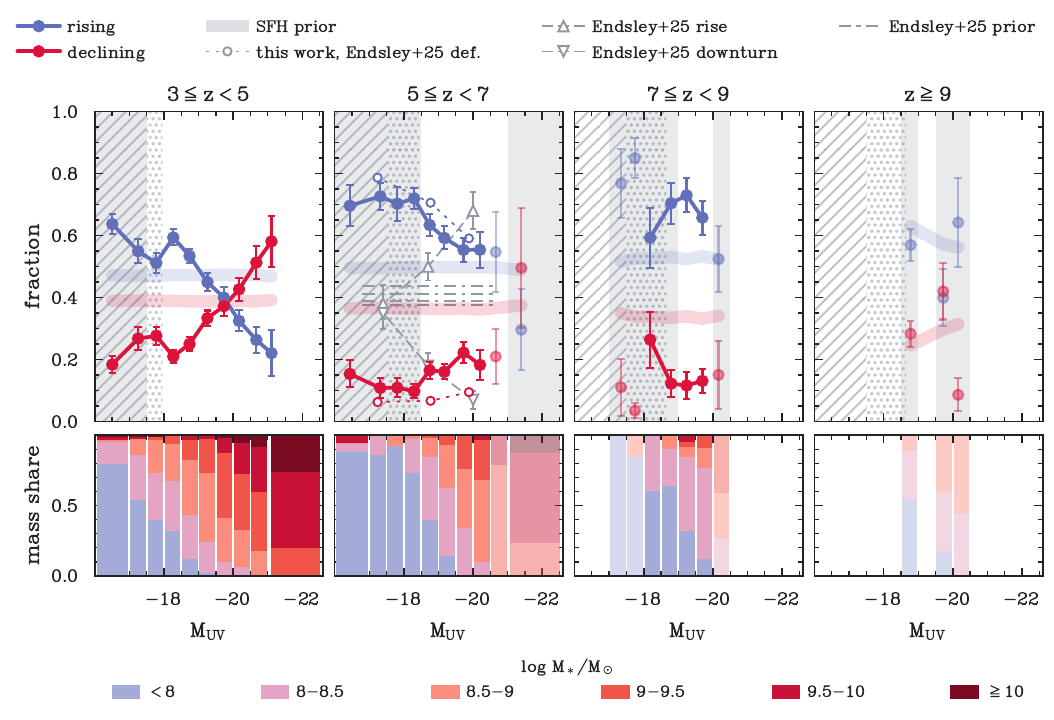}
   \caption{Fractions of galaxies with rising (blue) and declining (red) SFHs as a function of \muv\ in four redshift slices (upper row), calculated from each posterior draw, and the contribution of each stellar-mass bin to each \muv\ bin (lower row). Points are $w'$-weighted above the completeness limit, while fainter galaxies (diagonal hatching) are unweighted. Wide translucent lines show the fractions expected based on the galaxies' SFH priors. In the $5\leq z<7$ column, grey open triangles with dashed lines show the fractions reported by \protect\citet{Endsley:2025aa} at $z\sim6$ for their definitions of a rise (upward triangles) and a strong downturn (downward triangles), grey dash-dotted lines show the expectations that we compute from their SFH prior alone, and small open circles with dotted lines show our galaxies evaluated using their definitions, without weights. In the upper row, dark grey shading marks \muv\ bins dominated by unreliable mass--redshift bins, dotted hatching indicates where the selection becomes strongly incomplete, and semi-transparent symbols indicate bins with few galaxies and all bins at $z\geq9$. At $z=3$--5, the brightest galaxies are predominantly more massive than $10^9\,\Msun$, and about half have declining SFHs.}
    \label{fig:fraction_muv}
\end{figure*}

\begin{figure}
    \centering
    \includegraphics[width=\columnwidth]{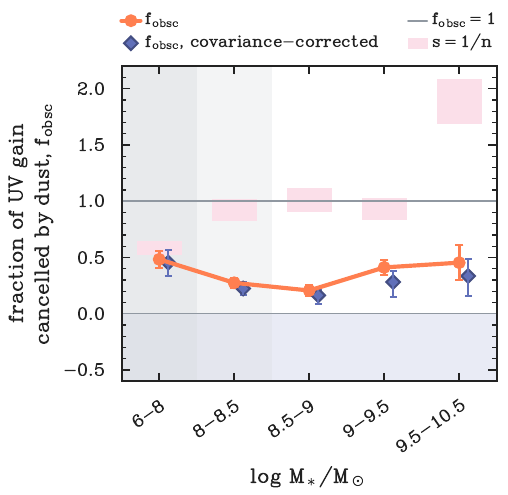}
  \caption{Fraction of the unattenuated UV brightening associated with increased \sfrten\ that is cancelled by dust, $f_{\rm obsc}$ (Eq.~\ref{eq:fobsc}), in each stellar-mass bin. Coral circles show the selection-aware fit (treatment S), while blue diamonds show the $w'$-weighted fit after subtracting the mean within-galaxy posterior covariance between dust and \sfrten. The grey line indicates complete cancellation ($f_{\rm obsc}=1$), the blue region indicates decreasing attenuation with increasing \sfrten, and the pink boxes show the expectation for $s=1/n$, assuming a Kennicutt--Schmidt index of $n=1.4\pm0.15$ \citep{kennicutt_global_1998}. Grey columns indicate unreliable (dark) and cautionary (light) mass bins. Dust offsets between one-fifth and one-half of the UV brightening but does not fully cancel it.}
    \label{fig:selfobscured}
\end{figure}

\subsection{Rising star formation in the UV-bright tail at fixed stellar mass}
\label{sec:bright_tail}

At fixed stellar mass, \sfrten\ is the strongest positive predictor of UV brightness (\S~\ref{sec:fixedmass}), while rising SFHs become less common with increasing stellar mass (\S~\ref{sec:variability}). We therefore investigate whether UV-brighter galaxies are more likely to have rising SFHs when controlling for stellar mass and redshift.

Within each stellar-mass and redshift bin, we define the UV-bright tail as the brightest 10~per~cent of galaxies, based on the weighted \muv\ distribution. We compare the fraction of rising galaxies in this tail, $f_{10\%}$, with that in the full bin, $f_{\rm all}$:
\begin{align}
    f_{10\%}
    &= \frac{\sum_{i\in\mathcal{T}} w'_i\,P_i(b>0.2)}
           {\sum_{i\in\mathcal{T}} w'_i},
    \qquad
    f_{\rm all}
    = \frac{\sum_i w'_i\,P_i(b>0.2)}
           {\sum_i w'_i},\\[3pt]
    \Delta f &= f_{10\%}-f_{\rm all},
    \label{eq:excess}
\end{align}
where $\mathcal{T}$ denotes the UV-bright tail, $P_i(b>0.2)$ is the posterior probability that galaxy $i$ has a rising SFH, and $w'_i$ is its selection weight (Eq.~\ref{eq:wmass}). A positive $\Delta f$ indicates an excess of rising SFHs among the UV-brightest galaxies at fixed stellar mass and redshift. However, this measure is limited when the rising fraction is already high, since $\Delta f\leq1-f_{\rm all}$.

We therefore also examine how the rising fraction varies across the full UV-luminosity distribution (Fig.~\ref{fig:bright_tail}a). For each galaxy, we define $\Delta M_{\rm UV}$ relative to the weighted median of its mass and redshift bin. We quantify the trend using a weighted logistic regression,
\begin{equation}
    \ln\frac{P}{1-P} = c + k\,(-\Delta M_{\rm UV}),
    \qquad P=P(b>0.2),
    \label{eq:logit_rise}
\end{equation}
where $e^k$ gives the factor by which the odds of a rising SFH increase per magnitude of UV brightening (Fig.~\ref{fig:bright_tail}b). The regression uses each galaxy's posterior rising probability rather than a binary classification, accounting for uncertainty in \bhun. We also include uncertainty in the selection weights and use a weak Gaussian prior to regularise sparsely populated bins.

\paragraph*{Rising SFHs are more common among UV-brighter low-mass galaxies.}
At fixed stellar mass and redshift, UV-brighter galaxies tend to have a higher probability of rising star formation (Fig.~\ref{fig:bright_tail}a). The trend is strongest at $8\leq\log_{10}(M_\ast/\Msun)<8.5$, where the odds of a rising SFH increase by a factor of approximately 2.3 per magnitude at both $z=3$--5 and $z=5$--7. However, these mass bins require caution because of their limited reliability. At $8.5\leq\log_{10}(M_\ast/\Msun)<9$, the trend is weaker: at $z=3$--5, the odds increase by approximately 26~per~cent per magnitude ($k=0.23\pm0.11$), while at $z=5$--7 the inferred trend is only marginally stronger than expected from correlated posterior errors. No significant trend is found at higher stellar masses, where rising SFHs remain uncommon across the UV-luminosity distribution.

\paragraph*{Validation and selection effects.}
The correlation between UV brightness and rising SFHs is partly expected from the SED model: at fixed stellar mass, higher recent SFRs increase UV luminosity, while the longer-timescale SFR is more closely tied to the accumulated stellar mass. As an observational check, we find that the fraction of galaxies with low dust-corrected H$\alpha$/UV ratios, indicative of declining recent star formation, also decreases towards brighter \muv\ at fixed stellar mass and $z<5$. However, the small number of such galaxies and the SED-based dust correction limit the independence of this test (Table~\ref{tab:robustness}).

The strength of the inferred trend also depends on sample selection. At $z=3$--5, the correlation in the $8.5\leq\log_{10}(M_\ast/\Msun)<9$ bin disappears when a brighter \muv\ limit is imposed, indicating that it is driven partly by faint galaxies with low inclusion probabilities near the original selection limit (\S~\ref{sec:muv_completeness}). Spectroscopic redshift success may also favour galaxies with strong emission lines and rising SFHs, an effect not explicitly included in our selection weights. However, it is unclear whether this would preferentially enhance the rising fraction in the UV-bright tail.

Consistent with these limitations, the brightest 10~per~cent of galaxies at $8.5\leq\log_{10}(M_\ast/\Msun)<9$ show an excess of rising SFHs at $z=5$--7 ($\Delta f=0.18\pm0.08$), but not at $z=3$--5 ($\Delta f=0.04\pm0.05$). The SFH prior alone produces a negligible excess (Table~\ref{tab:robustness}).

\paragraph*{The UV-bright tail is distinct from the absolute bright end.}
An excess of rising SFHs among UV-brighter galaxies at fixed stellar mass does not necessarily imply that rising SFHs dominate the absolute bright end ($M_{\rm UV}<-20$). At $z=3$--5, most galaxies at the absolute bright end have stellar masses above $10^9\,\Msun$, and approximately half have declining SFHs. At $z=5$--7, the rising fraction is consistent with the SFH prior (Fig.~\ref{fig:fraction_muv}).

At $z\sim6$, \citet{Endsley:2025aa} find that the fraction of galaxies with rising SFHs decreases towards fainter UV magnitudes, consistent with our trend at fixed stellar mass. However, their analysis does not control for stellar mass. Applying their definition to our sample, we instead find higher rising fractions towards fainter \muv, where their measurements approach the expectation from their SFH prior (Fig.~\ref{fig:fraction_muv}, $5\leq z<7$). Their SFHs are inferred from photometry and validated against NIRSpec prism spectra for a subset of galaxies, whereas our fits include spectroscopic redshifts and emission lines for every galaxy. However, our faintest comparison point lies below our completeness limit, and the next lies in a regime of strong incompleteness. Since spectroscopic selection may preferentially miss faint galaxies in post-burst phases, the difference could partly reflect selection effects as our reliability analysis highlights.

Furthermore, removing galaxy-to-galaxy scatter in the fitted SFHs changes the number of galaxies with $M_{\rm UV}<-20$ by at most 0.17~dex at $z<9$, and can even increase the counts at intermediate redshifts. Thus, SFH variability contributes relatively little to the abundance of galaxies at the absolute bright end in our sample. We examine this counterfactual in detail in \S~\ref{sec:uvlf_parameters}.

Overall, rising SFHs are associated with greater UV brightness at fixed stellar mass, particularly in low-mass galaxies. However, the strongest trends occur in less reliable bins, while the evidence in reliable bins is sensitive to selection and posterior covariance. At $z=3$--5, rising SFHs therefore contribute to the UV-bright tail at fixed stellar mass but do not dominate the absolute bright end.
\section{Dust and the UV-bright population}\label{sec:dust}

We have shown in \S~\ref{sec:results} that at fixed stellar mass, UV-bright galaxies have a high \sfrten\ and little dust, but also that galaxies with a higher \sfrten\ are dustier. In this section we aim to measure the dust response to recent star formation (\S~\ref{sec:selfobscured}--\ref{sec:balmer}), investigate how dust attenuation changes after a burst in star formation (\S~\ref{sec:burstphase}) and explore how sensitive the number of UV-bright galaxies is to the ingredients of our model (\S~\ref{sec:uvlf_parameters}).

\subsection{Partial self-obscuration of bursts}
\label{sec:selfobscured}

A burst of star formation increases a galaxy's UV luminosity, but may also increase its dust attenuation, partly offsetting the brightening. If the additional attenuation fully cancels the UV gain, bursts would not preferentially populate the UV-bright tail. We therefore investigate how much of the unattenuated UV brightening remains observable by separating the observed UV magnitude into its unattenuated component and dust attenuation (Eq.~\ref{eq:a1500}).

We regress the unattenuated UV magnitude and attenuation against $\log{\rm SFR}_{10}$ at fixed stellar mass and redshift:
\begin{align}
    M_{\rm UV}^{0}
    &= a_0 + a\,\log{\rm SFR}_{10} + a_M\,\Delta\!\log M_\ast + a_z\,\Delta z + \epsilon_{0}, \notag\\
    A_{1500}
    &= b_0 + b_A\,\log{\rm SFR}_{10} + b_M\,\Delta\!\log M_\ast + b_z\,\Delta z + \epsilon_A.
    \label{eq:bivariate}
\end{align}
We model $(\epsilon_{0}, \epsilon_A)$ as a bivariate normal distribution with dispersions $\sigma_{0}$ and $\sigma_A$ and correlation $\rho$, and apply the selection model of Eq.~\ref{eq:sel_like} to the sum $M_{\rm UV}^{0}+A_{1500}$. From the fitted slopes, we define the fraction of the unattenuated UV brightening cancelled by dust, $f_{\rm obsc}$, and the corresponding scaling of optical depth with star-formation rate, $s$, as
\begin{equation}
    \begin{aligned}
        f_{\rm obsc}
        &\equiv -\frac{b_A}{a}
        = -\frac{{\rm d}A_{1500}/{\rm d}\log{\rm SFR}_{10}}
        {{\rm d}M_{\rm UV}^{0}/{\rm d}\log{\rm SFR}_{10}}, \\
        s
        &\equiv \frac{{\rm d}\ln\tau_{1500}}{{\rm d}\ln{\rm SFR}_{10}}
        = \frac{b_A}{\ln 10\,\tilde{A}_{1500}},
    \end{aligned}
    \label{eq:fobsc}
\end{equation}
where $\tilde{A}_{1500}$ is the weighted median attenuation of the bin. Unlike the coefficient in \S~\ref{sec:fixedmass}, the unattenuated UV slope $a$ does not hold age and metallicity fixed. The unattenuated magnitude also includes nebular continuum emission, but its contribution to $a$ is negligible, and the nebular emission lines inside the wavelength window used to measure \muv\ (C\,\textsc{iv}~$\lambda1548$ and N\,\textsc{iv}]~$\lambda1486$) have zero flux in our nebular grid.

Attenuation increases with \sfrten\ in every mass bin, but the unattenuated UV magnitude brightens more rapidly (combining all galaxies in the reliable mass range). Consequently, the observed \muv\ still brightens by at least half a magnitude per dex in \sfrten.

Across the five mass bins, we find $f_{\rm obsc}$ between one fifth and one half (Fig.~\ref{fig:selfobscured}; treatment S; Table~\ref{tab:robustness}), while $w'$-weighted least squares gives higher values in four of the five bins, and the unweighted fit including $M_{\rm UV}>-17.5$ reaches $f_{\rm obsc}\approx1$ in the most massive bin. Although the lowest mass bin is unreliable and the next requires caution (Fig.~\ref{fig:reliability_map}), the reliable mass range gives the same result.

Two fitting effects bias $f_{\rm obsc}$ in opposite directions. Posterior covariance between dust and SFR contributes to the inferred relation, so subtracting the mean within-galaxy covariance lowers $f_{\rm obsc}$ (Fig.~\ref{fig:selfobscured}, diamonds). However, injection tests show that the fits dilute an input dust--\sfrten\ relation, with further dilution introduced by the covariance correction. 
Bracketing the effects of posterior covariance and fit dilution gives \(f_{\rm obsc}=0.16\)--0.73: among UV-visible galaxies, attenuation offsets an estimated 16--73~per~cent of the unattenuated UV brightening associated with increased recent star formation. The observed Balmer decrement provides an additional check: it excludes complete cancellation, although it cannot independently establish the partial effect (\S~\ref{sec:balmer}).

Our UV selection imposes a further limitation. Bursts obscured by $A_{1500}\gtrsim2$--3 mag fall below our UV limit, so the inferred $f_{\rm obsc}$ applies only to bursts that remain UV-visible. Including fully obscured bursts could therefore increase the obscured fraction.

The inferred optical-depth scaling also suggests that dust responds less strongly to increased star formation than expected if it directly followed the gas column. In every mass bin, $s$ lies below the $s=1/n=0.65$--0.80 expected for a Kennicutt--Schmidt index $n=1.4\pm0.15$ \citep{kennicutt_global_1998} if the optical depth scales with the gas surface density (Fig.~\ref{fig:selfobscured}, boxes). The lowest, unreliable mass bin comes closest. This is consistent with the gradual evolution of attenuation across the burst cycle found in \S~\ref{sec:burstphase}.

Overall, dust partially obscures the UV brightening associated with increased star formation, but does not fully cancel it. Although the precise fraction is affected by fitting uncertainties and UV selection, the remaining UV gain allows galaxies with rising SFHs to preferentially populate the UV-bright tail at fixed stellar mass.

\begin{figure*}
    \centering
    \includegraphics[width=\textwidth]{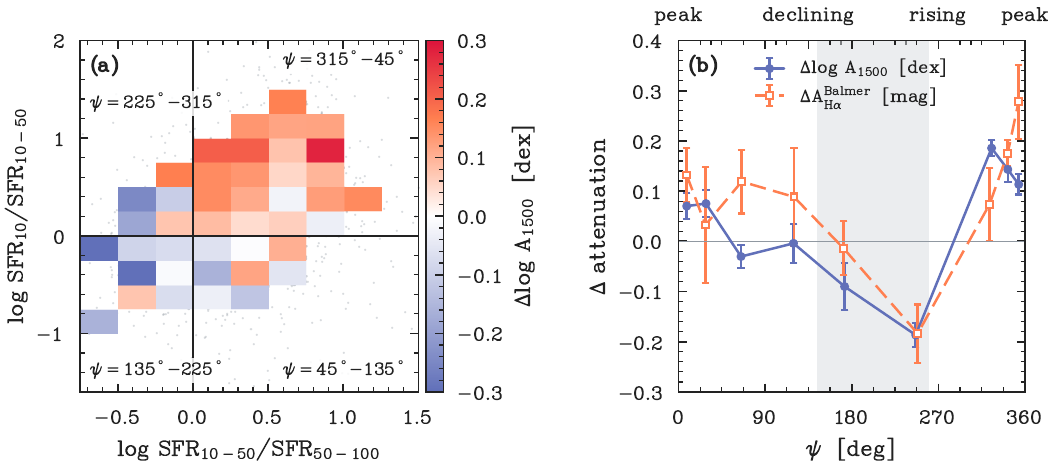}
    \caption{Dust attenuation across the star-formation burst cycle. The burst phase $\psi$ (Eq.~\ref{eq:psi}) is defined from the recent SFH, with $\psi\approx0^\circ$ marking the SFR peak, $\psi\approx90^\circ$--$180^\circ$ declining star formation, and $\psi\approx270^\circ$ renewed star formation. All galaxies are included without selection weights and divided into nine equally populated phase bins.
    (a) Median attenuation residuals at fixed stellar mass and redshift, obtained by subtracting a quadratic surface in $\log_{10}M_\ast$ and $z$ (\S~\ref{sec:burstphase}). Filled circles show $\Delta\log A_{1500}$ (dex) from SED fitting, and open squares show $\Delta A_{{\rm H}\alpha}$ (mag) from the observed Balmer decrement (H$\alpha$ and H$\beta$ at ${\rm S/N}\geq3$). The grey band marks phases whose stacked SFHs peaked 50--100~Myr before observation.
    (b) Burst-phase plane, showing the change in SFR from 10--50 to 0--10~Myr against that from 50--100 to 10--50~Myr. Colours indicate median $\Delta\log A_{1500}$ in cells with at least 10 galaxies; grey points show individual galaxies. The quadrants represent continuously rising (upper right), rising-to-declining (lower right), continuously declining (lower left), and declining-to-rising (upper left) SFHs, with the corresponding phase ranges labelled.
    Attenuation is lowest at $\psi\approx200^\circ$--$300^\circ$, after the SFR has declined and around the onset of renewed star formation.}
    \label{fig:burstphase}
\end{figure*}

\subsection{The Balmer-decrement check}
\label{sec:balmer}
The dust--\sfrten\ relation inferred from the SED fits is partly built into the model: the birth-cloud component attenuates only stars younger than about 10~Myr (\S~\ref{sec:method}; \citealt{charlot_simple_2000}), so increasing \sfrten\ raises the effective attenuation even at fixed dust content (\S~\ref{sec:limitations}). We therefore use the observed, dust-uncorrected Balmer decrement as an independent check, since it does not depend on this assumption. Only the grouping by \sfrten, which the fits tie to H$\alpha$, shares measurement noise with the decrement.
Across the reliable mass range, the Balmer decrement increases only weakly with \sfrten, consistent in sign with the fitted $b_A$. Nevertheless, it excludes complete obscuration of the burst at $3.8\sigma$, where the prediction converts $A_{1500}$ into $A_{{\rm H}\alpha}$ with the fitted ratio $A_{{\rm H}\alpha}/A_{1500}\approx0.28$. This exclusion is conservative because H$\alpha$ measurement noise, shared by the decrement and \sfrten, can only steepen the observed relation. However, the partial obscuration inferred from the SED fits predicts a slope comparable to that produced by measurement noise alone, so the Balmer decrement cannot independently confirm it.

\subsection{Dust attenuation following a burst}
\label{sec:burstphase}

\paragraph*{Burst phase.}
To investigate how dust attenuation evolves throughout a burst, we assign each galaxy a phase $\psi$ based on its mean SFRs over the past 0--10, 10--50, and 50--100~Myr:
\begin{align}
    &u = \log_{10}\frac{{\rm SFR}_{0-10}}{{\rm SFR}_{10-50}}, \quad v = \log_{10}\frac{{\rm SFR}_{10-50}}{{\rm SFR}_{50-100}}, \\
    &\psi = \left\{360^\circ - \left[\left(\arctan\!2(u, v) - 45^\circ\right) \bmod 360^\circ\right]\right\} \bmod 360^\circ,
    \label{eq:psi}
\end{align}
where $\psi$ increases with the time since the last SFR maximum.$\psi$ is the polar angle of a galaxy in the plane of its earlier ($v$) and more recent ($u$) change in SFR, measured clockwise from the diagonal $u=v>0$. Over one burst cycle a galaxy goes once around this plane: near $\psi=0^\circ$ it is rising in both intervals, so its SFR is at its maximum at the time of observation; near $\psi=90^\circ$ it has just passed the peak ($u<0<v$); near $\psi=180^\circ$ it is declining in both intervals; and near $\psi=270^\circ$ it begins to rise again ($v<0<u$). Galaxies whose stacked SFHs peak 50--100~Myr ago occupy the oldest phase interval resolved by $\psi$.

We calculate $\psi$ for each posterior draw and adopt the circular mean as the representative phase of each galaxy. Similarly, we calculate $A_{1500}$ (Eq.~\ref{eq:a1500}) for each draw and adopt its posterior median (Appendix~\ref{app:derived}). To remove trends with stellar mass and redshift, we define $\Delta y$ as the residual from a quadratic surface in $(\log_{10}M_\ast,z)$, including all second-order and cross terms, fitted to the sample by least squares. Although we cannot directly observe how attenuation evolves within an individual galaxy, we can compare galaxies at different inferred burst phases to investigate how attenuation varies throughout the burst cycle. Our sample covers the full range of burst phases, including galaxies whose SFR peaked 30--70~Myr before observation, allowing us to investigate how attenuation varies with time since the inferred SFR peak (\S~\ref{sec:burstphase}, Fig.~\ref{fig:burstphase}b).

\paragraph*{Attenuation across the burst phases.}
Attenuation varies moderately with burst phase (Fig.~\ref{fig:burstphase}a). It is highest near the SFR peak and gradually decreases towards later phases, varying by a factor of 2.3 in $A_{1500}$ across the cycle. The lowest attenuation occurs in galaxies observed late in the cycle ($\psi\approx200^\circ$--$300^\circ$), whose SFR has declined since the burst and may have begun to rise again in the last 10~Myr. The Balmer-decrement attenuation shows a similar gradual phase dependence (Fig.~\ref{fig:burstphase}a). In stacked SFHs, galaxies in the two phase bins nearest the peak reached their maximum SFR a few Myr before observation, and those at $\psi\approx90^\circ$--$300^\circ$ 30--70~Myr before.

\paragraph*{Amplitude and timing of the decline.}
Our burst phase is defined using SFRs averaged over intervals of at least 10~Myr, potentially smoothing over shorter events (\S~\ref{sec:limitations}). Injection tests show that it recovers most of an attenuation decline occurring over a few tens of Myr, but only about a third of a decline completed within a few Myr of the SFR peak. Defining the phase on shorter timescales recovers nearly all of such a rapid decline, yet produces an even smaller attenuation variation in the observed sample. Predicted dust-clearing timescales range from $\approx3$--10~Myr for the outflow velocities in the attenuation-free model of \citet{Ferrara:2024aa} to 5--25~Myr for three $z\gtrsim10$ galaxies interpreted as being in an outflow phase by \citet{Nakazato:2025aa}. Since the phase bins nearest the peak correspond to SFR maxima occurring a few to a few tens of Myr earlier, a sufficiently large decline on these timescales should produce a noticeable change between adjacent bins.

We instead observe a gradual decline in attenuation across the inferred post-burst phases, with no clear evidence for an abrupt, large reduction. This does not necessarily imply dust clearing: the trend could reflect dust removal or destruction \citep{Garaldi:2026aa}, changes in dust geometry \citep{Narayanan:2018aa,Mushtaq:2023aa}, or young stars emerging from their birth clouds as the burst ages \citep{charlot_simple_2000}(\S~\ref{sec:disc_dust}). We therefore cannot distinguish these mechanisms or exclude dust clearing that occurs gradually or on timescales too short for our SFH reconstruction to resolve \citep{Tsuna:2023aa,Ferrara:2024ab}.

\begin{figure*}
    \centering
    \includegraphics[width=0.9\textwidth]{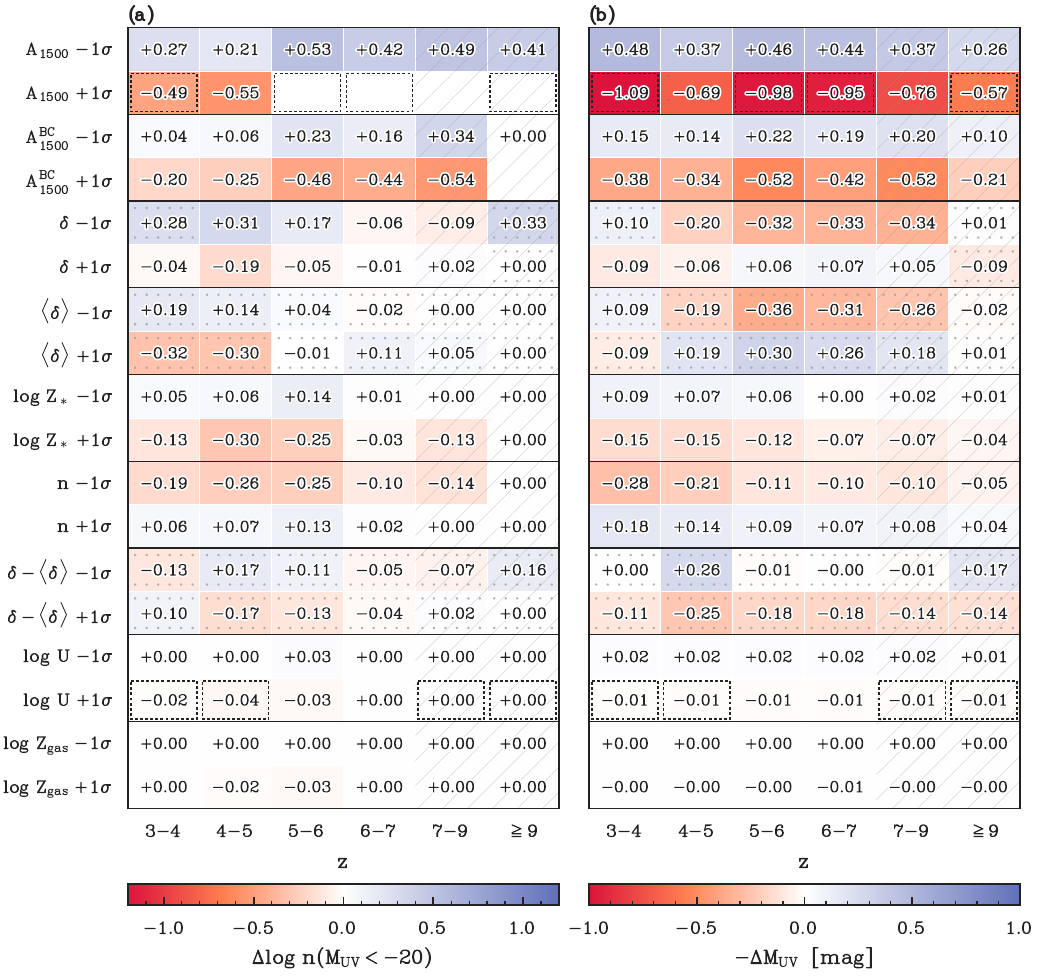}
    \caption{Sensitivity of the UVLF bright end to individual model ingredients at equal amplitude. Panel (a) shows the change in the number of galaxies brighter than $M_{\rm UV}=-20$, $\Delta\log n(<-20)$ (Eq.~\ref{eq:dlogn}), and panel (b) the median UV brightening, $-\Delta M_{\rm UV}$. Each ingredient is varied by $\pm\sigma_{\rm pop}$, its population spread within each redshift bin, while all other parameters are held fixed. These changes measure model sensitivity, not physical contributions. Dotted cells mark ingredients whose largest computed counterfactual shifts them by less than $\sigma_{\rm pop}$, so that the value is a lower limit, dashed outlines indicate extrapolation, and blank cells contain fewer than three bright galaxies. Hatched columns mark low-number regimes at $z\geq7$. At equal amplitude, dust has the strongest effect on the UVLF bright end at $3\leq z<6$ of the ingredients that reach $\sigma_{\rm pop}$; at $z<5$ the SFH changes the counts by about half as much as an increase in the attenuation.}
    \label{fig:ranking}
\end{figure*}

\subsection{Sensitivity of the bright end to the model ingredients}
\label{sec:uvlf_parameters}

To identify which model ingredients most strongly affect the bright end of the UV luminosity function, we examine how the predicted \muv\ changes when individual parameters are varied. For each posterior draw from \ceri, we modify one parameter at a time, keeping all others fixed, and recompute the spectrum and \muv. We then calculate the difference between the posterior-median \muv\ before and after the modification. These tests measure the sensitivity of the inferred UVLF to individual model ingredients, not their their physical contributions. In particular, extreme parameter changes may produce unrealistic combinations of galaxy properties.

For a counterfactual in which one ingredient $X$ is replaced by $X'$ (\S~\ref{sec:quantities}), the change in each galaxy's magnitude at fixed stellar mass is
$\Delta M_{{\rm UV},i}=M_{{\rm UV},i}(X\to X')-M_{{\rm UV},i}$.
We summarise the population response by the weighted median $\Delta M_{{\rm UV},i}$ and the change in the selection-corrected number of galaxies brighter than $M_{\rm UV}=-20$:
\begin{equation}
    \Delta \log n(<-20)
    = \log_{10} n(<-20;\,X\to X')
    - \log_{10} n(<-20),
    \label{eq:dlogn}
\end{equation}
where $n(<-20)=\sum_i w'_i\,\mathbbm{1} [M_{{\rm UV},i}<-20]$ is proportional to the number density. We retain the original selection weights $w'_i$ for the counterfactual magnitudes, so the survey volume cancels in $\Delta\log n$.

We consider two types of counterfactual. First, we apply fixed prescriptions to individual ingredients: removing or doubling dust attenuation, removing only the birth-cloud component, changing the slope of the attenuation curve, and setting stellar metallicity, gas-phase metallicity, or the ionisation parameter to their prior limits. For the SFH, we decompose the residuals $\delta_i$ of Eq.~\ref{eq:resid} into a population mean $\langle\delta\rangle(M_\ast,z)$, described by a weighted quadratic surface fitted separately for each ratio, and the galaxy-to-galaxy scatter around it. Setting $\delta\to0$ places galaxies on the prior main-sequence track, $\langle\delta\rangle\to0$ removes only the mean offset, and $\delta\to\langle\delta\rangle$ removes only the scatter. Because these interventions differ in amplitude, their effects cannot be directly ranked. We therefore also compare ingredients at equal amplitude, shifting each by $\pm\sigma_{\rm pop}$, the population spread of its own coordinate in each redshift slice. We compute each counterfactual at fixed amplitudes (steps, such as removing or doubling the SFH residual) and interpolate each galaxy's \muv\ between them to the amplitude $\pm\sigma_{\rm pop}$ (Fig.~\ref{fig:ranking}). For the total SFH residual, the largest step shifts \bhun\ by at least $\sigma_{\rm pop}$ in nine of the ten cases at $z<9$ (five redshift slices, removing or doubling), whereas the steps of its mean and scatter alone fall short of $\sigma_{\rm pop}$, so the latter equal-amplitude effects are lower limits (dotted entries in Fig.~\ref{fig:ranking}).

At equal amplitude, dust has the largest effect on the bright end of the ingredients that reach $\sigma_{\rm pop}$ (Fig.~\ref{fig:ranking}). Below $z=6$, increasing the total attenuation by $\sigma_{\rm pop}$ lowers $n(<-20)$ by about 0.5~dex, and decreasing it raises $n(<-20)$ by 0.2--0.5~dex. At $z<5$, the SFH changes $n(<-20)$ by up to 0.3~dex at $\sigma_{\rm pop}$, as much as a decrease and about half as much as an increase in the attenuation. At $z=5$--7, birth-cloud attenuation has a comparable effect. The full counterfactuals reinforce the importance of dust: removing or doubling all attenuation produces the largest change in the bright-end counts in every redshift slice, while steepening or flattening the attenuation curve has an effect similar to increasing or decreasing dust, respectively.

Below $z=5$, the SFH changes $n(<-20)$ by about 0.3~dex at $\sigma_{\rm pop}$, comparable to or larger than that of stellar metallicity. Its effect depends, however, on whether we change the population mean or the scatter. Placing every galaxy on the prior main-sequence track increases the number of bright galaxies at low redshift, where massive galaxies dominate the bright end and typically have declining SFHs (\S~\ref{sec:variability}). The intervention raises their recent star formation and therefore brightens them. At $z=6$--9, by contrast, it reduces the number of bright galaxies, although this rests on few galaxies (20 and 10 brighter than $M_{\rm UV}=-20$ at $z=6$--7 and 7--9). Removing only the mean SFH offset likewise increases $n(<-20)$ at $z<5$, but leaves it nearly unchanged at higher redshift.

The galaxy-to-galaxy SFH scatter, which includes bursts, has a much smaller effect on the absolute bright end. Removing or doubling this scatter changes $n(<-20)$ by at most 0.17~dex, and removing it leaves the counts nearly unchanged at $z=6$--9. These interventions do not reach the amplitudes required for a complete equal-$\sigma_{\rm pop}$ comparison. Nevertheless, their small effect is consistent with \S~\ref{sec:bright_tail}: bursts preferentially populate the UV-bright tail at fixed stellar mass, particularly among lower-mass galaxies, but do not dominate the absolute bright end, which at $z<5$ is largely populated by more massive galaxies.

Stellar metallicity has a smaller effect than dust at equal amplitude (Fig.~\ref{fig:ranking}). In the full counterfactuals, setting it to the high-metallicity prior limit reduces the number of bright galaxies by up to 0.5~dex, whereas setting it to the low-metallicity limit changes the counts by at most 0.20~dex. These changes span several $\sigma_{\rm pop}$ and should therefore not be compared directly with the equal-amplitude results. Gas-phase metallicity and the ionisation parameter have no significant effect on the UVLF even when set to their prior limits, because they barely affect the UV continuum.

The relative effects also depend on the quantity used to measure them: in the full counterfactuals, setting $\delta\to0$ changes the bright-end counts by about 0.3~dex but the median magnitude by at most 0.2~mag at $z<5$, whereas at $5\leq z<9$ it shifts the median by about 0.5~mag but the counts by at most 0.31~dex. Dust has the largest effect by either measure. Overall, of the ingredients that reach $\sigma_{\rm pop}$, , dust has the strongest influence on the bright end of the UVLF, with birth-cloud attenuation having a comparable effect at $z=5$--7. Changes to the SFH have a smaller but non-negligible effect, particularly on bright-end counts at $z<5$, where their impact is about half that of increased attenuation. However, removing burst-driven SFH scatter has a much smaller effect on the counts, while nebular parameters have negligible influence

\section{Discussion}
\label{sec:discussion}
Our results distinguish between the processes that make galaxies unusually UV-bright at fixed stellar mass and those associated with the absolute bright end of the UV luminosity function. We first discuss this distinction and the interplay between recent star formation and dust attenuation (\S~\ref{sec:roles}). We then discuss the evolution of attenuation across the burst cycle (\S~\ref{sec:disc_dust}), what our results imply for UV-bright galaxies at $z>10$ (\S~\ref{sec:highz}), and the main limitations of our analysis (\S~\ref{sec:limitations}).

\subsection{The roles of stellar mass, recent star formation, and dust}
\label{sec:roles}

Our results show that UV brightness depends on both stellar mass and recent star formation, with their relative importance changing across the mass range. Below $10^9\,\Msun$, recent star formation is the strongest predictor of \muv\ at fixed stellar mass (\S~\ref{sec:fixedmass}). At $8\leq\mass<9$, UV-brighter galaxies are also more likely to have rising SFHs (\S~\ref{sec:bright_tail}). Although the associated increase in dust attenuation partly offsets the UV brightening, it does not cancel it (\S~\ref{sec:selfobscured}). Moreover, dust contributes less to the variation in UV brightness at low masses because the spread in attenuation is smaller than in more massive galaxies (\S~\ref{sec:fixedmass}). Recent bursts can therefore make low-mass galaxies unusually UV-bright for their stellar mass.

At higher masses, the picture changes. Above $10^9\,\Msun$, stellar age and dust attenuation contribute as much as, or more than, \sfrten\ to the variation in UV brightness at fixed mass (\S~\ref{sec:fixedmass}), and UV-brighter galaxies show no significant excess of rising SFHs relative to other galaxies of the same mass (\S~\ref{sec:bright_tail}). At $z=3$--5, the absolute bright end of the UVLF is dominated by these massive galaxies, approximately half of which have declining SFHs (\S\S~\ref{sec:naive}, \ref{sec:bright_tail}). Their high UV luminosities are therefore more closely associated with their stellar mass than with recent bursts. Consistent with this, removing galaxy-to-galaxy SFH scatter has little effect on the absolute bright-end counts. Among the fitted model ingredients varied by comparable amplitudes, dust attenuation has the largest effect on these counts. At $z<5$, changing the SFH alters the counts by about 0.3~dex, approximately half the effect of increased attenuation (\S~\ref{sec:uvlf_parameters}).

\subsection{Interpreting the attenuation decline after bursts}
\label{sec:disc_dust}

The gradual decline in attenuation after the SFR peak (\S~\ref{sec:burstphase}) could have several physical explanations. Dust may be removed by outflows or destroyed during bursts \citep{Ferrara:2024aa,Garaldi:2026aa}, its spatial distribution may change \citep{Narayanan:2018aa,Mushtaq:2023aa}, or young stars may emerge from their birth clouds as the burst ages \citep{charlot_simple_2000}. Our measurements cannot distinguish between these possibilities.
Another possible explanation is the attenuation-free model (AFM) \citep{Ferrara:2024aa,Ferrara:2024ab,Ferrara:2025aa}, in which intense star formation drives radiation-pressure outflows that clear dust within a few Myr. We observe a gradual decline and no abrupt drop in attenuation, although our temporal resolution may smooth over brief clearing events (\S~\ref{sec:burstphase}). The overall trend is consistent with the sequence illustrated by \citet[][fig.~5]{Ferrara:2026ab} despite the seemingly longer clearing timescale. A direct test of the AFM would also require constraints on galaxy size, metallicity, and dust geometry, which influence the conditions for radiation-pressure-driven outflows but are not included in our analysis. The AFM therefore remains a possible explanation for the observed attenuation trend.

\subsection{Implications for $z>10$}
\label{sec:highz}

Our sample contains only 11 galaxies at $z>10$, so our results primarily provide a lower-redshift reference for interpreting the UV-bright population at these early epochs. Proposed explanations for its abundance (\S~\ref{sec:introduction}) include greater stellar mass at fixed halo mass \citep{Dekel:2023aa, Li:2024aa}, higher UV luminosity per unit stellar mass \citep{Inayoshi:2022aa, Shen:2023aa, Sun:2023ab, Yung:2024aa, Liu:2025aa}, and reduced dust attenuation \citep{Ferrara:2023aa, Ferrara:2024aa, Ferrara:2025aa}. Our results highlight the importance of distinguishing galaxies that are unusually UV-bright for their stellar mass from those at the absolute bright end. At $z=3$--5, enhanced recent star formation can make galaxies unusually UV-bright for their mass, whereas the absolute bright end is dominated by more massive galaxies (\S~\ref{sec:roles}). Whether the UV-bright population at $z>10$ is similarly dominated by massive galaxies or by galaxies temporarily brightened by recent star formation remains uncertain, particularly as galaxy sizes, gas fractions, metallicities, and dust properties may evolve.

The contribution of bursty star formation to the UVLF depends on both the amplitude and duration of SFR fluctuations. In JADES photometry, the intrinsic scatter of the SFMS is larger on 10~Myr than on 100~Myr timescales \citep{Simmonds:2025aa}. However, the UV continuum traces star formation over a longer, age-dependent interval \citep{Kennicutt:2012aa, Caplar:2019aa}, reducing its sensitivity to short bursts (\citealt{Furlanetto:2022aa}; \citealt{Pallottini:2023aa}). Halo-based models also consider the scatter in UV luminosity at fixed halo mass, which can substantially increase bright-end counts because of the steep halo mass function \citep{Shen:2023aa}. This scatter includes differences in stellar mass at fixed halo mass and in long-term SFHs, not only bursts \citep{Simmonds:2025aa}, and differs from the scatter in \muv\ at fixed stellar mass measured here (\S~\ref{sec:fixedmass}).

Simulations predict different effects of variability on the UVLF. In FIRE-2, bursty SFHs increase the abundance of galaxies with $M_{\rm UV}\lesssim-20$ at $z\gtrsim10$ by up to an order of magnitude relative to temporally smoothed SFHs \citep{Sun:2023ab}. In contrast, SFR fluctuations in \textsc{serra} at $z\approx7.7$ are too small to explain the observed abundance at $z\gtrsim10$ \citep{Pallottini:2023aa}. \citet{Kravtsov:2024aa} reproduce the UVLF at $z>10$ by increasing the amplitude of SFR fluctuations with redshift, predicting younger stellar populations among the brightest galaxies and greater scatter in ${\rm SFR}_{10}/{\rm SFR}_{100}$. Our results do not yet constrain this evolution, as the rising-SFH fraction at the highest reliable redshifts remains consistent with the prior (\S~\ref{sec:variability}). Moreover, our counterfactual removes galaxy-to-galaxy differences in the fitted SFHs at fixed stellar mass, while preserving each galaxy's stellar mass and dust attenuation. By contrast, \citet{Sun:2023ab} smooth individual SFHs over time while conserving their total formed stellar mass. The small response of our bright-end counts at $z<9$ (\S~\ref{sec:uvlf_parameters}) therefore does not necessarily contradict the larger effects predicted at $z>10$.

The role of dust is similarly uncertain. In \textsc{Shark}, most star formation in starbursts remains obscured to $z\approx13$ or beyond, depending on the dust-to-metal scaling \citep{Lagos:2026aa}. By contrast, bursts in \textsc{thesan-zoom} destroy and eject dust, producing a deficit at high sSFR \citep{Garaldi:2026aa}. \citet{Tsuna:2023aa} predict that radiation pressure clears the dust of galaxies below a certain stellar mass in bursts shorter than $\approx20$~Myr, while supernovae replenish it in bursts longer than $\approx30$~Myr. The attenuation-free model likewise predicts an obscured phase followed by outflow-driven dust clearing for JADES-GS-z14-0 \citep{Ferrara:2024ab}. However, the amount and shape of attenuation also depend on the geometry of dust relative to young stars \citep{Mushtaq:2023aa, Narayanan:2018aa}. Observations show increasing attenuation above the star-forming main sequence in JADES \citep{Looser:2025aa}, while at least half of the star formation in massive $z\simeq7$ galaxies with steeply rising SFHs is obscured \citep{Fisher:2026aa}. Our observed increase in attenuation with \sfrten\ and its gradual decline after bursts (\S\S~\ref{sec:selfobscured} and \ref{sec:burstphase}) could therefore reflect changes in dust content, geometry, or both.

Recent star formation likely remains important for UV luminosity at fixed stellar mass at $z>10$. However, the strength of the dust response, its gradual decline after bursts, and the limited effect of removing SFH scatter cannot necessarily be extrapolated to these redshifts, as they depend on the galaxy population and model assumptions.

For the 21 galaxies at $z\geq9$, attenuation at fixed stellar mass is about a factor of two lower than at $7\leq z<9$, comparable to the galaxy-to-galaxy scatter there. The small sample and limited sensitivity to very low attenuation (Fig.~\ref{fig:injection_recovery}f) prevent us from establishing whether attenuation evolves. We also cannot exclude rapid dust clearing or heavily obscured phases missed by UV selection, which favours galaxies during periods of enhanced star formation \citep{Sun:2023aa}. Larger samples at $z>10$, with constraints on stellar masses, SFHs over multiple timescales, obscured star formation, and galaxy sizes, are needed to establish how variability and dust shape the bright end of the UVLF, and whether UV-luminous galaxies are unusually massive, unusually bright for their mass, weakly attenuated, or a combination of these.

\subsection{Limitations}
\label{sec:limitations}

Our conclusions are subject to six main limitations. Most of them are common to studies that infer SFHs from SED fitting, and we test each one directly with selection weights, prior-only fits, injection--recovery tests and refits with a different stellar population model.

\begin{enumerate}[label=(\roman*)]

\item \textbf{Selection.}
Population statistics depend on the selection weights $w'$ and the adopted \muv\ limit (\S~\ref{sec:muv_completeness}). The weights account for differences in inclusion probability across $M_{\rm UV}$, redshift, and stellar mass, but cannot recover populations absent from the sample, such as mini-quenched or fully obscured galaxies. We therefore restrict population conclusions to the reliable and caution bins of Fig.~\ref{fig:reliability_map}. In particular, the strength of the rising-SFH trend in the UV-bright tail depends on faint galaxies with low inclusion probabilities near the selection limit (\S~\ref{sec:bright_tail}). Tests of mechanisms in individual galaxies instead use unweighted samples (treatment N).

\item \textbf{SFH prior.}
The inferred burst ratio \bhun\ is particularly sensitive to the SFH prior. When a galaxy formed much of its stellar mass recently, the prior can also determine \sfrtenfifty\ (flag M, tested by flag Q). Consequently, the increase of $b_{10-100}$ with redshift is likely driven by the prior, and the rising fraction at the highest reliable redshift is consistent with its prior expectation (flag Q; \S~\ref{sec:variability}). Population-wide SFH offsets are likewise defined relative to the prior (\S~\ref{sec:uvlf_parameters}). We assess the sensitivity of these results to alternative priors in Table~\ref{tab:robustness}.

\item \textbf{Shared fit inputs and model assumptions.}
H$\alpha$ constrains \sfrten\ in the SED fits, so it is not independent of the H$\alpha$/UV comparison or of the grouping used to examine the Balmer decrement as a function of \sfrten. The birth-cloud attenuation prescription also introduces part of the dust--\sfrten\ relation, while posterior covariance between dust and \sfrten\ can strengthen it (\S\S~\ref{sec:selfobscured}--\ref{sec:burstphase}). However, dust and the non-parametric SFH are both fitted freely, so the dependence of attenuation on burst phase is not imposed. Where possible, we use the observed Balmer decrement or H$\alpha$/UV ratio as additional checks, and account for covariance and dilution when bounding $f_{\rm obsc}$.
The assumed IMF is another model limitation. 
However, we find no excess in H$\alpha$/UV or H$\alpha$ equivalent width among the brightest galaxies at fixed inferred recent SFH, consistent with both an unchanged and a moderately top-heavy IMF. A strongly top-heavy IMF is disfavoured at only $0.9\sigma$. This test remains indirect because the SFHs were inferred assuming a Chabrier IMF (Appendix~\ref{sec:imf}).

\item \textbf{Temporal resolution.}
Our burst ratios and burst phase use SFRs averaged over at least 10~Myr. The burst-phase clock can therefore smooth over events lasting only a few Myr (\S~\ref{sec:burstphase}), including rapid dust clearing of the kind predicted by \citet{Ferrara:2024aa}, which takes $\approx3$--10~Myr for his outflow velocities of 65-300~km\,s$^{-1}$.

\item \textbf{Limited high-redshift statistics.}
Only 15 galaxies at $z\geq7$ have $M_{\rm UV}<-20$, too few to distinguish the effects of the population mean SFH offset from its scatter. The $z\geq9$ sample is also too small to establish whether the trends measured at lower redshift continue unchanged.

\item \textbf{Stellar population model dependence.}
The choice of stellar population model can have a greater effect on inferred recent star formation than the SFH prior. Refitting with BPASS v2.2 binary models \citep{Eldridge:2017aa, Stanway:2018aa} for 256 representative galaxies, under the same forward model including priors, lowers \sfrten\ by about 0.2~dex and \bhun\ by about 0.5, and reduces the fraction of rising SFHs.

\end{enumerate}

Several observations could address these limitations. Longer-wavelength dust-continuum measurements, for example with ALMA, could reveal fully obscured bursts and heavily obscured phases missed by UV selection. Deeper spectroscopy of fainter galaxies and larger samples at $z>9$ would improve population completeness and high-redshift statistics. Stellar-population models with variable IMFs, combined with nebular grids that include harder ionising sources and He\,\textsc{ii} emission \citep{Garofali:2024aa, Li:2025aa}, would enable a more direct IMF test, while star-formation tracers sensitive to timescales shorter than 10~Myr, such as Wolf--Rayet features, which are present mainly within $\approx3$--6~Myr of a burst \citep{Schaerer:1998aa}, would better resolve the evolution of dust across individual bursts.

\section{Conclusions}
\label{sec:summary}

The unexpectedly high number density of UV-bright galaxies discovered by \textit{JWST} at $z>10$ has motivated explanations involving bursty star formation, enhanced star-formation efficiency, reduced dust attenuation, and a top-heavy initial mass function (IMF). We investigate the roles of recent star formation, stellar mass and dust attenuation using a spectroscopically confirmed JADES sample at $3\leq z\leq9$, combining NIRCam photometry and NIRSpec emission lines with the SED-fitting code \ceri. After assessing the mass and redshift ranges in which our sample provides reliable population constraints, we examine which galaxy properties are associated with UV brightness and how the inferred bright end responds to changes in the fitted model.

Our main findings are:

\begin{enumerate}

    \item \textbf{The prevalence of rising star-formation histories depends on stellar mass and redshift.}
    At $z<5$, low-mass galaxies are more likely to have rising SFHs, whereas galaxies above $10^{9.5}\,\Msun$ are predominantly declining (\S~\ref{sec:variability}). The rising fraction at fixed stellar mass appears to increase with redshift, but at $z=6$--7 it is consistent with the expectation from our SFH prior alone, so we cannot establish a physical redshift trend. The intrinsic scatter of the burst ratio \bhun\ = $\log_{10}(\burst)$ at fixed stellar mass shows no monotonic mass dependence at $3\leq z<5$ and $10^{8.5}$--$10^{10}\,\Msun$: low-mass galaxies are more often rising, but we find no evidence that they are more variable.

    \item \textbf{Recent star formation is most closely associated with UV brightness at fixed stellar mass, but does not necessarily dominate the absolute bright end.}
    At fixed stellar mass, \sfrten\ is the strongest predictor of \muv. UV-brighter galaxies generally have higher recent SFRs, younger stellar populations, and less dust attenuation at fixed \sfrten\ (\S~\ref{sec:fixedmass}). At $8\leq\mass<9$, they are also more likely to have rising SFHs, with the strongest trend at $8\leq\mass<8.5$, where the odds of having a rising SFH increase by a factor of 2.3 for each magnitude brighter in \muv. However, at $8.5\leq\mass<9$, the trend is weak at $z=3$--5 and disappears for brighter selection cuts, while at $z=5$--7 it only marginally exceeds the effect of correlated fitting uncertainties (\S~\ref{sec:bright_tail}). The absolute bright end is instead dominated by more massive galaxies at $z=3$--5, about half of which have declining SFHs. In this redshift range, UV brightness increases with stellar mass up to $\sim10^{10}\,\Msun$, beyond which increasing dust attenuation causes the median \muv\ to become fainter again (\S~\ref{sec:naive}).

    \item \textbf{Dust reduces the UV brightening associated with bursts.}
    At fixed stellar mass and redshift, attenuation increases with recent star formation, cancelling an estimated 16--73~per~cent of the unattenuated UV gain for bursts that remain visible in the UV. This range brackets the results of our covariance and dilution tests and does not represent a statistical confidence interval (\S~\ref{sec:selfobscured}). The observed Balmer decrement independently excludes complete cancellation, although it cannot confirm the magnitude of the partial effect (\S~\ref{sec:balmer}). Attenuation also declines gradually after the inferred SFR peak, varying by a factor of approximately 2.3 across the burst phases (\S~\ref{sec:burstphase}).

    \item \textbf{The inferred absolute bright end is highly sensitive to dust, and only weakly to burst-driven SFH scatter, within our fitted model.}
    In counterfactual calculations at fixed stellar mass, removing galaxy-to-galaxy SFH scatter, including bursts, changes the number of galaxies brighter than $M_{\rm UV}=-20$ by at most 0.17~dex, whereas removing all dust, a much larger perturbation, produces the largest change of all full counterfactuals (\S~\ref{sec:uvlf_parameters}). When model ingredients are varied by equal amplitudes relative to their population spreads, dust attenuation has the strongest effect of the ingredients that reach this amplitude at $z<6$, and birth-cloud attenuation becomes comparable at $z=5$--7. At $z<5$, the SFH changes the number of bright galaxies by about 0.3~dex, about half as much as an increase in the attenuation. These experiments quantify model sensitivity rather than the physical contributions of individual mechanisms.

    \item \textbf{Our results do not require a top-heavy IMF.}
    The brightest galaxies show no excess in H$\alpha$/UV or H$\alpha$ equivalent width beyond that expected from their inferred recent SFHs. Conditional on the fixed Chabrier IMF used in the fits, neither a moderately nor a strongly top-heavy IMF is excluded; the latter is disfavoured at only $0.9\sigma$ (Appendix~\ref{sec:imf}).

\end{enumerate}

These findings are limited by the SFH prior, the stellar population model, the shared observational constraints entering the SED fits, and the UV and spectroscopic selection of our sample. In particular, the small number of galaxies at $z\geq9$ prevents a direct test of whether the same relationships hold at $z>10$. Larger and deeper spectroscopic samples, together with longer-wavelength observations of obscured star formation, will be needed to establish how these mechanisms operate in the earliest galaxies.

Our results suggest that the processes shaping the UV-bright tail at fixed stellar mass are not necessarily those that determine the absolute bright end of the UVLF. Bursts can make low-mass galaxies unusually UV-bright despite partial dust obscuration, but at $z=3$--5 the most UV-luminous galaxies are primarily massive, and their inferred abundance is particularly sensitive to dust attenuation and to their mean SFH. Explaining the excess of UV-bright galaxies at $z>10$ will require measuring the contributions of stellar mass, recent star formation and dust attenuation together.

\section*{Acknowledgements}

The paper manuscript is partly supported by generative artificial intelligence: \textit{Claude code Opus~5.5} helped produce the figures, revise the text for clarity and check if the citations are correct.
AS thanks Andrea Ferrara and Stephane Charlot for comments and discussion.
AS acknowledges support from the STFC Centre for Doctoral Training in Data Intensive Science at the University of Cambridge (ST/W006812/1), partly funded by the UKRI Frontier Research grant RISEandFALL, and a DiRAC Seedcorn award (project dp476), which enabled the model testing in this work.
ST acknowledges support by the Royal Society Research Grant G125142.
BER acknowledges support from the NIRCam Science Team contract to the University of Arizona, NAS5-02105, and JWST Program 3215.
AJB acknowledges funding from the ``FirstGalaxies'' Advanced Grant from the European Research Council (ERC) under the European Union's Horizon 2020 research and innovation programme (Grant) agreement No. 789056).
ECL acknowledges support of an STFC Webb Fellowship (ST/W001438/1). ALD acknowledges support from the Hintze Family Charitable Foundation.
FDE acknowledges support by the Science and Technology Facilities Council (STFC), by the ERC through Advanced Grant 695671 ``QUENCH'', and by the UKRI Frontier Research grant RISEandFALL.

This work is based on observations made with the NASA/ESA/CSA \textit{James Webb Space Telescope}. The data were obtained from the Mikulski Archive for Space Telescopes at the Space Telescope Science Institute, which is operated by the Association of Universities for Research in Astronomy, Inc., under NASA contract NAS 5-03127 for \textit{JWST}. These observations are associated with programmes 1180, 1181, 1210, 1286, 1287 and 3215 (NIRSpec; JADES DR4) and 1176, 1180, 1181, 1210, 1264, 1283, 1286, 1287, 1895, 1963, 2079, 2198, 2514, 2516, 2674, 3215, 3577, 3990, 4540, 4762, 5398, 5997, 6434, 6511 and 6541 (NIRCam; JADES DR5). We thank the teams of programmes 1895 \citep[FRESCO;][]{Oesch:2023aa}, 1963 \citep[JEMS;][]{Williams:2023aa}, 2079, 2514, 3215 \citep[JOF;][]{Eisenstein:2025aa}, 3577, 3990, 6434 and 6541 for designing their observations with a zero-exclusive-access period.

This work used the DiRAC Extreme Scaling service (Tursa) at the University of Edinburgh, managed by the EPCC on behalf of the STFC DiRAC HPC Facility (www.dirac.ac.uk). The DiRAC service at Edinburgh was funded by BEIS, UKRI and STFC capital funding and STFC operations grants. DiRAC is part of the UKRI Digital Research Infrastructure.

This work made use of \textsc{JAX} \citep{jax2018github}, \textsc{BlackJAX} \citep{Cabezas:2024aa}, \textsc{NumPy} \citep{harris2020array}, \textsc{SciPy} \citep{Virtanen:2020aa}, \textsc{Matplotlib} \citep{Hunter:2007}, \textsc{FSPS} \citep{conroy_propagation_2009, Conroy:2010aa}, \textsc{python-fsps}, \textsc{pandas}, \textsc{h5py}, \textsc{smplotlib}, and \textsc{Astropy} \citep{Astropy-Collaboration:2013aa, Astropy-Collaboration:2018aa, Astropy-Collaboration:2022aa}.

\section*{Data Availability}

The JADES DR5 photometric and DR4 spectroscopic catalogues are publicly available at MAST (doi:10.17909/8tdj-8n28); the JEMS medium-band imaging \citep{Williams:2023aa} is available at doi:10.17909/fsc4-dt61. We use the DR5 photometric catalogue v5.1 and the R1000 line fluxes of the DR4 spectroscopic catalogue (release v5.1.1, \texttt{Combined\_DR4\_external\_v1.2.1}), merged by field and NIRCam ID (merged catalogue v5.1). The SED-fitting code \ceri\ \citep{Stoffers:2026ab} is publicly available, and the stellar-population grids used here are archived on Zenodo (doi:10.5281/zenodo.22937956). The per-galaxy posterior summaries (stellar masses, star-formation rates on all baselines, burst ratios, and the derived quantities used in the figures) will be released as a value-added catalogue on publication; until then they are available from the corresponding author on request. Machine-readable versions of the appendix tables and further measurements quoted in the text are available on Zenodo (doi:10.5281/zenodo.23600428). The Zenodo record also contains six further tables: the mass dependence of the selection in each redshift slice, the applications of the selection-aware likelihood, the intrinsic scatter in $M_{\rm UV}$ at fixed stellar mass and redshift in each reliable and caution bin, the null and consistency tests, the summary statistics of the injection--recovery test, and the fits of the stellar mass--UV magnitude relation.

\bibliographystyle{mnras}
\bibliography{example}

\appendix
\renewcommand{\floatpagefraction}{0.5}\renewcommand{\dblfloatpagefraction}{0.5}%

\section{Definitions of derived quantities}
\label{app:derived}
\label{sec:quantities}
Every per-galaxy quantity used in this paper is computed for each posterior draw $k = 1, \ldots, K$ ($K = 400$) of the fiducial \texttt{bursty} fit and summarised by the median over the draws; its uncertainty is the 16--84~per~cent range of the draws. We therefore work with the full posterior until we arrive at the cited quantities.

\paragraph*{Time-averaged star-formation rates.} With the piecewise-constant SFH (\S~\ref{sec:sfhprior}) on bins with lookback-time edges $t_{i-1} < t_i$ ($t_0 = 0$), the star-formation rate averaged over the last $t$ is
\begin{equation}
    {\rm SFR}_t = \frac{1}{t} \sum_{i=1}^{N} {\rm SFR}_i\, \max\!\left[0,\; \min(t_i, t) - \min(t_{i-1}, t)\right],
    \label{eq:sfrt}
\end{equation}
evaluated for $t = 10$, 20, 30, 50 and 100~Myr. The mean SFR between 10 and 50~Myr before observation is $\sfrtenfifty=(50\,{\rm SFR}_{50}-10\,{\rm SFR}_{10})/40$, i.e. the window $[10,50]$~Myr of Eq.~(\ref{eq:sfrt}). The specific star-formation rate is ${\rm sSFR}_t = {\rm SFR}_t/M_\ast$, with $M_\ast$ the sampled stellar mass (total mass formed).

\paragraph*{Burst ratio and SFH classes.} The burst ratio on a baseline $t$ is
\begin{equation}
    b_{10-t} \equiv \log_{10}\!\left(\frac{{\rm SFR}_{10}}{{\rm SFR}_{10-t}}\right),\qquad b \equiv \bhun,
    \label{eq:burst}
\end{equation}
where ${\rm SFR}_{10-t}=(t\,{\rm SFR}_t-10\,{\rm SFR}_{10})/(t-10)$ is the mean SFR between 10 and $t$~Myr before observation, with $t=20$, 50 and 100~Myr. In 21 of the 546\,000 posterior draws of the sample (three galaxies), ${\rm SFR}_{10-20}$, \sfrtenfifty\ and ${\rm SFR}_{10-100}$ are all zero while ${\rm SFR}_{10}>0$, so $b=+\infty$; these draws count as rising, since star formation has resumed after a pause of at least 90~Myr. A galaxy is called \emph{rising} if $b > +0.2$, \emph{declining} if $b < -0.2$ and constant otherwise.

\paragraph*{Age.} $t_{50}$ is the lookback time within which the most recent half of the stellar mass formed,
\begin{equation}
    \int_0^{t_{50}} {\rm SFR}(t')\,{\rm d}t' = \frac{1}{2} \int_0^{t_N} {\rm SFR}(t')\,{\rm d}t',
    \label{eq:t50}
\end{equation}
interpolated linearly inside the bin in which the cumulative mass crosses one half.

\paragraph*{UV magnitude.} $M_{\rm UV}$ is measured on the posterior model spectrum (stellar and nebular continuum plus lines, attenuated) as the mean flux density in a rest-frame top-hat window,
\begin{equation}
    M_{\rm UV} = -2.5 \log_{10} \left\langle \frac{f_\nu(\lambda)}{1+z} \left(\frac{d_L(z)}{10~{\rm pc}}\right)^{2} \right\rangle_{1450-1550\,\text{\AA}} - 48.6,
    \label{eq:muv}
\end{equation}
with $f_\nu$ the observed-frame flux density in erg\,s$^{-1}$\,cm$^{-2}$\,Hz$^{-1}$ and $d_L$ the luminosity distance.

\section{Model implementation}\label{app:implementation}

\subsection{Birth-cloud slope $\alpha_{\rm BC}$}

The birth-cloud attenuation law, $\tau_{\rm BC}(\lambda)=\hat\tau_{\rm BC}\,(\lambda/5500\,\text{\AA})^{\alpha_{\rm BC}}$, has a fixed slope of $\alpha_{\rm BC}=-1$ in the fiducial model, consistent with the \textsc{Prospector} setups of \citet{tacchella_stellar_2022} and the \textsc{FSPS} default. \citet{charlot_simple_2000} adopt a shallower slope, while steeper slopes have also been used \citep{da-Cunha:2008aa, Wild:2011aa}. In the \texttt{bursty\_freealpha} variant, $\alpha_{\rm BC}$ is instead sampled with a uniform prior (Table~\ref{tab:sfms12}).

\subsection{Metallicity prior}
The fiducial prior on stellar metallicity is uniform in $[{\rm Fe/H}]=\log_{10}(Z_\ast/\Zsun)$ from $-2.5$ to $+0.5$, spanning the \textsc{MIST}/\textsc{MILES} grid (Table~\ref{tab:sfms12}). In the \texttt{bursty\_mzr} variant, we instead couple $Z_\ast$ to the sampled stellar mass using the mass--metallicity relation of \citet{Gallazzi:2005aa}, adopting a normal prior centred on the median relation, with a standard deviation equal to the full 16th--84th percentile width of the relation; draws outside the grid are set to the grid limits. Since this relation was measured for local SDSS galaxies with $\log_{10}(M_\ast/\Msun)\gtrsim8.9$, we hold it constant below its lowest tabulated mass and apply it at the redshifts of our sample. 
\section{Fit validation and observables that trace the star-formation history}
\label{app:validation}

\begin{figure*}
    \centering
    \includegraphics[width=\linewidth]{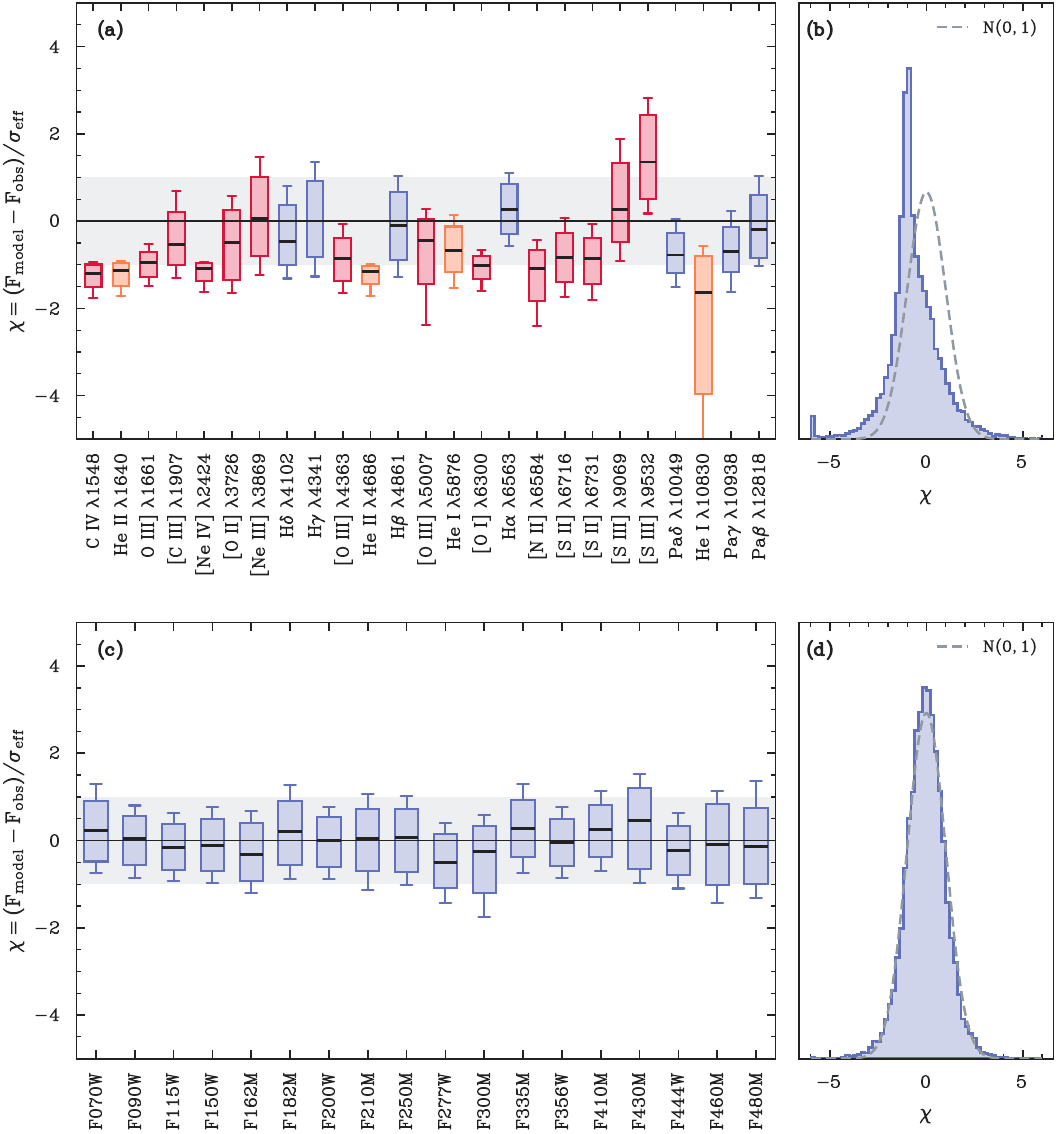}
    \caption{Pulls of the posterior-median model. Panels (a, c) show the distributions for individual emission lines and photometric filters, and (b, d) their stacked distributions. Boxes span the interquartile range and whiskers the 16th--84th percentiles, black lines mark the medians, and grey shading marks $|\chi|\leq1$. Pulls beyond $\pm6$ are collected in the outermost bins of (b, d). In (a), hydrogen, helium, and collisionally excited lines are shown in blue, coral, and red, respectively, and in (c) all filters are shown in blue. The histograms in (b, d) are normalised to unit area, and the grey dashed curve shows a standard normal distribution. Pulls include the 5 and 10~per~cent error floors of the photometry and emission lines, and no selection weights are applied. The photometry is reproduced within its uncertainties, while the model slightly underpredicts the emission-line fluxes (\S~\ref{sec:fit_quality}).}
    \label{fig:stacked_pulls}
\end{figure*}

\begin{figure*}
    \centering
    \includegraphics[width=\textwidth]{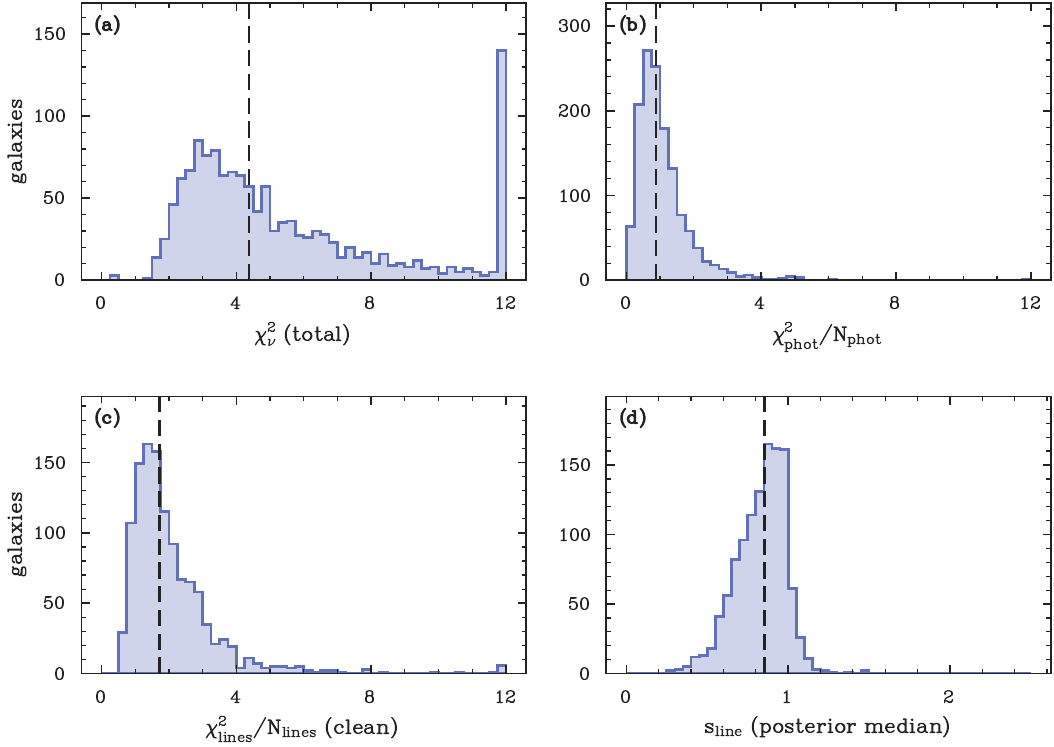}
    \caption{Fit-quality distributions. Panels show (a) the total reduced $\chi^2_\nu$, (b) $\chi^2_{\rm phot}/N_{\rm phot}$, (c) $\chi^2_{\rm lines}/N_{\rm lines}$ excluding model-limited lines (\S~\ref{sec:fit_quality}), and (d) the posterior-median line scaling $s_{\rm line}$. Panels (a, b) include all 1365 galaxies, while (c, d) include the 1168 with emission-line measurements. Dashed lines mark the medians, and the final bin includes values beyond the plotted range. In (a), the degrees of freedom are floored at one for galaxies with at most 20 data points. No selection weights are applied. The photometry is well fitted, with a median $\chi^2_{\rm phot}/N_{\rm phot}=0.88$, while the median line scaling is $s_{\rm line}=0.85$.}
    \label{fig:fit_quality}
\end{figure*}

\paragraph*{Robustness.} Table~\ref{tab:robustness} repeats the main results with the \texttt{continuity} and \texttt{hyperburst} priors, without the photometry-only galaxies, and for brighter $M_{\rm UV}$ cuts.

\begin{table*}
    \centering
    \caption{Robustness of the main results. The fiducial \texttt{bursty} fits are compared with the \texttt{continuity} and \texttt{hyperburst} fits (Table~\ref{tab:variants}; same data, code and sampler; $A_{1500}$ from the corresponding fit and $M_{\rm UV}^{0}=M_{\rm UV}-A_{1500}$), with the sample without the 197 photometry-only galaxies (1168), and with brighter $M_{\rm UV}$ cuts ($-18.0$ and $-18.5$: 1050 and 860 galaxies, $n_{\rm eff}=749$ and 743; $w'$ recalculated and the selection-aware likelihood truncated for each cut, $C(M_{\rm UV},z)$ unchanged). Sample, weights and observed $M_{\rm UV}$ are otherwise fixed; treatments follow \S~\ref{sec:muv_completeness}. $N$ is the size of the fiducial sample entering each statistic. Uncertainties are given for the fiducial and where they were computed; shifts for the cuts are in units of the fiducial bootstrap uncertainty. Removing in addition the galaxy fitted with two bands changes no central value by more than 0.01. The UVLF counterfactuals, the injection--recovery test and the covariance corrections are not repeated, because they need per-draw quantities of the fiducial fits. The full tables, with the sample sizes of each column, are available on Zenodo (doi:10.5281/zenodo.23600428).}
    \label{tab:robustness}
    \scriptsize\setlength{\tabcolsep}{2.5pt}
    \begin{tabularx}{\textwidth}{@{}>{\raggedright\arraybackslash}p{3.3cm}>{\raggedright\arraybackslash}p{1.0cm}*{6}{>{\raggedright\arraybackslash}X}>{\raggedright\arraybackslash}p{2.0cm}@{}}
        \toprule
        Result & $N$ & \texttt{bursty} (fiducial) & \texttt{continuity} & \texttt{hyperburst} & without photometry-only & $M_{\rm UV}<-18.0$ & $M_{\rm UV}<-18.5$ & Stable? \\
        \midrule
        \multicolumn{9}{@{}l}{\textit{Stellar mass and $M_{\rm UV}$}} \\
        Spearman $\rho(-M_{\rm UV},\log M_\ast)$, all; reliable bins (N) & 1163; 613 & 0.58; 0.48 & 0.60; 0.47 & 0.54; 0.40 & 0.60; 0.50 & -- & -- & yes \\
        Truncated slope $d\log M_\ast/dM_{\rm UV}$, reliable bins, $z=3$--4 / 4--5 / 5--6 (N) & 352 / 150 / 76 & $-0.38\pm0.04$ / $-0.63\pm0.16$ / $-0.28\pm0.08$ & $-0.36$ / $-0.62$ / $-0.47$ & $-0.33$ / $-0.49$ / $-0.23$ & $-0.42\pm0.05$ / $-0.66\pm0.18$ / $-0.34\pm0.10$ & -- & -- & sign; $z=5$--6 value prior-dependent \\
        \midrule
        \multicolumn{9}{@{}l}{\textit{What sets $M_{\rm UV}$ at fixed stellar mass (S, five mass bins, Fig.~\ref{fig:fixed_mass}); per $\sigma_x$, per unit of $x$ in parentheses}} \\
        $-\beta$ of $\log{\rm SFR}_{10}$ (mag) & 99--348 & $+0.43$ to $+0.64$ ($+0.70$ to $+1.87$) & $+0.46$ to $+0.65$ & $+0.35$ to $+0.66$ & $+0.39$ to $+0.67$ & $+0.43$ to $+0.61$ ($+0.70$ to $+1.91$) & $+0.44$ to $+0.54$ ($+0.73$ to $+2.10$) & sign; weaker above $\log M_\ast=9$ for \texttt{hyperburst}; per dex $\leq2.6\sigma$ for cuts \\
        $-\beta$ of $A_{1500}$ (mag) & 99--348 & $-0.50$ to $-1.06$ ($-0.98$ to $-1.04$) & $-0.41$ to $-1.07$ & $-0.52$ to $-1.02$ & $-0.49$ to $-1.08$ & $-0.41$ to $-1.06$ ($-0.96$ to $-1.04$) & $-0.42$ to $-1.05$ ($-0.97$ to $-1.06$) & yes; per mag $\leq1.4\sigma$ for cuts \\
        $-\beta$ of $\log t_{50}$ (mag) & 99--348 & $-0.08$ to $-0.61$ ($-0.18$ to $-2.00$) & $-0.01$ to $-0.43$ & $-0.06$ to $-0.75$ & $-0.08$ to $-0.62$ & $-0.06$ to $-0.57$ ($-0.15$ to $-1.99$) & $-0.02$ to $-0.58$ ($-0.05$ to $-2.03$) & sign; amplitude prior-dependent \\
        $\sigma(M_{\rm UV}\,|\,M_\ast,z)$ (mag), reliable and caution bins & -- & 0.46--0.83 & -- & -- & -- & 0.45--0.84 & 0.40--0.98 & yes ($\leq2.0\sigma$; 40 bootstraps for cuts) \\
        \midrule
        \multicolumn{9}{@{}l}{\textit{Burstiness against mass and redshift (W)}} \\
        Slope $d\bhun/d\log M_\ast$, $z=3$--4 / 4--5, reliable and caution bins & 444 / 150 & $-0.66\pm0.06$ / $-0.46\pm0.09$ & $-0.63\pm0.07$ / $-0.48\pm0.08$ & $-0.63\pm0.08$ / $-0.37\pm0.15$ & $-0.68\pm0.07$ / $-0.49\pm0.09$ & -- & -- & yes \\
        Median \bhun\ at $8\leq\log M_\ast<8.5$, $z=4$--5 / 5--6 / 6--7 & 63 / 54 / 41 & $+0.22$ / $+0.29$ / $+0.41$ & $+0.21$ / $+0.28$ / $+0.32$ & $+0.47$ / $+0.38$ / $+0.61$ & $+0.22$ / $+0.32$ / $+0.42$ & -- & -- & sign \\
        Rising fraction ($b>0.2$) at $8.8\leq\log M_\ast<9.5$, $z=3$--4 / 4--5 / 5--6 / 6--7 & 201 / 80 / 48 / 21 & $0.29\pm0.02$ / $0.36\pm0.03$ / $0.40\pm0.04$ / $0.46\pm0.08$ & $0.21$ / $0.22$ / $0.34$ / $0.44$ & $0.39$ / $0.47$ / $0.46$ / $0.42$ & $0.29\pm0.02$ / $0.36\pm0.03$ / $0.39\pm0.04$ / $0.47\pm0.08$ & $0.29\pm0.02$ / $0.37\pm0.03$ / $0.40\pm0.04$ / $0.46\pm0.08$ & $0.29\pm0.02$ / $0.37\pm0.03$ / $0.40\pm0.04$ / $0.46\pm0.08$ & level prior-dependent; no rise with $z$ for \texttt{hyperburst}; $2.6\sigma$ and $0.6\sigma$ below the prior at $z=5$--6, 6--7 \\
        Rising fraction, reliable bins $8.5\leq\log M_\ast<9$, $z=3$--4 / 4--5 (prior 0.47 / 0.48) & 177 / 76 & $0.40\pm0.02$ / $0.44\pm0.03$ & -- & -- & -- & $0.46\pm0.03$ / $0.44\pm0.03$ & $0.47\pm0.03$ / $0.42\pm0.04$ & no: moves to the prior ($3.0$, $1.3\sigma$ below it at $-17.5$; $0.1$, $1.6\sigma$ at $-18.5$) \\
        Same, $9\leq\log M_\ast<9.5$ & 128 / 51 & $0.24\pm0.02$ / $0.37\pm0.03$ & -- & -- & -- & $0.24\pm0.02$ / $0.37\pm0.03$ & $0.23\pm0.02$ / $0.37\pm0.03$ & yes \\
        $\sigma_b$ (dex), reliable bins $z=3$--4, $\log M_\ast=8.5$--9 / 9--9.5 / 9.5--10 & 177 / 128 / 47 & $0.37\pm0.04$ / $0.14\pm0.07$ / $0.51\pm0.08$ & -- & -- & -- & $0.38\pm0.06$ / $0.16\pm0.08$ / $0.51\pm0.08$ & $0.38\pm0.06$ / $0.18\pm0.07$ / $0.51\pm0.08$ & yes ($\leq0.6\sigma$) \\
        Same, $z=4$--5 & 76 / 51 / 23 & $0.19\pm0.08$ / $0.03\pm0.03$ / $0.48\pm0.14$ & -- & -- & -- & $0.19\pm0.08$ / $0.03\pm0.03$ / $0.48\pm0.15$ & $0.25\pm0.09$ / $0.03\pm0.04$ / $0.48\pm0.15$ & yes ($\leq0.8\sigma$) \\
        \midrule
        \multicolumn{9}{@{}l}{\textit{The UV-bright tail (W)}} \\
        Logistic slope $k$ (per mag brighter), $8.5\leq\log M_\ast<9$, $z=3$--5 / 5--7 & 253 / 72 & $0.23\pm0.11$ / $0.59\pm0.21$ & $0.46\pm0.12$ / $0.67\pm0.23$ & $-0.07\pm0.10$ / $0.48\pm0.28$ & $0.17\pm0.11$ / $0.55\pm0.23$ & $-0.05\pm0.12$ / $0.53\pm0.32$ & $0.03\pm0.15$ / $0.80\pm0.29$ & no at $z=3$--5: sign changes for \texttt{hyperburst}; consistent with zero for brighter cuts \\
        Same, $9\leq\log M_\ast<9.5$, $z=3$--5 & 179 & $-0.10\pm0.10$ & $-0.05\pm0.12$ & $-0.42\pm0.10$ & $-0.31\pm0.11$ & -- & -- & sign; amplitude prior- and subset-dependent ($-2.1\sigma$) \\
        Rising excess of the brightest 10\%, $\Delta f$, $8.5\leq\log M_\ast<9$, $z=3$--5 / 5--7 & 253 / 72 & $+0.04\pm0.05$ / $+0.18\pm0.08$ & $+0.10$ / $+0.25$ & $-0.03$ / $+0.26$ & $+0.05\pm0.05$ / $+0.15\pm0.09$ & $0.00\pm0.05$ / $+0.15\pm0.08$ & $+0.01\pm0.05$ / $+0.22\pm0.08$ & $z=5$--7 yes; $z=3$--5 consistent with zero \\
        Logistic slope of the odds of a low H$\alpha$/UV ratio on $M_{\rm UV}$, $z=3$--5 (16--84\%) & 609 & $0.75$ ($0.29$--$1.20$) & $0.76$ ($0.29$--$1.18$) & $0.69$ ($0.23$--$1.10$) & $0.75$ ($0.29$--$1.20$) & -- & -- & yes \\
        \midrule
        \multicolumn{9}{@{}l}{\textit{Mechanisms (treatment as stated in each row)}} \\
        Burst-phase amplitude of $\log A_{1500}$ (dex) & 1365 & $0.35\pm0.03$ & $0.29\pm0.03$ & $0.30\pm0.03$ & $0.36\pm0.03$ & $=$ & $=$ & yes \\
        Cancelled fraction $f_{\rm obsc}$ of the unattenuated UV gain (S) & 99--348 & $0.21$--$0.48$ & $0.26$--$0.52$ & $0.18$--$0.40$ & $0.25$--$0.49$ & $0.25$--$0.48$ & $0.19$--$0.44$ & yes ($\leq2.2\sigma$ for cuts) \\
        $\Delta\log\xi_{\rm ion}$ of the 20 brightest at fixed \bhun, $b_{10-20}$, $t_{50}$; exclusion of $\alpha_3=1.7$ from ${\rm EW}({\rm H}\alpha)$ & 20 & $+0.06\pm0.14$; $0.9\sigma$ & $+0.12\pm0.11$; $1.6\sigma$ & $+0.12\pm0.10$; $1.1\sigma$ & $+0.06\pm0.14$; $0.9\sigma$ & -- & -- & yes (null) \\
        \bottomrule
    \end{tabularx}
\end{table*}

\begin{table}
    \centering
    \caption{Model variants used in this work. Each differs from the fiducial \texttt{bursty} model in one prior component: the SFH prior width $\sigma$, stellar metallicity prior, or birth-cloud attenuation slope $\alpha_{\rm BC}$.}
    \label{tab:variants}

    \footnotesize
    \renewcommand{\arraystretch}{1.2}

    \begin{tabular*}{\columnwidth}{
        @{\extracolsep{\fill}}
        lccc
        @{}
    }
        \toprule
        \textbf{Variant}
        & \textbf{SFH $\sigma$}
        & \textbf{$Z_\ast$ prior}
        & \textbf{$\alpha_{\rm BC}$} \\
        \midrule
        \texttt{bursty} (fiducial)
        & $0.5$
        & $\Un$ (grid)
        & $-1$ \\

        \texttt{continuity}
        & $0.3$
        & $\Un$ (grid)
        & $-1$ \\

        \texttt{hyperburst}
        & $1.0$
        & $\Un$ (grid)
        & $-1$ \\

        \texttt{bursty\_mzr}
        & $0.5$
        & MZR
        & $-1$ \\

        \texttt{bursty\_freealpha}
        & $0.5$
        & $\Un$ (grid)
        & $\Un[-2,0]$ \\
        \bottomrule
    \end{tabular*}

    \vspace{4pt}
    \parbox{\columnwidth}{%
        \footnotesize
        \raggedright
        \textit{Note.} $\Un$ denotes a uniform prior over the indicated range; MZR denotes the mass--metallicity relation of \citet{Gallazzi:2005aa}. All variants are fitted to the same 1388 galaxies (1365 in the final sample; \S~\ref{sec:data}). All other model components, including the SSPs, dust prescription, nebular grid, likelihood, and sampler, are unchanged, allowing direct comparison of the per-galaxy Bayesian evidence $\ln Z$.
    }
\end{table}

\begin{table}
    \centering
\caption{Intrinsic scatter of the burst ratio \bhun\ in reliable and caution bins at $z<9$, using $M_{\rm UV}<-17.5$ and selection weights $w'$ (\S~\ref{sec:variability}). Reliability flags follow Fig.~\ref{fig:reliability_map}. The columns give the sample size $N$, effective sample size $n_{\rm eff}$, weighted 16th--84th percentile half-range of the galaxy posterior medians ($\Delta_{\rm med}$), and median posterior half-width ($\delta_{\rm post}$). The population mean $\mu_b$ and intrinsic scatter $\sigma_b$ account for posterior uncertainties and are reported with 16th--84th percentile bootstrap intervals. Italics indicate $n_{\rm eff}<25$ or an unconstrained $\sigma_b$. In the well-constrained bins ($3\leq z<5$, $8.5\leq\log_{10}(M_\ast/\Msun)<10$), the intrinsic scatter is $\sigma_b=0.14$--$0.51$~dex, except at $4\leq z<5$ and $9\leq\log_{10}(M_\ast/\Msun)<9.5$, where $\sigma_b$ is not resolved, with no monotonic trend with stellar mass.}
    \label{tab:sigma_b}
    \setlength{\tabcolsep}{4.8pt}
    \begin{tabular}{llrrcccc}
    \toprule
    $z$ & $\log M_\ast$ & $N$ & $n_{\rm eff}$ & $\Delta_{\rm med}$ & $\delta_{\rm post}$ & $\mu_b$ & $\sigma_b$ \\
    \midrule
    3--4 & 8--8.5 (fpM) & 92 & 73 & 0.55 & 0.48 & $+0.44$ & $0.44^{+0.06}_{-0.10}$ \\
    3--4 & 8.5--9 (M) & 177 & 142 & 0.50 & 0.44 & $-0.00$ & $0.37^{+0.03}_{-0.05}$ \\
    3--4 & 9--9.5 & 128 & 118 & 0.39 & 0.43 & $-0.31$ & $0.14^{+0.06}_{-0.10}$ \\
    3--4 & 9.5--10 & 47 & 46 & 0.49 & 0.39 & $-0.53$ & $0.51^{+0.06}_{-0.10}$ \\
    \midrule
    4--5 & 8.5--9 (M) & 76 & 65 & 0.37 & 0.42 & $+0.06$ & $0.19^{+0.05}_{-0.14}$ \\
    \textit{4--5} & \textit{9--9.5} & 51 & 48 & 0.35 & 0.46 & $-0.05$ & $0.03^{+0.00}_{-0.02}$ \\
    \textit{4--5} & \textit{9.5--10} & 23 & 22 & 0.53 & 0.47 & $-0.65$ & $0.48^{+0.09}_{-0.20}$ \\
    5--6 & 8--8.5 (pM) & 54 & 47 & 0.35 & 0.51 & $+0.26$ & $0.27^{+0.10}_{-0.20}$ \\
    \textit{5--6} & \textit{8.5--9} (M) & 51 & 24 & 0.27 & 0.44 & $+0.05$ & $0.12^{+0.04}_{-0.11}$ \\
    \textit{5--6} & \textit{9--9.5} (M) & 25 & 24 & 0.40 & 0.41 & $+0.02$ & $0.18^{+0.08}_{-0.17}$ \\
    6--7 & 8--8.5 (pM) & 41 & 29 & 0.35 & 0.53 & $+0.49$ & $0.34^{+0.08}_{-0.11}$ \\
    \textit{6--7} & \textit{8.5--9} (M) & 21 & 14 & 0.54 & 0.45 & $-0.12$ & $0.35^{+0.03}_{-0.12}$ \\
    \textit{6--7} & \textit{9--9.5} (M) & 14 & 12 & 0.36 & 0.35 & $+0.12$ & $0.25^{+0.03}_{-0.14}$ \\
    \textit{7--9} & \textit{8--8.5} (pM) & 39 & 14 & 0.68 & 0.57 & $-0.07$ & $0.55^{+0.15}_{-0.27}$ \\
    \bottomrule
    \end{tabular}
\end{table}

\begin{figure*}
    \centering
    \includegraphics[width=\textwidth]{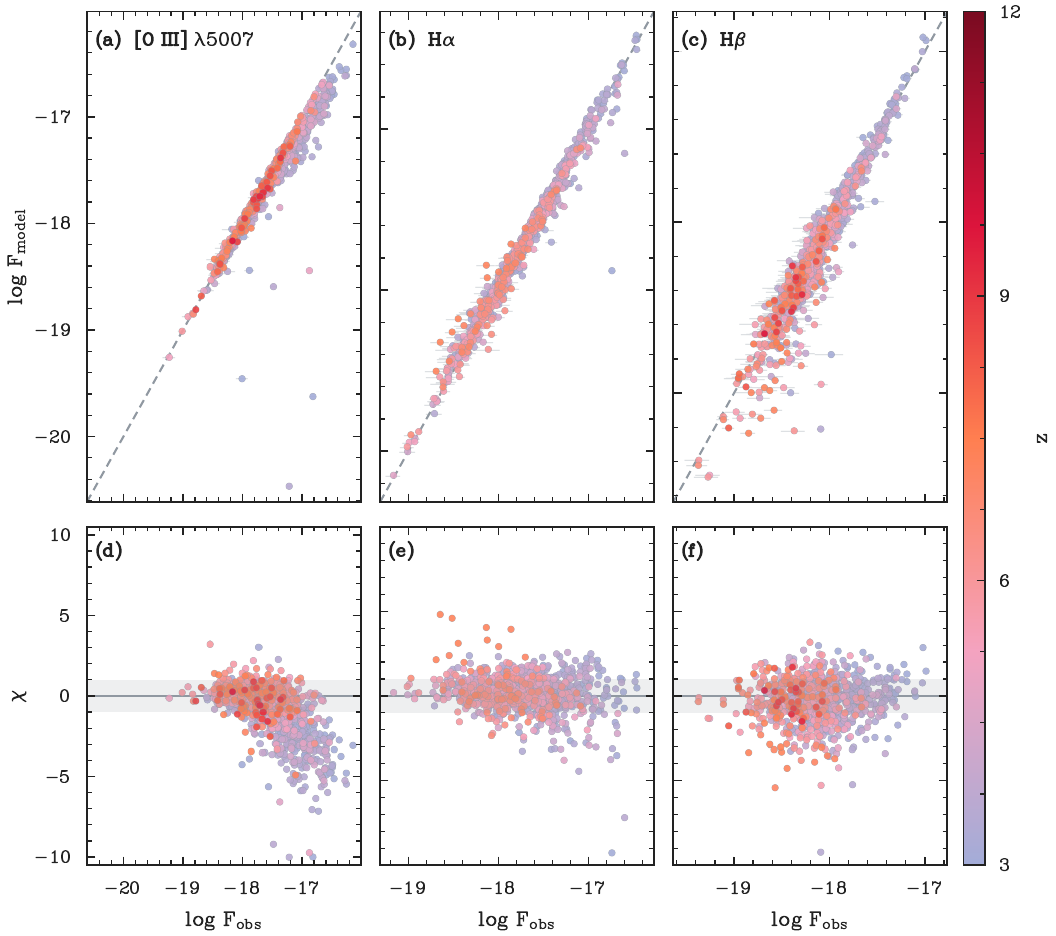}
    \caption{Comparison of observed and modelled emission-line fluxes for galaxies with catalogue ${\rm S/N}\geq3$. Panels (a--c) show the posterior-median model flux against the R1000 catalogue flux for [O\,\textsc{iii}]$\lambda5007$, H$\alpha$, and H$\beta$, coloured by redshift. Light grey horizontal error bars indicate the observed $1\sigma$ uncertainties, and dashed lines mark the one-to-one relation. Panels (d--f) show the corresponding pulls $\chi$ (Fig.~\ref{fig:stacked_pulls}) against the observed flux, also coloured by redshift. The grey line marks $\chi=0$, grey shading marks $|\chi|\leq1$, and pulls beyond $\pm10$ are plotted at $\pm10$. No selection weights are applied. The Balmer-line fluxes are recovered with small offsets, while [O\,\textsc{iii}] is systematically underpredicted (\S~\ref{sec:fit_quality}).}
    \label{fig:lines}
\end{figure*}

\begin{figure*}
    \centering
    \includegraphics[width=\textwidth]{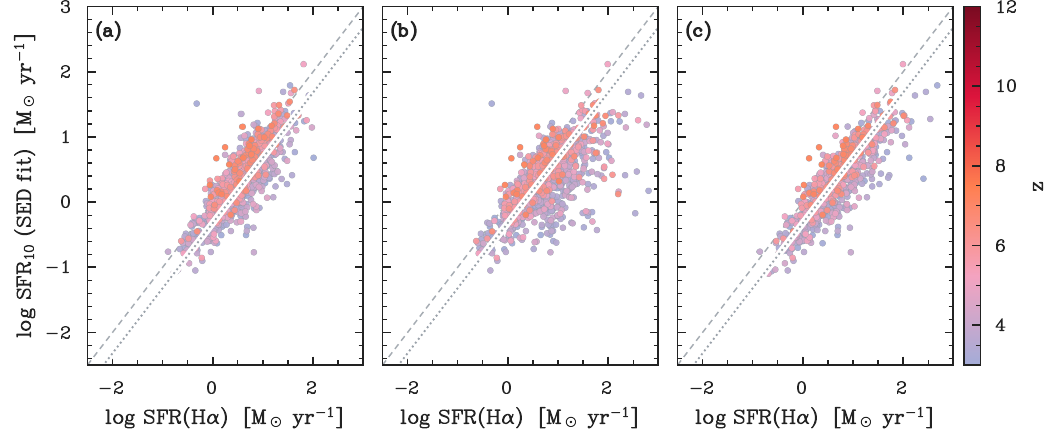}
    \caption{Comparison of ${\rm SFR}_{10}$ from the SED fit (Eq.~\ref{eq:sfrt}) with the SFR inferred from H$\alpha$ \citep{Kennicutt:2012aa} for galaxies with ${\rm S/N(H\alpha)}\geq3$. The H$\alpha$ SFR is (a) uncorrected for dust, (b) corrected using the Balmer decrement, and (c) corrected using the posterior attenuation at H$\alpha$. Galaxies are coloured by redshift, and dashed lines mark the one-to-one relation, and dotted lines the one-to-one relation for the metal-poor calibration of \protect\citet{Clarke:2025aa} ($C=-41.59$, 0.32~dex below \protect\citet{Kennicutt:2012aa}) Four galaxies in (b) lie outside the plotted range. No selection weights are applied. H$\alpha$ fluxes are scaled to the Kron photometry. The Balmer-decrement correction in (b) adds the H$\beta$ uncertainty, and because negative $E(B-V)$ are set to zero, this noise can only raise ${\rm SFR}({\rm H}\alpha)$, which broadens the relation towards high ${\rm SFR}({\rm H}\alpha)$; panel (c) instead uses the smooth dust model of the fit. The two SFR estimates differ by 0.15--0.30~dex in the median, providing a consistency check rather than an independent validation, since H$\alpha$ is included in the SED fit.}
    \label{fig:sfr10_halpha}
\end{figure*}

\begin{figure}
    \centering
    \includegraphics[width=\columnwidth]{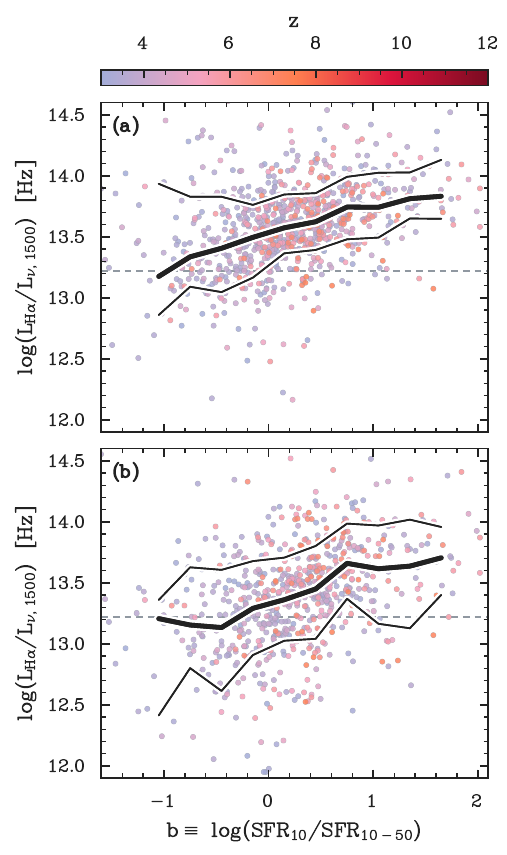}
    \caption{H$\alpha$-to-UV luminosity ratio versus the burst ratio \bhun\ (Eq.~\ref{eq:burst}) for galaxies with H$\alpha$ detected at ${\rm S/N}\geq3$. Panel (a) shows the observed ratio ($N=954$); panel (b) shows the dust-corrected ratio ($N=729$), using the Balmer decrement for H$\alpha$ and $A_{1600}=4.43+1.99\beta$ for the UV \citep{Meurer:1999aa}, requiring ${\rm S/N}\geq3$ in both Balmer lines. Galaxies are coloured by redshift. Black lines show the running median (thick) and 16th--84th percentiles (thin) in bins containing at least ten galaxies. The dashed line marks the ratio expected for constant star formation, $\log_{10}(L_{{\rm H}\alpha}/L_{\nu,1500})=13.22$, from the H$\alpha$ and FUV calibrations of \citet{Kennicutt:2012aa}. No selection weights are applied. H$\alpha$ fluxes are scaled to the Kron photometry. The ratio increases with \bhun\ in both panels (Spearman $\rho=0.40$ and 0.40), providing a consistency check rather than independent validation because H$\alpha$ and the UV photometry are included in the SED fits.}
    \label{fig:halpha_uv}
\end{figure}

\begin{figure*}
    \centering
    \includegraphics[width=\textwidth]{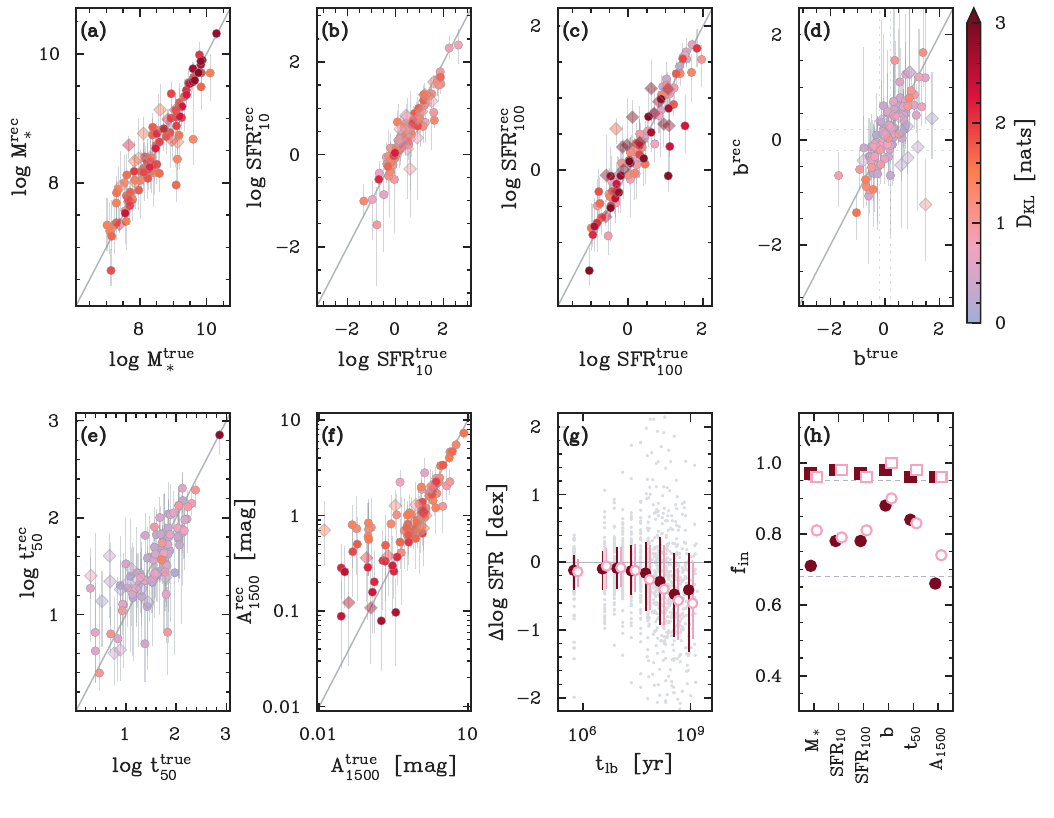}
    \caption{Injection--recovery test of the SFH inference for 100 galaxies (\S~\ref{sec:fit_quality}), including 35 photometry-only galaxies and 15 at $z\geq9$. Panels (a--f) compare posterior medians and 16th--84th percentile intervals of the noisy mocks with the input values for stellar mass, ${\rm SFR}_{10}$, ${\rm SFR}_{100}$, \bhun, $t_{50}$, and $A_{1500}$, coloured by information gain $D_{\rm KL}$ (Eq.~\ref{eq:kl}). Semi-transparent diamonds mark galaxies at $z\geq9$, the colour scale saturates at $D_{\rm KL}=3$, the grey line indicates equality, and dotted lines in (d) mark $b=\pm0.2$. Panel (f) shows $A_{1500}$ (Eq.~\ref{eq:a1500}) on logarithmic axes. Panel (g) shows the difference between recovered and input $\log{\rm SFR}$ in each age bin, with individual bins of the noisy mocks as grey points and medians and 16th--84th percentiles for noisy (filled, dark red) and noiseless (open, pink) mocks. Panel (h) shows the fraction of inputs within the 68~per~cent (circles) and 95~per~cent (squares) posterior intervals for the same two sets of mocks, and dashed lines mark 0.68 and 0.95. No selection weights are applied. Stellar masses and SFRs are recovered with small offsets, \bhun\ with moderate scatter, and $A_{1500}$ least reliably. In (d), the upper error bars of seven galaxies extend beyond the plotted range. Recovery is poorer at $z\geq9$, although nearly all inputs remain within the 95~per~cent intervals.}
    \label{fig:injection_recovery}
\end{figure*}

\begin{figure*}
    \centering
    \includegraphics[width=\textwidth]{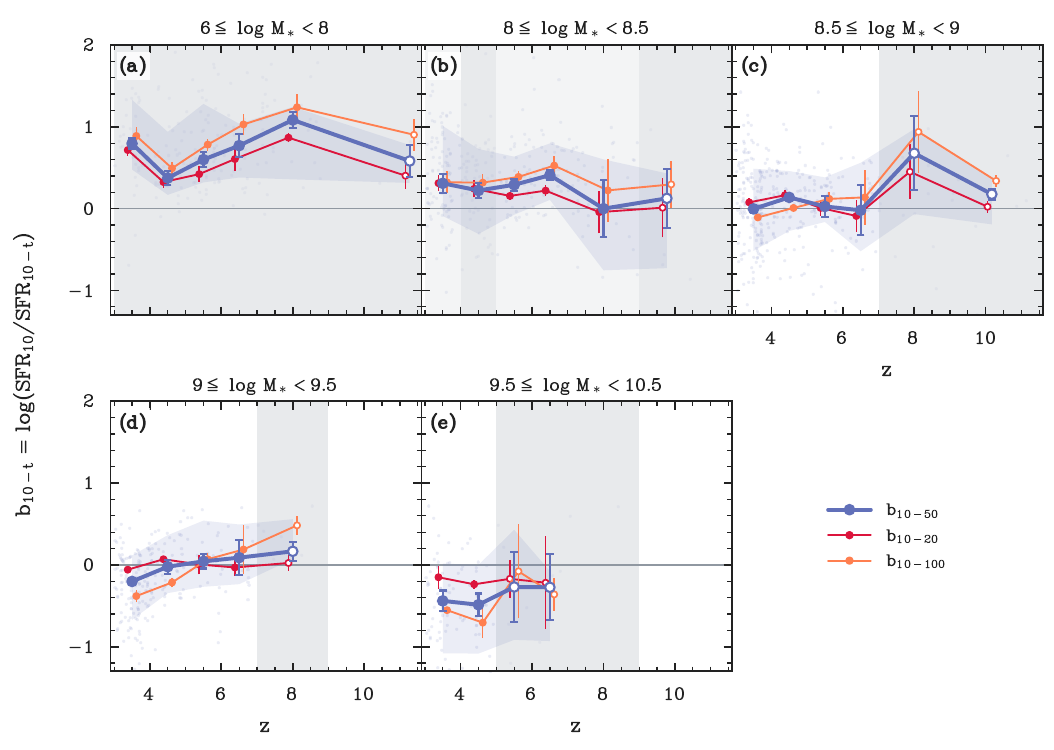}
    \caption{Redshift evolution of the burst ratios $b_{10-20}$, \bhun\ and $b_{10-100}$ in five stellar-mass bins. Symbols, bands, and individual galaxies are shown as in Fig.~\ref{fig:variability}, with open symbols for bins with $N<10$ or $n_{\rm eff}<10$ and the bin at $z\geq9$. Dark and light grey shading marks unreliable and caution redshift ranges in each mass bin. Above $\log_{10}(M_\ast/\Msun)=8.5$, $b_{10-100}$ increases with redshift more strongly than \bhun, and $b_{10-20}$ shows no clear trend. Most of the increase in $b_{10-100}$ occurs where the fraction of stellar mass formed in the past 100~Myr becomes consistent with the SFH prior and is therefore likely driven by the prior rather than the data (\S~\ref{sec:variability}).}
    \label{fig:variability_z}
\end{figure*}

\begin{figure}
    \centering
    \includegraphics[width=\columnwidth]{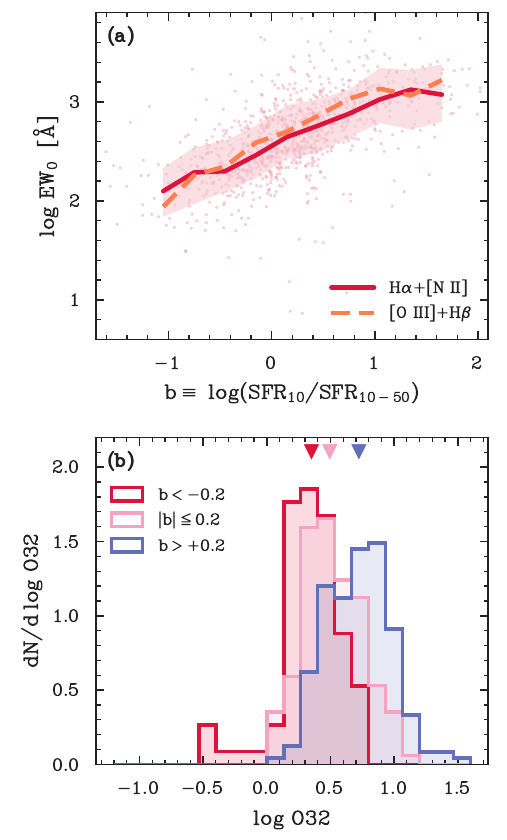}
    \caption{Emission-line tracers of the burst ratio \bhun\ (Eq.~\ref{eq:burst}) in the sample. (a) Observed rest-frame ${\rm EW}_0$(H$\alpha$+[N\,\textsc{ii}]) versus \bhun\ ($N=994$). Small light-red points show individual galaxies, and the red line and shaded band show the running median and 16th--84th percentile range; the coral dashed line shows the running median of ${\rm EW}_0$([O\,\textsc{iii}]+H$\beta$) ($N=840$). (b) Distributions of $\log_{10}{\rm O32}$, where ${\rm O32}=F_{[{\rm O\,III}]\lambda5007}/F_{[{\rm O\,II}]\lambda3727}$, for galaxies classified by their posterior-median \bhun\ as declining ($\bhun<-0.2$, red), approximately constant ($|\bhun|\leq0.2$, pink), and rising ($\bhun>0.2$, blue). The distributions are normalised to unit area, and triangles at the top of the panel mark their medians. Panel (b) includes galaxies with $8\leq\log_{10}(M_\ast/\Msun)<9.5$ and ${\rm S/N}\geq3$ in both lines. No selection weights are applied. Both equivalent widths and O32 correlate with \bhun, providing a consistency check rather than independent validation because the emission lines are included in the SED fits.}
    \label{fig:observable}
\end{figure}

\paragraph*{H$\alpha$ star-formation rates.}
We compare ${\rm SFR}_{10}$ with the H$\alpha$-based SFR calibrated by \citet{Kennicutt:2012aa} for galaxies with ${\rm S/N(H\alpha)}\geq3$ (Fig.~\ref{fig:sfr10_halpha}). This cut also excludes galaxies whose H$\alpha$ line falls within or near the gap between the NIRSpec detectors, where its flux is unconstrained and therefore does not contribute to the fit. The calibration is derived from \textsc{Starburst99} models \citep{leitherer_starburst99_1999} for solar metallicity and a Kroupa IMF, whereas our fits adopt a \citet{Chabrier:2003aa} IMF. We do not apply an IMF conversion because its effect is small compared with the observed scatter. At lower metallicity, the ionising luminosity per unit SFR is higher, so the calibration overestimates the SFR of metal-poor galaxies \citep{Kennicutt:2012aa, Reddy:2022aa}. For the Balmer-decrement correction (Figs~\ref{fig:sfr10_halpha}b and \ref{fig:halpha_uv}b), we derive a non-negative $E(B-V)$ from H$\alpha$/H$\beta$, assuming a case-B ratio of 2.86 \citep{osterbrock_astrophysics_1989} and the \citet{Cardelli:1989aa} extinction curve with $R_V=3.1$. Alternatively, we correct H$\alpha$ using the attenuation predicted by each galaxy's posterior dust model, including both birth-cloud and diffuse attenuation (Fig.~\ref{fig:sfr10_halpha}c).

\section{The sample}\label{app:sample}
\begin{figure*}
    \centering
    \includegraphics[width=\textwidth]{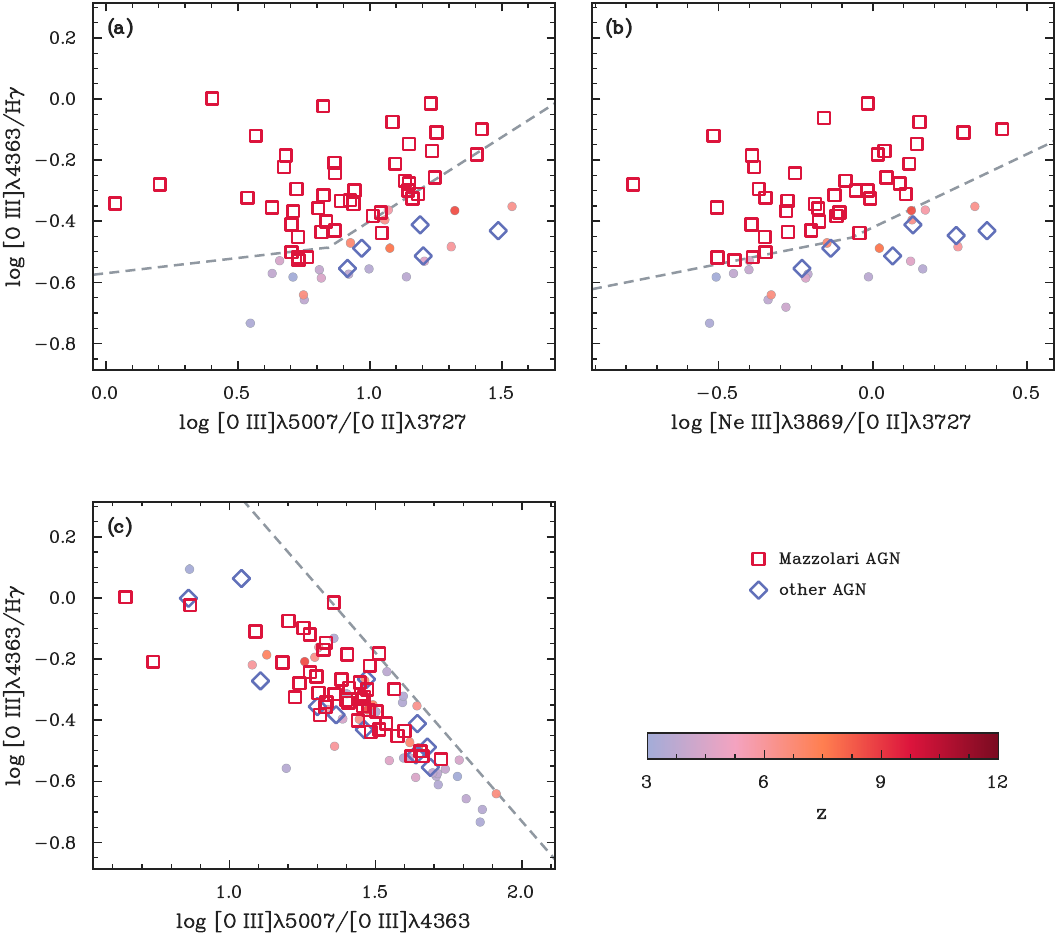}
    \caption{AGN selection using the [O\,\textsc{iii}]$\lambda4363$ diagnostics of \citet{Mazzolari:2024aa}. The panels show [O\,\textsc{iii}]$\lambda4363$/H$\gamma$ against (a) [O\,\textsc{iii}]$\lambda5007$/[O\,\textsc{ii}]$\lambda3727$ ($N=70$), (b) [Ne\,\textsc{iii}]$\lambda3869$/[O\,\textsc{ii}]$\lambda3727$ ($N=68$), and (c) [O\,\textsc{iii}]$\lambda5007$/[O\,\textsc{iii}]$\lambda4363$ ($N=95$), with logarithmic ratios on both axes. The parent sample comprises 1495 galaxies at $z\geq3$ with spectroscopic redshift flags A--C and NIRCam photometry; only galaxies with the required emission lines detected at ${\rm S/N}\geq3$ appear in each panel. Grey dashed lines mark the AGN demarcations of \citet{Mazzolari:2024aa}. Red open squares identify the 46 galaxies above at least one demarcation, while blue open diamonds indicate galaxies excluded by other AGN criteria (\S~\ref{sec:data}). Remaining galaxies are coloured by redshift. No selection weights are applied. In total, 107 AGN and LRD candidates are excluded before SED fitting.}
    \label{fig:agn_diagnostics}
\end{figure*}

\begin{figure*}
    \centering
    \includegraphics[width=\textwidth]{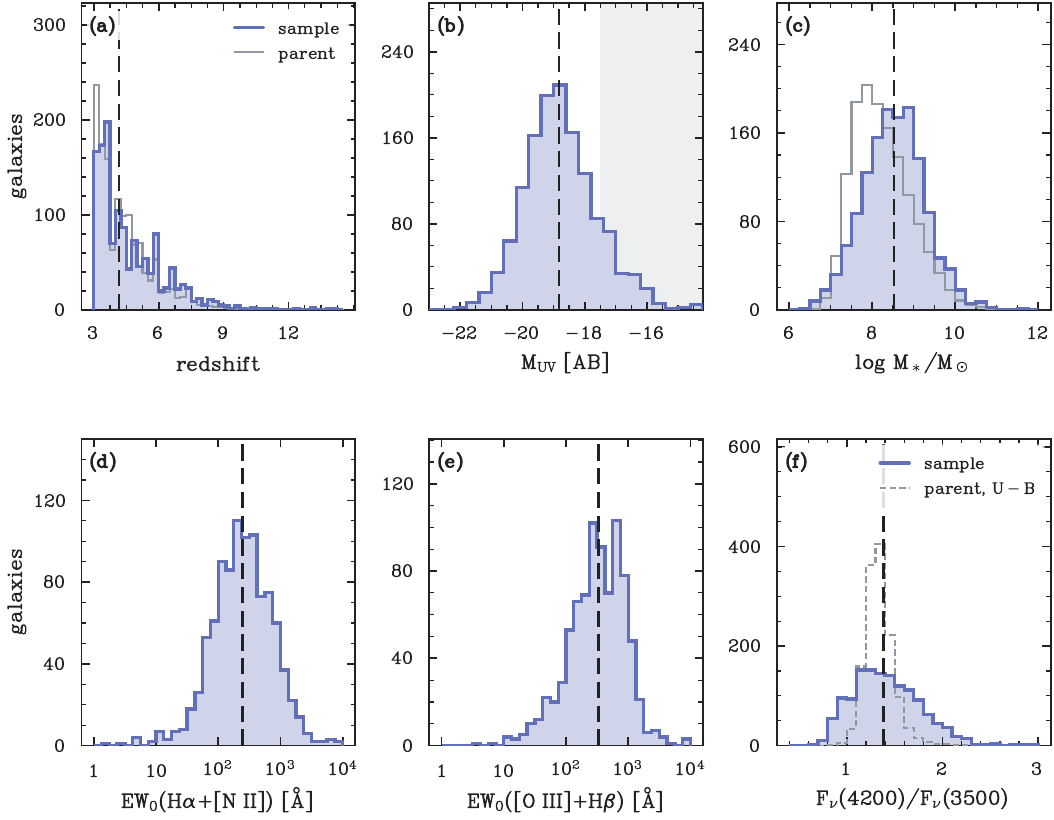}
    \caption{Distributions of (a) redshift, (b) UV magnitude $M_{\rm UV}$, (c) stellar mass, (d) observed ${\rm EW}_0$(H$\alpha$+[N\,\textsc{ii}]), (e) observed ${\rm EW}_0$([O\,\textsc{iii}]+H$\beta$), and (f) the posterior-model continuum ratio $F_\nu(4200\,\text{\AA})/F_\nu(3500\,\text{\AA})$. The sample (1365 galaxies; blue) is compared with the NIRCam parent catalogue \citep{Duan:2026aa} (29\,267 galaxies; grey), scaled to the sample histogram area where comparable measurements are available. Panels (d) and (e) contain 994 and 840 galaxies, respectively; in (f), the parent catalogue's rest-frame $U-B$ colour is shown as an approximate proxy (dashed). Black dashed lines mark the sample medians, and shading in (b) indicates $M_{\rm UV}>-17.5$, below the adopted completeness limit. No selection weights are applied. The sample is more massive than the NIRCam parent catalogue.}
    \label{fig:sample_histograms}
\end{figure*}

\begin{figure*}
    \centering
    \includegraphics[width=\textwidth]{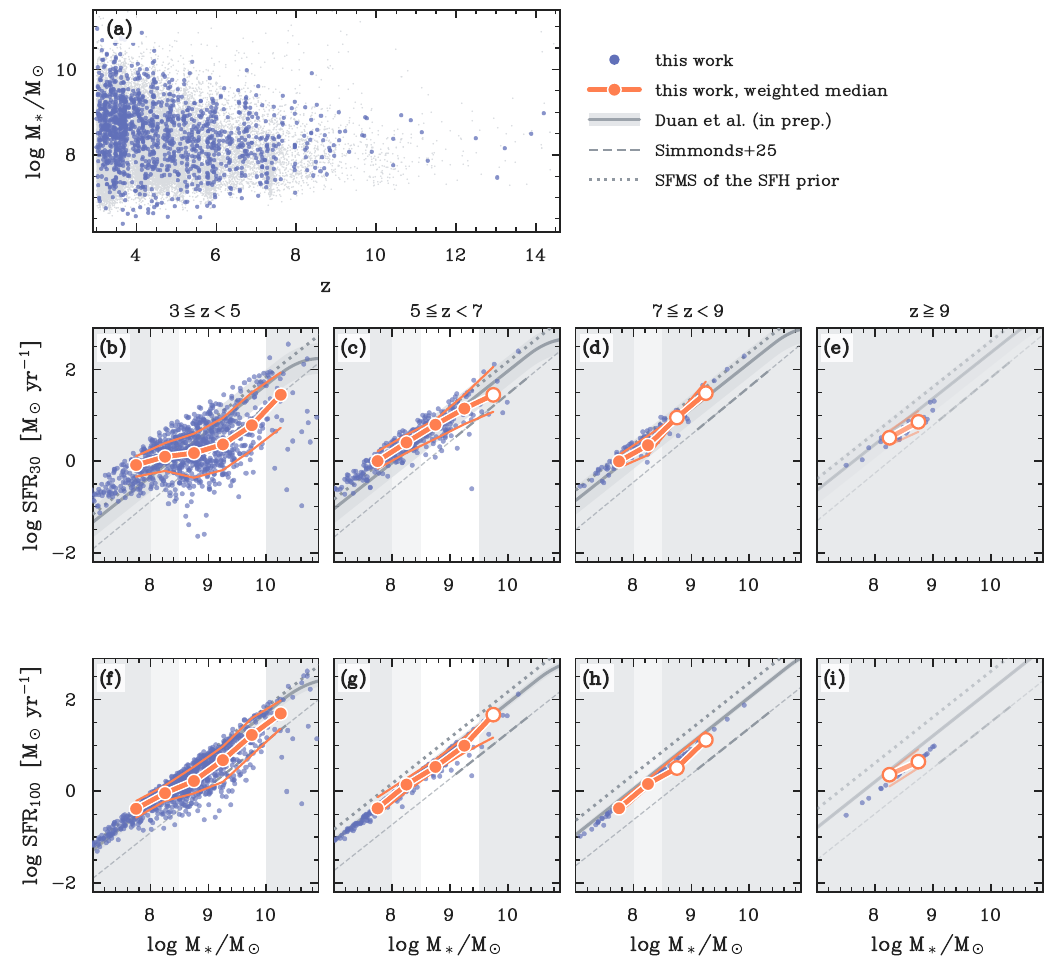}
    \caption{Stellar mass and star formation rates of the sample. (a) Stellar mass versus spectroscopic redshift for the 1365 galaxies (blue), compared with the JADES NIRCam parent catalogue \citep{Duan:2026aa} (grey). Panels (b--e) and (f--i) show ${\rm SFR}_{30}$ and ${\rm SFR}_{100}$ (Eq.~\ref{eq:sfrt}), respectively, versus stellar mass in four redshift bins. Points show individual galaxies; coral symbols and thin lines indicate the $w'$-weighted medians and 16th--84th percentiles in 0.5-dex mass bins for galaxies with $M_{\rm UV}<-17.5$. Open symbols mark bins with $N<10$ or $n_{\rm eff}<10$, and all bins at $z\geq9$, where the medians and reference relations are drawn semi-transparently. Grey curves show the star-forming main sequence (SFMS) of Duan et al.\ (in prep., Eq.~19), with dark bands indicating the 16th--84th percentiles of posterior predictions and light bands their intrinsic scatter; these relations are extrapolated at $z>9$. Grey dashed curves show the SFMS of \citet{Simmonds:2025aa}, highlighted over their fitted mass range. The relations of Duan et al. (in prep.) and \citet{Simmonds:2025aa} use the SFR timescale of the corresponding panel and formed stellar masses. Grey dotted curves show the SFMS underlying our SFH prior \citep{Duan:2026aa}, so agreement with this relation is not independent evidence. Dark and light shading marks unreliable and caution mass ranges, respectively (\S~\ref{sec:reliability}). The sample preferentially probes the massive end of the NIRCam parent population.}
    \label{fig:mass_z}
\end{figure*}

This appendix shows the [O\,\textsc{iii}]$\lambda4363$ diagnostics used to remove AGN candidates before fitting (Fig.~\ref{fig:agn_diagnostics}; \S~\ref{sec:data}) and compares the sample with the NIRCam parent catalogue of \citet{Duan:2026aa} in redshift, UV magnitude, stellar mass, emission-line equivalent widths and continuum colour (Fig.~\ref{fig:sample_histograms}), and in stellar mass and SFR (Fig.~\ref{fig:mass_z}).

\section{The stellar mass--UV magnitude relation}
\label{app:mass_muv}

UV magnitude is strongly correlated with stellar mass (\S~\ref{sec:naive}; Fig.~\ref{fig:naive}). Here we quantify this relation, its redshift evolution, and its agreement with the model of \citet{Tacchella:2018ab}. The fixed-mass analyses in the main text do not depend on these fits.

\paragraph*{Fitting method.}
Following \citet[][Eq.~6]{Tacchella:2018ab}, we describe the relation as
\begin{equation}
    \log_{10}(M_\ast/\Msun)
    = a(M_{\rm UV}+19.5)+\log_{10}M_{\ast,0}(z),
    \label{eq:mass_muv}
\end{equation}
where $a$ is the slope and $M_{\ast,0}$ is the stellar mass at $M_{\rm UV}=-19.5$. In each redshift slice, we fit this relation by least squares to the median stellar masses in bins of \muv\, both with $a$ free and with $a=-0.5$, the slope found by \citet{Tacchella:2018ab}.

We model the redshift evolution of the fixed-slope normalisation as
\begin{equation}
    \log_{10}M_{\ast,0}(z)
    =c_1\log_{10}(1+z)+c_0,
    \label{eq:mass_muv_evol}
\end{equation}
fitting the five redshift-bin normalisations at their median redshifts by inverse-variance weighted least squares. We also jointly fit Eqs.~\ref{eq:mass_muv} and \ref{eq:mass_muv_evol} to all bin medians, allowing a common free slope. To test for slope evolution, we introduce
\begin{equation}
    a(z)=a_5+a_1\log_{10}\!\left(\frac{1+z}{6}\right),
    \label{eq:mass_muv_slope}
\end{equation}
where $a_5$ is the slope at $z=5$ and $a_1=0$ corresponds to a constant slope. We compare the constant- and evolving-slope models using an $F$-test on the residuals and the Akaike information criterion (AIC).

The fits are unweighted to match the treatment of the volume-complete model population in \citet{Tacchella:2018ab}. We quantify scatter using the normalised median absolute deviation of $\log_{10}M_\ast$ about the fitted relation. The 21 galaxies at $z\geq9$ are shown in Fig.~\ref{fig:naive} but excluded from the fits. We also convert the zero-points of \citet{Tacchella:2018ab} from their stellar-mass definition (stars and remnants, Salpeter IMF) to ours (total mass formed, Chabrier IMF), using their model predictions.

\paragraph*{Comparison with previous work.}
The fitted slope agrees with \citet{Tacchella:2018ab}, but the normalisation is lower. The joint fit gives $a=-0.48\pm0.03$, consistent with their fitted value of $-0.5$. However, our fixed-slope zero-points lie up to 0.3~dex below their predictions, even after correcting for the IMF and mass definition. This difference is unlikely to arise from the mass-dependent selection bias: galaxies are preferentially more massive than the parent population at fixed \muv\ (\S~\ref{sec:muv_completeness}), and applying the mass-corrected weights lowers the inferred zero-points further.

\paragraph*{Redshift evolution of the slope.}
The apparent slope evolution is sensitive to sample selection. In the full sample, the slope steepens slightly but non-monotonically with redshift, and the evolving-slope model is preferred over a constant slope ($F$-test, $p=8\times10^{-5}$). However, restricting the analysis to reliable bins and accounting for the resulting mass truncation removes any clear redshift trend. The slope is consistent with $-0.5$ in one of the three testable redshift slices and shallower in the other two (Table~\ref{tab:robustness}). In particular, at $5\leq z<6$, where the full-sample slope is steepest, the result is driven by galaxies in unreliable bins, primarily at low mass. At $z\geq6$, the reliable bins cover too narrow a mass range to constrain the slope. We therefore find no robust evidence for redshift evolution of the stellar mass--UV magnitude slope.

\section{A test for a top-heavy IMF}\label{sec:imf}

\begin{figure}
    \centering
    \includegraphics[width=\columnwidth]{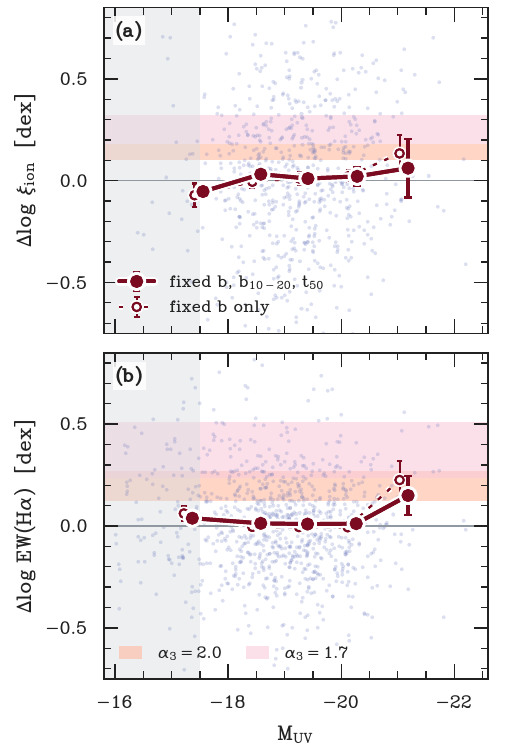} 
    \caption{
    Test for a top-heavy IMF at fixed inferred recent SFH. The panels show residuals in (a) the dust-corrected ionising photon production efficiency, $\log\xi_{\rm ion}$, and (b) the observed $\log{\rm EW}({\rm H}\alpha+[{\rm N}\,\textsc{ii}])$, as functions of $M_{\rm UV}$ (galaxies with $M_{\rm UV}<-16$; treatment N; $N=727$ and 985, respectively). Filled symbols show median residuals about a quadratic surface in \bhun, $b_{10-20}$ and $\log t_{50}$; open symbols show residuals about \bhun\ alone. Medians are calculated in magnitude bins containing at least eight galaxies, with bootstrap uncertainties; small points show individual galaxies for the quadratic-surface residuals. The shaded bands indicate the changes predicted by \textsc{FSPS} for high-mass IMF slopes $\alpha_3=2.0$ (coral) and 1.7 (pink), relative to $\alpha_3=2.3$ (Appendix~\ref{sec:imf}). The grey region marks $M_{\rm UV}>-17.5$, below the adopted completeness limit. The brightest galaxies are consistent with both an unchanged IMF and a moderate flattening to $\alpha_3=2.0$.}
    \label{fig:imf}
\end{figure}

A top-heavy IMF produces a larger fraction of massive stars and therefore more ionising photons at fixed UV luminosity. If the brightest galaxies owe their luminosity to a top-heavy IMF, they should have higher H$\alpha$-to-UV luminosity ratios and H$\alpha$ equivalent widths than other galaxies. However, recent bursts of star formation produce similar signatures. We therefore compare galaxies with similar recent SFHs to test whether the brightest galaxies show an additional excess in either observable. This test is conditional on SFHs inferred with a fixed Chabrier IMF. Because H$\alpha$ constrains \sfrten\ in these fits, a top-heavy IMF could partly appear as a higher burst ratio. Comparing galaxies at fixed recent SFH may therefore absorb part of the IMF signal. The predicted signals below are thus upper limits on the residual excess, and any evidence against a strongly top-heavy IMF may be overstated.

\ceri\ uses pre-computed simple stellar population grids with a fixed IMF, so we cannot vary the IMF directly in the fits. Instead, we use \textsc{FSPS} \citep{conroy_propagation_2009}, with the same MIST/MILES libraries, to predict the effect of a top-heavy IMF. We adopt Kroupa-type IMFs \citep{kroupa_variation_2001} with three slopes above one solar mass: $\alpha_3=2.3$, matching the high-mass slope of the \citet{Chabrier:2003aa} IMF used in our fits; $\alpha_3=2.0$ (moderately top-heavy); and $\alpha_3=1.7$ (strongly top-heavy).

For constant star formation, with no dust or escape of ionising photons, a flatter high-mass slope increases both the ionising photon production efficiency $\xi_{\rm ion}$ (ionising photons per unit UV luminosity) and the H$\alpha$ equivalent width. For the strongly top-heavy IMF, these increases reach 0.3 and 0.5~dex, respectively (bands in Fig.~\ref{fig:imf}).

We measure both observables independently of the SED fit. We calculate
\begin{equation}
    \xi_{\rm ion}
    = \frac{L_{{\rm H}\alpha}}
    {1.36\times10^{-12}\,{\rm erg}\,L_{\nu,1500}},
\end{equation}
from the observed H$\alpha$-to-UV luminosity ratio. The conversion assumes case-B recombination \citep{Leitherer:1995aa}. We correct H$\alpha$ for dust using the Balmer decrement and the UV using the UV slope--attenuation relation of \citet{Meurer:1999aa}. We also use the observed rest-frame ${\rm EW}({\rm H}\alpha+[{\rm N}\,\textsc{ii}])$, measured from the catalogue line flux and line-free photometric continuum.

To account for recent SFH variations, we subtract a quadratic fit in \bhun, $b_{10-20}$, and $\log t_{50}$ from each observable. We then examine the residuals as a function of \muv, testing whether the brightest galaxies have higher H$\alpha$/UV ratios or equivalent widths than expected for their recent SFHs. We use all galaxies with $\muv<-16$ and apply no selection weights (treatment N). Using only \bhun\ to calculate the residuals gives similar results (open symbols in Fig.~\ref{fig:imf}).

The brightest galaxies show no significant excess in either observable (Fig.~\ref{fig:imf}). For the 20 galaxies with $\muv<-21$ and H$\alpha$ measurements, both median residuals are consistent with zero, and the largest residual occurs at the bright end. The brightest galaxies are therefore consistent with both an unchanged IMF and a moderately top-heavy one. A strongly top-heavy IMF ($\alpha_3=1.7$) is disfavoured at only $0.9\sigma$ by the H$\alpha$ equivalent width, assuming approximately constant star formation over at least the past few tens of Myr, and at only $0.8\sigma$ by $\xi_{\rm ion}$. The evidence against this IMF is stronger under the continuity SFH prior ($1.6\sigma$) and similar under the hyperburst prior ($1.1\sigma$; Table~\ref{tab:robustness}).

Three further limitations apply. First, the brightest bin contains only 20 galaxies, limiting the precision of the test. A forward model using top-heavy SSP grids would improve the predicted signal but not the statistical uncertainty. Second, we neglect the IMF dependence of nebular line ratios and nebular continuum emission. Third, the nebular grids used by \ceri\ \citep{Byler:2017aa} do not include He\,\textsc{ii}\,$\lambda4686$, preventing us from using this additional IMF diagnostic. Given these limitations, the brightest galaxies do not require a top-heavy IMF, although a moderate flattening of the high-mass slope remains consistent with the data.

\bsp	

\label{lastpage}
\end{document}